\documentclass{JHEP3}
\usepackage{epsfig,multicol,bbm}
\usepackage{makeidx}
\usepackage{amsmath}
\usepackage{amsfonts}
\usepackage{amssymb}
\usepackage{graphicx}
\usepackage{accents}
\input{tcilatex}

\newcommand\fverb{\setbox\fverbbox=\hbox\bgroup\verb}
\newcommand\fverbdo{\egroup\medskip\noindent%
            \fbox{\unhbox\fverbbox}\ }
\newcommand\fverbit{\egroup\item[\fbox{\unhbox\fverbbox}]}
\newbox\fverbbox

\title{\textbf{ \textbf{ Mutation Symmetries in BPS Quiver Theories: \\ \hspace
{0.5cm} Building the BPS Spectra }}}
\author{ \hspace{0.3cm} El Hassan Saidi\\
   \small {1.Lab Of High Energy Physics, Modeling and Simulation, Faculty Of Science,
   \\ University Of Mohammed V-Agdal, Rabat, Morocco,}\\
   \small {2.Centre Of Physics and Mathematics, CPM-CNESTEN, Rabat,
Morocco,} \\
   E-mails: \email{h-saidi@fsr.ac.ma}}

 \preprint{LPHE Preprint: March 2012}
\abstract {We study the basic features of BPS quiver mutations in 4D
$\mathcal{N}=2$ supersymmetric quantum field theory with $G=ADE$
gauge symmetries.\ We show, for these gauge symmetries, that there
is an isotropy group $\mathcal{G}_{Mut}^{G}$ associated to a set of
quiver mutations capturing information about the BPS spectra. In the
strong coupling limit, it is shown that BPS chambers correspond to
finite and closed groupoid orbits with an isotropy symmetry group
$\mathcal{G}_{strong}^{G}$ isomorphic to the discrete
dihedral groups $Dih_{2h_{G}}$ contained in Coxeter$\left( G\right) $ with $%
h_{G}$ the Coxeter number of G. These isotropy symmetries allow to
determine the BPS spectrum of the strong coupling chamber; and give
another way to count the total number of BPS and anti-BPS states of
$\mathcal{N}=2$ gauge
theories. We also build the matrix realization of these mutation groups $%
\mathcal{G}_{strong}^{G}$ from which we read directly the
electric-magnetic charges of the BPS and anti-BPS states of
$\mathcal{N}=2$ QFT$_{4}$ as well as their matrix intersections. We
study as well the quiver mutation symmetries in the weak coupling
limit and give their links with infinite Coxeter groups. We show
amongst others that $\mathcal{G}_{weak}^{su_{2}}$ is contained in
${GL}\left( {2,}\mathbb{Z}\right) $; and isomorphic to the
infinite Coxeter ${I_{2}^{\infty }}$. Other issues such as building $\mathcal{G}%
_{weak}^{so_{4}}$ and $\mathcal{G}_{weak}^{su_{3}}$ are also
studied.} \keywords{Electric magnetic duality in $\mathcal{N}=2$
QFT$_{4}$, BPS quiver theory, Quiver mutations, Building BPS
spectra, groupoids}

\begin{document}

\tableofcontents

\section{Introduction}

Recently a BPS quiver theory has been proposed in
\textrm{\cite{1A,1B}} to build the full set of BPS spectra in 4D
$\mathcal{N}=2$ supersymmetric quantum field theory (QFT$_{4}$) with
rank $r$ gauge invariance $G_{r}$ given by the standard simply laced
ADE symmetries. These massive and charged protected states of the
Hilbert space of the $\mathcal{N}=2$ QFT$_{4}$, which are
undetermined by the low energy theory alone; are remarkably
described in the approach of \textrm{\cite{1A,1B}} relying on
quantum mechanical dualities \textrm{\cite{1C}. }These dualities are
encoded by quiver mutations relating distinct quivers living at
different regions of the parameter space of the BPS theory. The
originality of the quiver method is that it gives a new way to deal
with general BPS states across the Coulomb branch $\mathcal{U}$; and
proposes an explicit algorithm to deduce
them. A key ingredient in the theory of \textrm{\cite{1A,1B}} is \emph{%
quiver mutations} essentially based on the two following things: \

\  \  \  \newline (\textbf{1}) start from the "\emph{primitive}" BPS
quiver, referred here to as $\mathfrak{Q}_{0}^{G}$, living at a
point $u$ of the Coulomb branch, and
made of the r \emph{elementary monopoles }$\left( \mathfrak{M}_{1},...,%
\mathfrak{M}_{r}\right) $ and the $r$ \emph{elementary dyons}
$\left( \mathfrak{D}_{1},...,\mathfrak{D}_{r}\right) $ of the
effective low-energy
solutions of $\mathcal{N}=2$ supersymmetric gauge theory \textrm{\cite{2A,2B}%
; }see also\textrm{\  \cite{02A}-\cite{3I}. }The\textrm{\ }$\mathfrak{M}_{i}$%
's and $\mathfrak{D}_{i}$'s are thought of as the elementary
building block of the BPS spectrum; and form a positive integral
basis of the entire BPS spectrum. These elementary particle
hypermultiplets have complex central
charges Z respectively denoted here as $X_{i}$ and $Y_{i}$ with $\func{Im}%
X_{i}>0,$ $\func{Im}Y_{i}>0$; their absolute values $\left \vert
X_{i}\right \vert $ and $\left \vert Y_{i}\right \vert $\ give the
masses of the BPS particles and their arguments $\arg X_{i}$, $\arg
Y_{i}$ for distinguishing BPS chambers. \newline The
$\mathfrak{M}_{i}$ and $\mathfrak{D}_{i}$ particle states live in
the upper half plane $\mathcal{H}$ ($\func{Im}Z>0$ of the complex
$Z$ plane); and have electric-magnetic (EM) charges respectively
given by the $2r$ components vectors $b_{i}=\left( 0,\alpha
_{i}\right) ,$ $c_{i}=\left( \alpha _{i},-\alpha _{i}\right) $ with
intersections $b_{i}\circ c_{j}$ remarkably encoded by the Cartan
matrix $K_{ij}=\alpha _{i}.\alpha _{j}$ of the Lie algebra of the
ADE gauge symmetries.

\  \  \  \newline (\textbf{2}) perform \emph{mutation
transformations} acting on the EM
charges $b_{i}$ and $c_{i}$ mapping therefore the \emph{primitive} $%
\mathfrak{Q}_{0}^{G}$ into mutated quivers $\mathfrak{Q}_{n}^{G}$,
$n\geq 0$
describing the same physics as $\mathfrak{Q}_{0}^{G}$. From the obtained $%
\mathfrak{Q}_{n}^{G}$s we can learn the EM charges of the BPS states of the $%
\mathcal{N}=2$ supersymmetric QFT$_{4}$. In \textrm{\cite{1A,1B},
t}he quiver mutations have been realized as rotations $\mathcal{H}$
$\rightarrow $ $e^{-i\theta }\mathcal{H}$ in the half plane of the
complex central charges; and have been interpreted \ in terms of
quantum mechanical dualities
relating the BPS spectra of the quivers $\left \{ \mathfrak{Q}%
_{n}^{G}\right \} $. This family of quivers lead to a chamber of BPS
states with EM charges given by positive integer combinations of the
charges of the elementary monopoles $b_{i}$ and dyons $c_{i}$.

\  \  \  \newline In this paper, we contribute to this matter by
studying\emph{\ the algebraic structure of the} \emph{quiver
mutations} which we use to get the BPS spectra of the
$\mathcal{N}=2$ QFT$_{4}$; and to interpret results obtained in
\textrm{\cite{1A,1B}}. This algebraic approach may be also viewed as
a way to deal with the complexity of the BPS content of the infinite
weak coupling chambers of these supersymmetric QFT$_{4}$s.\
\newline
More precisely, if thinking about generic BPS quivers $\mathfrak{Q}_{n}^{G}$%
, $n\in \mathbb{N}$, of the pure\textrm{\footnote{%
extension to implement fundamental matter is also possible; but it
is not considered here; see \textrm{\cite{030I}}.}} ADE gauge
theories as given by the pair
\begin{equation}
\begin{tabular}{llll}
$\mathbf{v}^{\left( n\right) }=\left( b^{\left( n\right) },c^{\left(
n\right) }\right) $ & , & $\mathcal{A}^{\left( n\right) }=v^{\left(
n\right) }\circ v^{\left( n\right) }$ &
\end{tabular}
\label{p}
\end{equation}%
with the $b_{i}^{\left( n\right) },$ $c_{i}^{\left( n\right) }$ entries of $%
\mathbf{v}^{\left( n\right) }$ describing the nodes of
$\mathfrak{Q}_{n}^{G}$
and $\mathcal{A}^{\left( n\right) }$ their intersection matrix, the \emph{%
successive} quiver mutations $\mathfrak{Q}_{0}^{G}$ $\rightarrow $ $%
\mathfrak{Q}_{1}^{G}$ $\rightarrow $ ... $\rightarrow
\mathfrak{Q}_{n}^{G}$,
with positive integer n, can be thought of as particular morphisms $\mathcal{%
M}_{n}$ that can be then realized in terms\ of invertible matrices $\mathcal{%
M}_{n}$ belonging to a set $\mathcal{G}_{Mut}^{G}$ $\subset GL\left( 4r^{2},%
\mathbb{Z}\right) $ and acting on the known elementary BPS quiver
$\left \{ \mathbf{v}^{\left( 0\right) },\mathcal{A}^{\left( 0\right)
}\right \} $ as follows
\begin{equation*}
\begin{tabular}{llll}
$\mathbf{v}^{\left( n\right) }=\mathcal{M}_{n}\mathbf{v}^{\left(
0\right) }$ & , & $\mathcal{A}^{\left( n\right)
}=\mathcal{M}_{n}\mathcal{A}^{\left( 0\right) }\mathcal{M}_{n}^{T}$
&
\end{tabular}%
\end{equation*}%
Clearly, the big $GL\left( 4r^{2},\mathbb{Z}\right) $ is not a
symmetry of the BPS quiver theory; only a particular subset
$\mathcal{G}_{Mut}^{G}$ of
it, to be explicitly built later, which is a symmetry. In this way of doing%
\emph{\ the data} about quiver mutations; in particular the ones
concerning the strong coupling BPS chamber considered in sections 3,
4, 5 and appendix II, are captured by the primitive
$\mathfrak{Q}_{0}^{G}$ and the set $\left \{ \emph{M}_{n};\text{
}n\in \mathbb{N}\right \} $ whose determination and their algebraic
properties are therefore of major interest.

\  \  \  \newline Now, if denoting by $L_{1}$ the mutation
transformation that maps the
elementary BPS quiver $\mathfrak{Q}_{0}^{G}$ into the mutated quiver $%
\mathfrak{Q}_{1}^{G}$, and by $L_{2}$ the mutation transformation mapping $%
\mathfrak{Q}_{1}^{G}$ into the mutated quiver
$\mathfrak{Q}_{2}^{G}$; and in
general by $L_{n}$ the mutation transformation that maps $\mathfrak{Q}%
_{n-1}^{G}$ into the mutated $\mathfrak{Q}_{n}^{G}$, the structure
of the
BPS spectra in the \emph{chamber based on} $\mathfrak{Q}_{0}^{G}$, \emph{%
generated by successive }$L_{n}$\emph{\ actions}, should be encoded
in the following typical set
\begin{equation}
\mathcal{G}_{Mut}^{G}=\left \{ \mathcal{M}_{m,0}:\mathfrak{Q}%
_{0}^{G}\rightarrow \mathfrak{Q}_{m}^{G};\text{ \  \ }m\in
\mathbb{N}\right \} \label{1}
\end{equation}%
with $\mathcal{M}_{m,0}\equiv \mathcal{M}_{m}$ realized by the
morphisms
product%
\begin{equation}
\mathcal{M}_{m}=L_{m}L_{m-1}....L_{2}L_{1}  \label{2}
\end{equation}%
In the above relation, the matrix generators $L_{n}$ are given by
some involution operators to be given later on; their expression
depends on the BPS chamber we are dealing with; i.e: strong or weak
coupling regimes. They
satisfy amongst others,%
\begin{equation}
\begin{tabular}{lll}
$\left( L_{n}\right) ^{2}$ & $=I_{id},\text{ \  \ }\forall n$ &  \\
$L_{0}$ & $\equiv I_{id}=\mathcal{M}_{0}$ &
\end{tabular}
\label{33}
\end{equation}%
Along with the constraints (\ref{33}), physical considerations
require also
that the mutation transformations $\mathcal{M}_{m}$ of the quiver $\mathfrak{%
Q}_{0}^{G}$ have to be invertible; and then (\ref{1}) should appear
as a set
of discrete symmetries of the BPS quiver theory of \textrm{\cite{1A,1B} }%
that turns out to share basic features of a groupoid\textrm{{\textrm{%
\footnote{%
We thank the referee for pointing to us this remarkable property;
the set of quiver mutations has a groupoid structure and the BPS
chambers are groupoid
orbits as exhibited in appendix I.}}}. }Indeed, the particular realization (%
\ref{1}) recalls the composition law of arrows $hg$ $\left( =\mathcal{M}%
_{m,n}\mathcal{M}_{n,0}\text{ in our study}\right) $ placing head of
the
second arrow $\left( h=\mathcal{M}_{m,n}\right) $ to tail of the first one $%
\left( g=\mathcal{M}_{n,0}\right) $; this feature shows that in
general the set of quiver mutations $\mathcal{G}_{Mut}^{G}$ is a
\emph{groupoid }and the BPS chambers are \emph{groupoid orbits}.
This somehow exotic structure is almost a group structure; except it
is defined only for certain pairs of elements as in the
eq(\ref{21}); see appendix I for explicit details.\ Notice in
passing that groupoids are also believed to describe physical
symmetries, often thought of as synonymous of groups and their
representations. The special class of Lie groupoids, involving
smooth manifolds, have been used in many occasions in physical
literature; in particular in dealing with the moduli space of flat
connections in
2-dimensional topological field theory\textrm{\  \cite{3II}; }see also\textrm{%
\  \cite{31II,32II,33II} }and refs therein\textrm{.}

\  \  \  \  \newline Furthermore, as far as the strong and the weak
coupling chambers of the BPS quiver theory are concerned, a careful
analysis of the set \textrm{(\ref{1})}
shows that BPS chambers are described by groupoid orbits defined by eqs(\ref%
{O}-\ref{S}); and allows to distinguish between two possible
situations:

\  \  \  \newline
\textbf{a}) either the groupoid orbit is finite, closed and having an \emph{%
isotropy symmetry group}\textrm{\ }$\mathbb{G}_{x}^{x}$ as given by eq(\ref%
{S}); the closure property is ensured by requiring a periodicity of
the successive mutations restricting therefore the set
$\mathbb{G}_{x}^{y}$ defined by (\ref{TT}) to $\mathbb{G}_{x}^{x}$.
This means that, if for instance we start from
$x=\mathfrak{Q}_{0}^{G}$, there exists a positive integer $N$ such
that\ the mutation\ transformation $\mathcal{M}_{N}$ is
equal to the identity map $I_{\mathfrak{Q}_{0}^{G}}$; i.e:%
\begin{equation*}
\mathcal{M}_{N}=I_{id},\qquad \mathcal{M}_{m+kN}=\mathcal{M}_{N}
\end{equation*}%
for any positive integer $k$. This constraint on $\mathcal{M}_{N}$
captures the property of mapping of the quiver
$\mathfrak{Q}_{0}^{G}$ into itself after performing $N$ successive
mutations; and allows moreover to determine
the expressions of the inverse of the $\mathcal{M}_{m,0}$'s by help of eq(%
\ref{2}) and which are nothing but $\mathcal{M}_{0,m}$. Moreover,
one expects that the order of the isotropy group
$\mathcal{G}_{Mut}^{G}$ depends on $N$ which in turns depends on the
nature of the ADE symmetry; see below eq(\ref{DR}).

\  \  \  \newline \textbf{b}) or the groupoid orbit is infinite;
that is the number $N$ of the mutation transformations
$\mathcal{M}_{m}$ is infinite. In this case the
quiver mutations are given by the set of arrows $\mathbb{G}_{x}^{y}$ of eq(%
\ref{TT}) with $x=\mathfrak{Q}_{0}^{G}$ and $y=\mathfrak{Q}_{\infty
}^{G}$.
Here, there is no cyclic property\textrm{\footnote{%
the groupoid orbit of the weak coupling chamber in the SU$\left(
2\right) $
model is somehow particular in the sense that it has an isotropy group $%
\mathbb{G}_{x}^{x}\equiv \mathcal{G}_{weak}^{su_{2}}$. This group
can be
thought of as given by the combination of the infinite open orbit $\mathbb{G}%
_{x}^{y}$ with $x=\mathfrak{Q}_{0}^{G}$, $y=\mathfrak{Q}_{\infty
}^{G}$; and the reciprocal $\mathbb{G}_{y}^{x}$ respectively
associated with left and right mutations. The composition of these
two open orbits leads to an infinite; but closed orbit with isotropy
group $\mathbb{G}_{x}^{x}\sim
\mathbb{G}_{x}^{\infty }\cup \mathbb{G}_{\infty }^{x}$ as shown on fig \ref%
{sw}.}} and so one expects that the quiver transformation $\mathcal{M}%
_{\infty }$ corresponding to the limit
\begin{equation}
\lim_{m\rightarrow \infty }\mathcal{M}_{m}=\mathcal{M}_{\infty }
\label{GM}
\end{equation}%
encodes some specific data on the infinite weak coupling chamber of the $%
\mathcal{N}=2$ QFT$_{4}$ with ADE symmetries. It happens that the limit $%
m\rightarrow \infty $ tends to degenerate morphisms that correspond
in the case of an $SU\left( 2\right) $ gauge symmetry, and fortiori
for any ADE gauge symmetry, exactly to the $W^{\pm }$ gauge
particles in the construction of \textrm{\cite{1A,1B}}. This
property will be studied in sections 6 and 7; see also figures
15-17.

\  \  \  \  \newline In the first part of the present study, we
focus on closed groupoid orbits; in particular on their isotropy
groups $\mathbb{G}_{x}^{x}$, to which we refer to as
$\mathcal{G}_{strong}^{G}$, and on the corresponding BPS content of
the $\mathcal{N}=2$ QFT$_{4}$ with an arbitrary ADE gauge symmetry.
Then,
we consider the case of the infinite set $\mathbb{G}_{x}^{y}$, with $x=%
\mathfrak{Q}_{0}^{G}$, $y=\mathfrak{Q}_{\infty }^{G}$ and refered
below to as $\mathcal{G}_{weak}^{G}$, and study three examples of
supersymmetric QFTs with spontaneously broken gauge symmetries. To
that purpose, we build matrix
realizations of the various mutations $\mathcal{M}_{m}$ for all $\mathcal{G}%
_{strong}^{G}$ with $G=ADE$ from which we determine directly the BPS
electric-magnetic charges that appear as the rows of the
$\mathcal{M}_{m}$ matrices. We show amongst others that the total
number of the BPS states in
the strong coupling chamber of a gauge symmetry G is given by%
\begin{eqnarray}
&&%
\begin{tabular}{l|l|l|l|l}
{\small G} & {\small rank}$_{{\small G}}$ & \  \ $\mathcal{G}_{{\small strong}%
}^{G}$ \  & $\left \vert \mathcal{G}_{{\small strong}}^{G}\right
\vert $ & {\small nbr of BPS + anti-BPS} \\ \hline {\small
SU}$_{r+1}$ & $r$ & $Dih_{2\left( r+1\right) }$ & ${\small 4}\left(
{\small r+1}\right) $ & ${\small 4r}\left( r+1\right) \left.
\begin{array}{c}
\\
\end{array}%
\right. $ \\
{\small SO}$_{2r}$ & $r$ & $Dih_{2\left( r-1\right) }$ & ${\small \
4}\left( {\small r-1}\right) $ & ${\small 4r}\left( r-1\right)
\left.
\begin{array}{c}
\\
\end{array}%
\right. $ \\
{\small E}$_{6}$ & $6$ & $Dih_{12}$ & ${\small \  \ 24}$ & ${\small
6\times 24=72+72}\left.
\begin{array}{c}
\\
\end{array}%
\right. $ \\
{\small E}$_{{\small 7}}$ & $7$ & $Dih_{18}$ & ${\small \  \ 36}$ & ${\small %
7\times 36=126+126}\left.
\begin{array}{c}
\\
\end{array}%
\right. $ \\
{\small E}$_{{\small 8}}$ & $8$ & $Dih_{30}$ & ${\small \  \ 60}$ & ${\small %
8\times 60=240+240}\left.
\begin{array}{c}
\\
\end{array}%
\right. $ \\ \hline
\end{tabular}
\label{DR} \\
&&  \notag
\end{eqnarray}%
with $\left \vert \mathcal{G}_{strong}^{G}\right \vert =2h$; twice
the Coxeter number of G. The order $2h=h+h$ captures the fact that
BPS chambers contains $h$ BPS and $h$ anti-BPS states and so are CPT
invariant. \newline In the case of infinite weak coupling chambers,
we build the mutation groupoids $\mathcal{G}_{weak}^{su_{2}}$ and
$\mathcal{G}_{weak}^{so_{4}}$ as
well as $\mathcal{G}_{weak}^{su_{3}}$ respectively associated with $\mathcal{%
N}=2$ QFT with SU$\left( 2\right) ,$ SO$\left( 4\right) $ and
SU$\left(
3\right) $ gauge symmetries. We show moreover that $\mathcal{G}%
_{weak}^{su_{2}}$ is isomorphic to the infinite Coxeter group
$I_{2}\left(
\infty \right) $; the groupoid $\mathcal{G}_{weak}^{so_{4}}$ to the limits $%
k,l\rightarrow \infty $ of the direct product $I_{2}\left( k\right)
\times I_{2}\left( l\right) $; and $\mathcal{G}_{weak}^{su_{3}}$ to
a generalization involving the infinite limits $k,l,n\rightarrow
\infty $ of 3 positive integers.

\  \  \  \  \  \newline The organization of this paper is as
follows: In section 2, we review some useful aspects on low energy
properties of $\mathcal{N}=2$ gauge theories. In section 3, we study
the BPS quiver theory of $\mathcal{N}=2$ QFT$_{4}$'s with gauge
symmetry G. To illustrate the construction, we focus on the
particular example $G=SU\left( 3\right) $; but the method applies
for any
ADE gauge symmetry. In sections 4, we study the properties of $\mathcal{G}%
_{strong}^{su_{N}}$ describing quiver mutations in the strong
coupling limit of $\mathcal{N}=2$ QFT$_{4}$ with $SU\left( N\right)
$ gauge symmetries. We construct the matrix realization of
$\mathcal{G}_{strong}^{su_{N}}$ and give
the link with the BPS spectra and the class of finite Coxeter groups $%
Dih_{2N}$. In section 5, we give the extension of the results
obtained in sections 4 to the SO$\left( 2N\right) $\ and the
exceptional E$_{r}$ gauge symmetries. In section 6, we study the
weak coupling limit of two examples
of supersymmetric QFTs with spontaneously broken SU$\left( 2\right) $ and SO$%
\left( 4\right) $ gauge symmetries. In section 7, we study the weak
coupling chamber of supersymmetric SU$\left( 3\right) $ gauge
theory; and in section 8, we give a conclusion and a comment. The
last sections are devoted to 4
appendices. In appendix I, we study the groupoid structure of $\mathcal{G}%
_{Mut}^{G}$ given by the set (\ref{1}). In appendix II, we extend
the results of section 3 on strong coupling of chamber to the
$SU\left( N\right) $ models. In appendix III, we give technical
details concerning the strong coupling chamber of the $E_{6}$ gauge
theory; and in appendix IV, we give some useful tools on Coxeter
groups.

\section{Low energy properties of $\mathcal{N}=2$ gauge theories}

The field content of the 4 dimensional pure $\mathcal{N}=2$
supersymmetric gauge theories includes in addition to the gauge
field $\mathcal{A}_{\mu }$, a complex scalar field $\phi $ that
plays a basic role here; and two chiral spinors $\lambda $ and $\psi
$, the superpartners. These are field matrices valued in the adjoint
representation of the rank $r$ gauge symmetry $G_{r}$ of the theory;
and so can be expanded like
\begin{equation}
\phi =\sum_{i=1}^{r}\phi _{i}h_{i}+\sum_{\beta \in \Delta }\phi
_{\beta }E_{\beta }  \label{EX}
\end{equation}%
with similar expansions for the other fields. In this relation
$\Delta $ is the set of roots of the gauge symmetry $G_{r}$, and
$h_{i}$, $E_{\beta }$ are respectively the usual Cartan charges and
the step operators generating the Lie algebra of $G_{r}$.

\subsection{Coulomb branch and residual symmetries}

Supersymmetric background solutions of this gauge theory are
obtained by solving the vanishing condition of the scalar potential
$\mathcal{V}\left(
\phi \right) $ of theory. As this potential is proportional to the trace of $%
\left[ \phi ,\phi \right] ^{2}$, the general supersymmetric solution
is
given by%
\begin{equation}
\left \langle \phi \right \rangle
=\sum_{i=1}^{r}a_{i}h_{i}=\vec{a}.\vec{h}
\end{equation}%
with complex numbers $a_{i}$ interpreted as the local coordinates of
the Coulomb branch $\mathcal{U}$ of the supersymmetric gauge theory.
On this branch, the scalar field matrix $\phi $ can develop
expectation values in the supersymmetric vacuum which break the
gauge symmetry spontaneously. For generic moduli $a_{i}$, the gauge
symmetry $G_{r}$ is broken down to the abelian Cartan sector
\begin{equation}
G_{r}\rightarrow U^{r}\left( 1\right)  \label{GB}
\end{equation}%
Along with this continuous and abelian residual symmetry, there is
moreover
a discrete group of gauge transformations containing the Weyl group $%
\mathcal{W}\left( G_{r}\right) $. The latter is generated by the
reflections
$R_{\beta }$ that act on the scalar moduli as follows%
\begin{equation}
R_{\beta }\left( a\right) =a-\left( \beta ^{v}.a\right) \beta
\end{equation}%
with $\beta $ a root and $\beta ^{v}=\frac{1}{2\beta ^{2}}\beta $
the corresponding coroot. For the case of simply laced Lie algebras
$\beta ^{2}=2 $, the above relation reduces to $R_{\beta }\left(
a_{i}\right) =R_{ij}a_{j}$ with
\begin{equation}
R_{ij}=\delta _{ij}-\beta _{i}\beta _{j}
\end{equation}%
In the perturbation theory region, the gauge breaking (\ref{GB})
generates masses for all degrees of freedom, except for those that
correspond to the residual invariance $U^{r}\left( 1\right) $. As a
result, there are $r$ massless Maxwell type supermultiplets
\begin{equation}
\left( \mathcal{A}_{\mu }^{i},\lambda ^{i},\psi ^{i},\phi
^{i}\right)
\end{equation}%
with no electric charges; but interacting with the electrically
charged objects of the gauge theory. The other fields of the
expansions (\ref{EX}) namely the $\mathcal{A}_{\mu }^{\alpha },$
$\lambda ^{\alpha },$ $\psi ^{\alpha },$ $\phi ^{\alpha }$
associated with roots of G have now acquired masses.

\subsection{Low energy properties}

The low energy properties of the supersymmetric spontaneously broken
gauge theory are described by one holomorphic function on the
Coulomb branch namely the prepotential
$\mathcal{F}=\mathcal{F}\left( a\right) $. For large
values of the scalar moduli $a_{i}$ with respect to some cut off parameter $%
\Lambda $; i.e: $\left \vert \vec{a}\right \vert >>\Lambda $, the
prepotential reads as
\begin{equation}
\mathcal{F}\left( a_{1},...,a_{r}\right) \simeq \frac{i}{8\pi
}\sum_{\alpha
\in \Delta }\left( \vec{\alpha}.\vec{a}\right) ^{2}\ln \frac{\left( \vec{%
\alpha}.\vec{a}\right) ^{2}}{\Lambda ^{2}}
\end{equation}%
This function play a major role in the study of low energy effective
supersymmetric gauge theory; and allows to define the dual moduli
\begin{equation}
\tilde{a}_{i}=\frac{\mathcal{F}\left( a\right) }{\partial a_{i}}
\end{equation}%
These complex numbers, which read explicitly as%
\begin{equation}
\tilde{a}_{i}\simeq \frac{i}{4\pi }\sum_{\alpha \in \Delta }\left( \vec{%
\alpha}.\vec{a}\right) \left( 1+\ln \frac{\left(
\vec{\alpha}.\vec{a}\right) ^{2}}{\Lambda ^{2}}\right) \alpha _{i}
\end{equation}%
are also holomorphic functions in the complex moduli
$a_{1},...,a_{r}$; and are generally used to express the complex
effective coupling constant
\begin{equation}
\tau _{ij}=\frac{\theta }{2\pi }+i\frac{4\pi }{g^{2}}=\frac{\partial ^{2}%
\mathcal{F}}{\partial a_{i}\partial a_{j}}
\end{equation}%
of the effective low energy theory like
\begin{equation}
\tau _{ij}=\frac{\partial \tilde{a}_{i}}{\partial a_{j}}
\end{equation}%
Before closing this section, notice the two following: first under
the
reflections $R_{\beta }$, the dual moduli transform as $\tilde{a}%
_{i}\rightarrow R_{\beta }\left( \tilde{a}_{i}\right) +\left( \beta
.a\right) \beta $ showing that $\tilde{a}_{i}$ has a monodromy due
to the
log dependence. Second in the large limit approximation of the moduli ($%
\left \vert \vec{a}\right \vert >>\Lambda $), we have $\ln \left( \vec{\alpha%
}.\vec{a}\right) ^{2}\simeq \ln \left( \vec{a}\right) ^{2}$. So the
dual
moduli $\tilde{a}_{i}$ can be put into the form%
\begin{equation}
\tilde{a}_{i}\simeq \frac{i}{2\pi }\tilde{h}\ln \frac{a^{2}}{\Lambda ^{2}}%
\text{ }a_{i}
\end{equation}%
with $\tilde{h}$ the dual Coxeter number defined as $\sum_{\alpha
\in \Delta }\alpha ^{i}\alpha ^{j}=2\tilde{h}\delta ^{ij}$.\ The
above relation is
important in the sense it leads to the diagonal expression%
\begin{equation}
\tau _{ij}\simeq 2\tilde{h}\tau \text{ }\delta _{ij},\qquad \tau =\frac{i}{%
4\pi }\ln \frac{a^{2}}{\Lambda ^{2}}
\end{equation}%
showing moreover that there is only one coupling constant $\tau $
with asymptotic behavior governed by twice the dual Coxeter number.

\section{Central charges in BPS quiver theory}

In this section, we describe basic properties of the elementary BPS
states in the $\mathcal{N}=2$ supersymmetric quantum field theories
in 4D space
time and in BPS quiver theory. This study will be illustrated on the $%
SU\left( 3\right) $ example; but applies to the full set of
$\mathcal{N}=2$ QFT$_{4}$ with finite dimensional ADE gauge
symmetries.

\subsection{BPS quivers: example of $SU\left( 3\right) $ model}

In BPS quiver theory of the $\mathcal{N}=2$ supersymmetric QFT with a $%
SU\left( 3\right) $ gauge symmetry \textrm{\cite{1A,1B}}, one deals
with many quivers belonging to several kinds of BPS chambers. These
quivers are defined at a generic point $u=\left( u_{1},u_{2}\right)
\in \mathbb{C}^{2}$ of the Coulomb branch of the gauge theory where
the $SU\left( 3\right) $ gauge symmetry is spontaneously broken down
to
\begin{equation}
U\left( 1\right) \times U\left( 1\right)
\end{equation}%
The electric magnetic duality together with the primitive BPS quiver $%
\mathfrak{Q}_{0}^{su_{3}}$ and its mutations are basic things that
play a central role in building the full BPS spectra of this
supersymmetric gauge theory. Below, we study these things.

\subsubsection{Monopoles and dyons}

Following Seiberg-Witten approach \textrm{\cite{2A,2B}}, the low
energy properties of the $\mathcal{N}=2$ supersymmetric theory at
strong coupling are described by light monopoles and light dyons.
The latters have both electric $\vec{q}$ and magnetic $\vec{g}$
charges; and then interact with the supersymmetric gauge field
multiplets $\left( \mathcal{A}_{\mu
}^{i},\lambda ^{i},\psi ^{i},\phi ^{i}\right) $ of the spontaneously broken $%
SU\left( 3\right) $ gauge theory.\ The $\vec{q}$ and $\vec{g}$
charges of these particles, believed to be BPS states of
$\mathcal{N}=2$ supersymmetry, are described by 2-dimensional
vectors respectively lying in the root and coroot lattices of the
Lie algebra of SU$\left( 3\right) $. We have
\begin{equation}
\begin{tabular}{ll}
$\vec{q}$ & $=m_{1}\vec{\alpha}_{1}+m_{2}\vec{\alpha}_{2}$ \\
$\vec{g}$ & $=n_{1}\vec{\alpha}_{1}^{v}+n_{2}\vec{\alpha}_{2}^{v}$%
\end{tabular}
\label{HM}
\end{equation}%
with $\vec{\alpha}_{1}$, $\vec{\alpha}_{2}$ the two simple roots of SU$%
\left( 3\right) $; and $\vec{\alpha}_{i}^{v}=\frac{2}{\alpha
_{i}^{2}}\alpha _{i}$ the two coroots which in present case coincide
precisely with $\alpha _{i}$ due to the relation $\alpha
_{i}^{2}=2$. These EM charges obey the Dirac-Schwinger-Zwanziger
condition which states that magnetic $\vec{g}_{i}$ and electric
$\vec{q}_{i}$ charges of any two dyons satisfy the following
quantization condition%
\begin{equation}
\vec{q}_{1}.\vec{g}_{2}-\vec{q}_{2}.\vec{g}_{1}=\vec{\gamma}_{1}\Omega \vec{%
\gamma}_{2}\in \mathbb{Z}  \label{EM}
\end{equation}%
with%
\begin{equation*}
\Omega =\left(
\begin{array}{cc}
0 & I \\
-I & 0%
\end{array}%
\right)
\end{equation*}%
The 4- dimensional vectors%
\begin{equation}
\vec{\gamma}_{i}=\left(
\begin{array}{c}
\vec{q}_{i} \\
\vec{g}_{i}%
\end{array}%
\right)
\end{equation}%
stand for generic charge vectors in the EM lattice $\Gamma
_{su_{3}}^{em}$ of the $\mathcal{N}=2$ supersymmetric QFT$_{4}$ with
$SU\left( 3\right) $\ gauge symmetry. As we are dealing with a
$\mathcal{N}=2$ supersymmetric pure gauge theory, we will refer to
the EM charges of the two elementary monopoles and the two
elementary dyons of this theory respectively as $b_{i}$ and $c_{i}$
with
\begin{equation}
\begin{tabular}{ll}
$b_{1}=\left(
\begin{array}{c}
0 \\
\alpha _{1}%
\end{array}%
\right) $, & $b_{2}=\left(
\begin{array}{c}
0 \\
\alpha _{2}%
\end{array}%
\right) $%
\end{tabular}
\label{bc}
\end{equation}%
and%
\begin{equation}
\begin{tabular}{ll}
$c_{1}=\left(
\begin{array}{c}
\alpha _{1} \\
-\alpha _{1}%
\end{array}%
\right) ,$ & $c_{2}=\left(
\begin{array}{c}
\alpha _{2} \\
-\alpha _{2}%
\end{array}%
\right) $%
\end{tabular}
\label{cb}
\end{equation}%
These EM charges extend those of the case of $\mathcal{N}=2$\
QFT$_{4}$ with spontaneously broken $SU\left( 2\right) $ gauge
symmetry of the elementary monopole and elementary dyon reads as
\begin{equation}
\begin{tabular}{lll}
$b=\left( 0,\alpha \right) $ & , & $c=\left( \alpha ,-\alpha \right) $%
\end{tabular}%
\end{equation}%
Below, we denote the electric-magnetic product $\vec{\gamma}_{1}\Omega \vec{%
\gamma}_{2}$ like $\gamma _{1}\circ \gamma _{2}$; so for SU$\left(
2\right) $ theory this symplectic product reads as $b\circ c$ and is
equal to $-\alpha ^{2}=-2$. In the case of SU$\left( 3\right) ,$ we
have $b_{i}\circ c_{j}=-K_{ij}$ with $K_{ij}$ the $2\times 2$ Cartan
matrix; the same thing is valid for ADE gauge symmetries.

\subsubsection{The primitive quiver $\mathfrak{Q}_{0}^{su_{3}}$}

Among the special features of the primitive
$\mathfrak{Q}_{0}^{su_{3}}$ of the BPS quiver theory of
$\mathcal{N}=2$\ QFT$_{4}$ with broken SU$\left( 3\right) $ gauge
invariance, we mention the three following:

\begin{itemize}
\item it involves only elementary BPS states: the two light monopoles and
the two light dyons with EM charges (\ref{bc}-\ref{cb}); we will
refer to it as the primitive BPS quiver.

\item it has no anti-BPS state; this property let understand that there
exist also a primitive anti-BPS quiver in any CPT invariant BPS
chamber; and should be generated by quiver mutations.

\item it is viewed as the leading element of sequence of BPS quivers
\begin{equation}
\mathfrak{Q}_{n}^{su_{3}},\text{ }n\in \mathbb{N}
\end{equation}%
related to the primitive $\mathfrak{Q}_{0}^{su_{3}}$ by mutation
transformations.
\end{itemize}

\  \  \  \newline These features are not specific for
$\mathfrak{Q}_{0}^{su_{3}}$; they are shared by all primitive
quivers $\mathfrak{Q}_{0}^{G}$ associated with any ADE gauge
symmetry.\newline The quiver $\mathfrak{Q}_{0}^{su_{3}}$ of the
supersymmetric pure SU$\left(
3\right) $ gauge theory is then made by 4 particular BPS particle states $%
\left \vert \mathrm{\vec{\gamma}}_{i}\right \rangle \equiv \left \vert \vec{p%
}_{i},\vec{g}_{i}\right \rangle $, $i=1,2,3,4$. Each state $\left
\vert
\mathrm{\vec{\gamma}}\right \rangle \equiv \left \vert \vec{p},\vec{g}%
\right \rangle $ is described by a massive supersymmetric short
multiplet preserving four supercharges of the underlying
$\mathcal{N}=2$ superalgebra. It carries an electric charge vector
$\vec{p}=\left( q_{1},q_{2}\right) $ and magnetic one
$\vec{g}=\left( g_{1},g_{2}\right) $; and has a mass
\textrm{m}$_{\gamma }$ that depends on the EM charges and on the
VEVs moduli u; i.e:
\begin{equation}
\mathrm{m}_{\gamma }=\mathrm{m}\left( q,g;u,\bar{u}\right)
\end{equation}%
In SU$\left( 3\right) $ gauge theory, the masses of the two monopoles $%
\left
\{ \mathfrak{M}_{1},\mathfrak{M}_{2}\right \} $ and two dyons $%
\left \{ \mathfrak{D}_{1},\mathfrak{D}_{2}\right \} $ are given by
the absolute value of the complex central charge $Z_{u}\left( \gamma
_{i}\right) $ saturating the BPS bound of the 4D $\mathcal{N}=2$
supersymmetric algebra. Geometrically, these central charges are
realized as complex integrals like \textrm{\cite{1A},}
\begin{equation}
Z_{u}\left( \mathrm{\gamma }_{i}\right) =\int_{\mathrm{\gamma
}_{i}}\lambda _{u},\qquad \left \vert Z_{u}\left( \mathrm{\gamma
}_{i}\right) \right \vert \sim \mathrm{m}_{\mathrm{\gamma }_{i}}
\label{z}
\end{equation}%
where here $\mathrm{\gamma }_{i}$ are thought of as 1-cycles of the
homology of some Riemann surface $\Sigma $; but can identified with
a the 4-component
charge vector $\gamma _{i}=\left( q_{i},g_{i}\right) $ in the EM lattice $%
\Gamma $ of the $\mathcal{N}=2$ quantum field theory. In (\ref{z}),
the
index $u$ carried by $Z_{u}\left( \gamma \right) $ and by the differential $%
\lambda _{u}$ is used to indicate that the complex central charge
$Z$ and Seiberg-Witten differential $\lambda $ are in fact
parametric functions depending on the coordinates of the Coulomb
branch of the moduli space of the $\mathcal{N}=2$ $SU\left( 3\right)
$ gauge theory. \newline The central charge $Z_{u}\left(
\mathrm{\gamma }\right) $ can be written in a more explicit manner
by using the VEVs $a_{1}$ and $a_{2}$ of the Higgs fields of the
spontaneously broken $SU\left( 3\right) $ gauge theory as well
as their symplectic duals $\tilde{a}_{i}$. We have\textrm{\footnote{%
In the presence of $N_{f}$ multiplets of fundamental matter with
Mass $M_{I}$ and with electric charges $q_{I}^{\prime }$, the
central charge $Z_{u}\left(
\gamma \right) $ of (\ref{zg}) gets an extra term $\sum_{i=1}^{N_{f}}q_{I}^{%
\prime }M_{I}.$}}
\begin{equation}
Z_{u}\left( \gamma \right) \sim \sum_{i=1}^{r}\left( q_{i}a_{i}-g_{i}\tilde{a%
}_{i}\right) ,\qquad r=2  \label{zg}
\end{equation}%
To fix the ideas, notice that the complex VEVs $a_{i}$ may be also
used to define the moduli of the theory, but up to identifications
under some
discrete monodromy symmetry. In fact, the $a_{i}$'s are related to the $%
u_{i} $ complex moduli by some relations $u_{i}=u_{i}\left( a\right)
$ that
can be inverted to $a_{i}=a_{i}\left( u\right) $ giving the dependence $%
Z=Z\left( u\right) $ exhibited on the left hand side of (\ref{zg});
for technical details see \textrm{\cite{3A}}. \newline The complex
central charge $Z_{u}\left( \gamma \right) $ exhibits some useful
features for studying the BPS states; in particular the three
following: First, $Z_{u}\left( \gamma \right) $ has a manifest
symplectic structure as shown on (\ref{zg}-\ref{EM}) and the
following expression
\begin{equation}
Z_{u}\left( \gamma \right) \sim \left( q,g\right) \left(
\begin{array}{cc}
0 & I \\
-I & 0%
\end{array}%
\right) \left(
\begin{array}{c}
a \\
\tilde{a}%
\end{array}%
\right)
\end{equation}%
Second, it is linear in the EM charges $\gamma $; so for a generic
vector charge $\gamma $ given by a linear combination of the charge
vectors $\gamma _{l}$ of the two elementary monopoles and two
elementary dyons namely
\begin{equation}
\gamma =\sum_{l=1}^{4}n_{l}\gamma _{l},\qquad n_{l}\in
\mathbb{Z}^{+}, \label{gn}
\end{equation}%
we have the linearity property
\begin{equation}
Z_{u}\left( \gamma \right) =\sum_{l}n_{l}Z_{u}\left( \gamma
_{l}\right) \label{zu}
\end{equation}%
Third, eqs(\ref{gn}-\ref{zu}) show that the BPS states $\left \vert
\gamma \right \rangle $ of the $\mathcal{N}=2$ supersymmetric theory
can be engineered by taking bound states of elementary BPS states
$\left \vert \gamma _{i}\right \rangle $ with central charges
$Z_{u}\left( \gamma _{i}\right) $. For later use, let us collect
below other useful features:

\  \

1) \emph{elementary BPS states}\newline
the EM charge vectors $\gamma $ of the monopoles $\left \{ \mathfrak{M}_{1},%
\mathfrak{M}_{2}\right \} $ and the dyons $\left \{ \mathfrak{D}_{1},%
\mathfrak{D}_{2}\right \} $ are respectively denoted as $\left \{
b_{1},b_{2}\right \} $ and $\left \{ c_{1},c_{2}\right \} $. The
entries of these charge vectors are as in eqs(\ref{bc}). \newline
The charge vectors $b_{i}$ and $c_{i}$ are then $4$- component
vectors with entries belonging to the root/coroot lattices of
$SU\left( 3\right) $; a property that makes the extension of the
present analysis to the supersymmetric field theories with ADE gauge
symmetries straightforward. \

\  \  \

\begin{figure}[tbph]
\begin{center}
\hspace{0cm} \includegraphics[width=10cm]{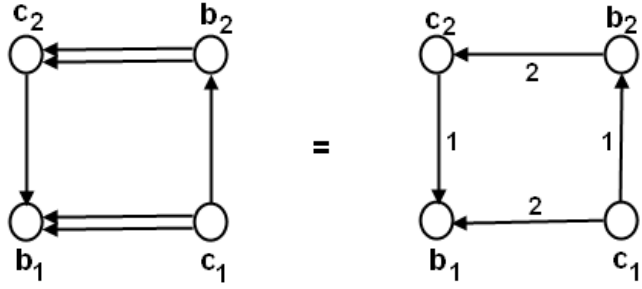}
\end{center}
\par
\vspace{-1 cm} \caption{the primitive quiver
$\mathfrak{Q}_{0}^{su_{3}}$ made completely of elementary BPS states
with central charges in the half plane $\mathcal{H}$. } \label{U3}
\end{figure}

2) \emph{the primitive quiver} $\mathfrak{Q}_{0}^{su_{3}}$\newline
the EM products of the charge vectors $b_{i}$ and $c_{i}$ (\ref{bc})
are given by
\begin{equation}
\begin{tabular}{lll}
$b_{i}\circ b_{j}=0$ & , & $c_{i}\circ b_{j}=+K_{ij}$ \\
$c_{i}\circ c_{j}=0$ & , & $b_{i}\circ c_{j}=-K_{ij}$%
\end{tabular}
\label{sp}
\end{equation}%
with $K_{ij}=\alpha _{i}.\alpha _{j}$ being the Cartan matrix of
$SU\left( 3\right) $. These relations describe the intersection
matrix of the BPS
states making the quiver $\mathfrak{Q}_{0}^{su_{3}}$ as depicted by \textrm{%
fig \ref{U3}}.

3) $\mathfrak{Q}_{0}^{su_{3}}$ \emph{as\ a symplectic
quiver}\newline
From the figure \textrm{\ref{U3}}, we learn that the graph $\mathfrak{Q}%
_{0}^{su_{3}}$ is completely specified by the charge vectors $\left(
b,c\right) $ and their EM products (\ref{sp}). Moreover, because of
the two following features:

\begin{itemize}
\item the linear dependence of the central charge into the EM charges of the
BPS states,

\item the appearance of the pair $\left( b_{i},c_{i}\right) $ of charge
vectors as a building block of $\mathfrak{Q}_{0}^{su_{3}}$,
\end{itemize}

\  \  \  \  \newline it is natural to group together these charges
into a large vector $\mathbf{v}
$ having $\left( 2r\right) ^{2}=16$ components\textrm{\footnote{%
For later use notice that in $SU\left( 3\right) $ the $b_{i}$'s and $c_{i}$%
's are real vectors with $4$ entries as in (\ref{bc}-\ref{cb}). For
a
generic rank r gauge symmetry, the number of the entries of each of the $%
b_{i}$'s and $c_{i}$'s is $2r$. So that the exact number of components of $%
v_{I}$ is $\left( 2r\right) ^{2}$. Below, we shall think of $v_{I}$
as a 2r vector with entries given by building blocks $\left(
b_{i},c_{i}\right) $ involving 2r component blocks. The details of
these sub-blocks are irrelevant for the study of mutation
symmetries, all we need to know is the
intersection matrix; see also eq(\ref{ps}).}} as follow%
\begin{equation}
\mathbf{v}_{I}=\left(
\begin{array}{c}
b_{1} \\
b_{2} \\
c_{1} \\
c_{2}%
\end{array}%
\right) ,\qquad r=2  \label{rep}
\end{equation}%
with $b_{i}$ and $c_{i}$ given by eq(\ref{bc}). In this way of
doing, the
quiver $\mathfrak{Q}_{0}^{su_{3}}$ can be then defined by the above vector $%
\mathbf{v}_{I}$ together with the antisymmetric intersection matrix $%
\mathcal{J}_{IJ}=\mathbf{v}_{I}\circ \mathbf{v}_{J}$ which, by using (\ref%
{sp}), reads as%
\begin{equation}
\mathcal{J}_{IJ}=\left(
\begin{array}{cc}
0 & -K_{ij}I_{4\times 4} \\
K_{ji}I_{4\times 4} & 0%
\end{array}%
\right)  \label{ps}
\end{equation}%
with $I_{4\times 4}$ the $4\times 4$ identity matrix and
\begin{equation}
K_{ij}=\left(
\begin{array}{cc}
2 & -1 \\
-1 & 2%
\end{array}%
\right)
\end{equation}%
Below, we denote the vector (\ref{rep}) by $\mathbf{v}^{\left(
0\right) }$ where the upper index refers to
$\mathfrak{Q}_{0}^{su_{3}}$. Later on we will also consider the
vectors
\begin{equation}
\mathbf{v}^{\left( n\right) }=\left(
\begin{array}{c}
b_{1}^{\left( n\right) } \\
b_{2}^{\left( n\right) } \\
c_{1}^{\left( n\right) } \\
c_{2}^{\left( n\right) }%
\end{array}%
\right) ,\qquad n\geq 0  \label{vn}
\end{equation}%
describing the mutated BPS quivers $\mathfrak{Q}_{n}^{su_{3}}$ that
are related to $\mathfrak{Q}_{0}^{su_{3}}$ by performing $n$
successive elementary mutations. \newline Notice that the matrix
representation (\ref{rep}) of $\mathbf{v}^{\left( 0\right) }$ is not
unique since it is defined up to permutations of the entries. So
there are different, but equivalent, choices of parameterizing the
vector $\mathbf{v}^{\left( 0\right) }$; each having an advantage and
a disadvantage; say a specific feature. For example the choice
(\ref{rep}) allows to exhibit explicitly the Cartan matrix in the
electric magnetic products as in (\ref{ps}), while the choice
\begin{equation}
\left(
\begin{array}{c}
b_{1} \\
c_{1} \\
b_{2} \\
c_{2}%
\end{array}%
\right)  \label{bb}
\end{equation}%
puts in front the $SU\left( 2\right) $ building block property of
the
quiver. In what follows, we consider the parametrization (\ref{rep}-\ref{ps}-%
\ref{vn}). The choice (\ref{bb}) will be used in section 6 and 7
when considering the weak coupling chamber.

\subsection{Quivers, Dynkin graphs and chambers}

BPS quivers in the $\mathcal{N}=2$ supersymmetric pure gauge theory
with a spontaneously broken gauge symmetry $G$ are intimately linked
to the Dynkin graph of $G$ depicted by the figures \ref{ADE}.
Moreover, as for Dynkin diagrams of simple Lie algebras, the
primitive $\mathfrak{Q}_{0}^{G}$'s of the BPS quiver theory have
also outer-automorphism symmetries that can be used other purposes
such as approaching the BPS spectra of $\mathcal{N}=2$ QFTs with non
simply laced gauge symmetries \textrm{\cite{4A,4B}}. As far BPS
quivers are concerned, these outer-automorphisms have to commute
with the mutation transformations.

\subsubsection{Link with Dynkin graphs}

The link between BPS quivers and Dynkin graphs of the underlying
gauge symmetries of the $\mathcal{N}=2$ supersymmetric QFT is
manifested through two channels: either by quiver folding to
compensate the orientation of BPS quivers; or by the shared
outer-automorphism symmetries.

\  \  \

\begin{figure}[tbph]
\begin{center}
\hspace{0cm} \includegraphics[width=10cm]{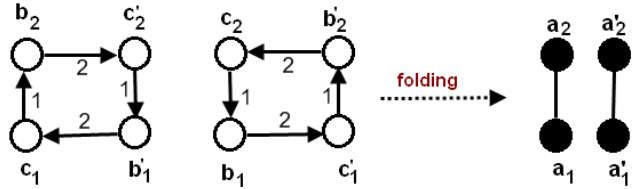}
\end{center}
\par
\vspace{-1 cm} \caption{the SU$\left( 3\right) $ Dynkin graph
obtained by folding the nodes of the BPS quiver
$\mathfrak{Q}^{su_{3}}$ interchanged by the outer automorphism
$\mathbb{Z}_{2}^{diag}$.} \label{23}
\end{figure}

\emph{folding}\newline Quiver folding is direct way to exhibit the
link between the BPS quivers of the $\mathcal{N}=2$ QFT and Dynkin
diagrams. In the case of the gauge symmetry $SU\left( 3\right) $,
the link is as depicted in fig \ref{23}.
There, one first considers two oriented primitive quivers $\mathfrak{Q}%
_{0}^{su_{3}}$ and $\mathfrak{P}_{0}^{su_{3}}$, with EM charges as follows:%
\begin{equation}
\begin{tabular}{ll|ll}
\multicolumn{2}{l}{$\  \  \  \  \mathfrak{Q}_{0}^{su_{3}}$ \ } &
\multicolumn{2}{|l}{$\  \  \  \  \mathfrak{P}_{0}^{su_{3}}$} \\
\hline
$\ b_{1},$ & $\ c_{1}^{\prime }$ & $\ b_{1}^{\prime },$ & $\ c_{1}$ \\
$\ b_{2}^{\prime },$ & $\ c_{2}$ & $\ b_{2},$ & $\ c_{2}^{\prime }$%
\end{tabular}%
\end{equation}%
then compensate the quiver orientations by folding the $b_{i},c_{i}$
nodes
among themselves; and do the same thing for $b_{i}^{\prime },c_{i}^{\prime }$%
. As such, the obtained quiver has the following nodes
\begin{equation}
\begin{tabular}{lll}
$w_{i}=b_{i}+c_{i}$ & , & $w_{i}^{\prime }=b_{i}^{\prime }+c_{i}^{\prime }$%
\end{tabular}%
\end{equation}%
In this way, one ends with two un-oriented diagrams describing the
Dynkin
graph of $SU\left( 3\right) \times SU\left( 3\right) $ with gauge particles $%
w_{i}$ and $w_{i}^{\prime }$.
The quiver doubling is the price to pay to kill the orientation.
This construction extends to the Dynkin graphs of all ADE Lie
algebras.

\  \  \

\begin{figure}[tbph]
\begin{center}
\hspace{0cm} \includegraphics[width=8cm]{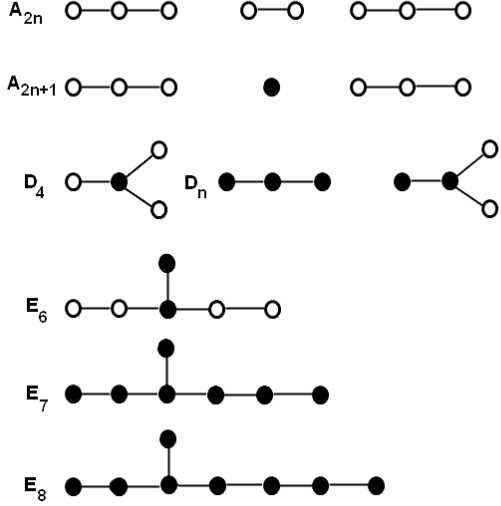}
\end{center}
\par
\vspace{-1 cm} \caption{Outer automorphisms of Dynkin graphs; red
nodes indicate those simple roots that are fixed by
outer-automorphisms.} \label{ADE}
\end{figure}

\emph{outer-automorphisms} \newline The BPS Quivers of the
supersymmetric broken SU$\left( 3\right) $ gauge theory has an
outer-automorphism symmetry isomorphic to the outer- automorphism
group of the $SU\left( 3\right) $ Dynkin graph. Here also this
property is not specific for $SU\left( 3\right) $; it exists as well
for BPS quivers of $\mathcal{N}=2$ QFTs associated with any simply
laced gauge symmetry G with Dynkin diagram of $G$ as in fig
\ref{ADE}.
Notice that the usual outer-automorphism symmetries $\Gamma ^{G}$
are very important in gauge theories; they are generally used to
approach non simply laced gauge symmetries from the ADE ones;
offering therefore a tricky way to get the BPS spectra of
$\mathcal{N}=2$ supersymmetric QFTs with non simply laced gauge
symmetries \textrm{\cite{5A,5B,5C}}. For example, the Dynkin graph
of the $SO\left( 2N+1\right) $ non simply laced series can be
obtained by folding the nodes the Dynkin graph of $SO\left(
2N+2\right) $ which are interchanged by the $\mathbb{Z}_{2}$ outer
automorphism symmetry. Similarly,
the Dynkin graph of G$_{2}$ can be obtained by folding three nodes of the SO$%
\left( 4\right) $ graph. In general, the outer automorphism groups
$\Gamma ^{G}$ for the Dynkin diagrams and the number of fixed of
simple roots are as
follows.%
\begin{eqnarray}
&&%
\begin{tabular}{l|l|l|l|l|l|l|l}
& $SU_{2N}$ & $SU_{2N+1}$ & $SO_{8}$ & $SO_{2N}$ & $E_{6}$ & $E_{7}$ & $%
E_{8} $ \\ \hline
$\  \  \  \  \  \  \  \Gamma ^{G}$ & $\mathbb{Z}_{2}$ & $\mathbb{Z}_{2}$ & $\mathbb{%
S}_{3}$ & $\mathbb{Z}_{2}$ & $\mathbb{Z}_{2}$ & $-$ & $-\left.
\begin{array}{c}
\\
\\
\end{array}%
\right. $ \\
{\small nbr of fixed roots} & $1$ & $0$ & $1$ & $N-2$ & $2$ & $7$ &
$8\left.
\begin{array}{c}
\\
\\
\end{array}%
\right. $ \\ \hline
\end{tabular}
\\
&&  \notag
\end{eqnarray}%
and the result on folding is:%
\begin{equation}
\begin{tabular}{lll}
$D_{N+1}/\mathbb{Z}_{2}$ & $\rightarrow $ & $B_{N}$ \\
$A_{2N-1}/\mathbb{Z}_{2}$ & $\rightarrow $ & $C_{N}$ \\
$E_{6}/\mathbb{Z}_{2}$ & $\rightarrow $ & $F_{4}$ \\
$D_{4}/\mathbb{S}_{3}$ & $\rightarrow $ & $G_{2}$%
\end{tabular}%
\end{equation}

\begin{figure}[tbph]
\begin{center}
\hspace{0cm} \includegraphics[width=10cm]{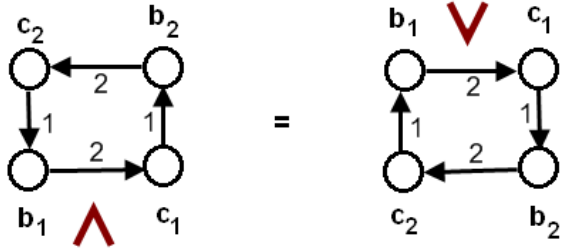}
\end{center}
\par
\vspace{-1 cm} \caption{the BPS quiver viewed from top and from
bottom. } \label{F}
\end{figure}

\emph{case of SU}$\left( 3\right) $\newline In the case of BPS
quivers of the $SU\left( 3\right) $ gauge symmetry, the outer
automorphisms act on the $SU\left( 3\right) $ Dynkin graph by
interchanging its two nodes $\alpha _{1},$ $\alpha _{2}$. On the
side of BPS quiver theory, this symmetry corresponds to permuting
the role of the two monopoles among themselves and the same thing
for the two dyons; this leaves
the quiver $\mathfrak{Q}_{0}^{su_{3}}$ invariant.%
\begin{equation}
\begin{tabular}{lll}
$b_{1}$ $\  \leftrightarrow $ $\ b_{2}$ & and & $c_{1}$ $\  \leftrightarrow $ $%
\ c_{2}$%
\end{tabular}
\label{bbb}
\end{equation}%
At the diagrammatic level, (\ref{bbb}) corresponds to rotate the
planar quiver $\mathfrak{Q}_{0}^{su_{3}}$ around its central axis by
an angle $\pi $ as depicted on the figure \ref{F}; this\ discrete
symmetry describes to the equivalence of viewing the quiver from top
or from bottom.
Using (\ref{rep}), one can represents the action of the outer
automorphisms on the nodes of $\mathfrak{Q}_{0}^{su_{3}}$ by a
linear representation on the vector $\mathbf{v}^{\left( 0\right) }$
that leaves invariant the
intersection matrix $\mathcal{A}^{\left( 0\right) }$. We have%
\begin{equation}
\begin{array}{ccc}
\mathbf{v}^{\left( 0\right) } & \rightarrow & \mathbf{v}^{\prime
\left(
0\right) }=M_{0}\mathbf{v}^{\left( 0\right) }%
\end{array}%
\end{equation}%
with
\begin{equation}
\mathcal{A}^{\prime \left( 0\right) }=M_{0}\mathcal{A}^{\left(
0\right) }M_{0}^{T}=\mathcal{A}^{\left( 0\right) }
\end{equation}%
More explicitly,
\begin{equation}
\mathbf{v}^{\prime \left( 0\right) }=\left(
\begin{array}{c}
b_{2} \\
b_{1} \\
c_{2} \\
c_{1}%
\end{array}%
\right)
\end{equation}%
from which we read
\begin{equation}
M_{0}=\left(
\begin{array}{cccc}
0 & I_{2} & 0 & 0 \\
I_{2} & 0 & 0 & 0 \\
0 & 0 & 0 & I_{2} \\
0 & 0 & I_{2} & 0%
\end{array}%
\right) =M_{0}^{T}  \label{MZ}
\end{equation}%
satisfying the properties $M_{0}^{2}=I_{id}$. Then, quivers $\mathfrak{Q}%
_{n}^{su_{3}}$ that are related under outer automorphism
transformations should be identified; so mutation transformations
should commute with outer automorphisms.

\subsubsection{the cone of BPS particles}

A chamber in the quiver theory of \textrm{\cite{1A,1B}} contains
many BPS states organized into several packages of BPS states
belonging to several quivers related among themselves by mutation
transformations.\newline In the case of $SU\left( 3\right) $ gauge
symmetry, the BPS states belonging to a chamber are arranged into
subsets of $\emph{4}$ BPS states each; related by mutation
transformations. To build a chamber in the $SU\left( 3\right) $ BPS
quiver theory, one proceeds as follows:

\begin{itemize}
\item start with 4 BPS particles with EM charges $\vec{\gamma}_{1},$ $\vec{%
\gamma}_{2},$ $\vec{\gamma}_{3},$ $\vec{\gamma}_{4}$ and
intersection matrix $\mathcal{J}_{ij}=\vec{\gamma}_{i}\circ
\vec{\gamma}_{j}$\ and central charges $Z_{i}=Z\left(
\vec{\gamma}_{i}\right) $.

\item think about the complex numbers $Z_{i}$ as points in the complex plane
$Z\sim \left( \func{Re}Z,\func{Im}Z\right) $ with absolute values $%
\left \vert Z_{i}\right \vert $ giving the masses $m_{\gamma _{i}}$
of the BPS particles and the arguments $\arg Z_{i}$ to make
\emph{partial ordering} from left to right or equivalently from
right to left in the upper half plane $\func{Im}Z>0$; see fig
\ref{CZ}.
\begin{figure}[tbph]
\begin{center}
\hspace{0cm} \includegraphics[width=10cm]{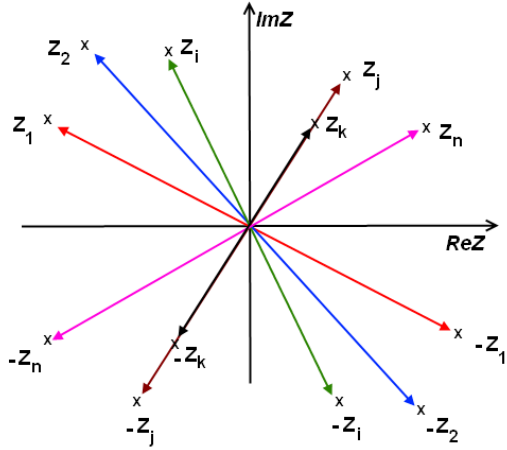}
\end{center}
\par
\vspace{-1 cm} \caption{Central charges of BPS and anti-BPS states
as points in the complex $Z$ plane. } \label{CZ}
\end{figure}
In general there are several ways to order these arguments as given
by the
cone,%
\begin{equation}
\pi >\arg Z_{i_{1}}>\arg Z_{i_{2}}>\arg Z_{i_{3}}>\arg Z_{i_{4}}>0
\label{zbc}
\end{equation}

\item introduce the vectors $\mathbf{v}^{\left( 0\right) }$ and $\Theta
^{\left( 0\right) }$
\begin{equation}
\begin{tabular}{lll}
$\mathbf{v}^{\left( 0\right) }=\left(
\begin{array}{c}
\vec{\gamma}_{1} \\
\vec{\gamma}_{2} \\
\vec{\gamma}_{3} \\
\vec{\gamma}_{4}%
\end{array}%
\right) $ & , & $\Theta ^{\left( 0\right) }=\left(
\begin{array}{c}
Z_{1} \\
Z_{2} \\
Z_{3} \\
Z_{4}%
\end{array}%
\right) $%
\end{tabular}%
\end{equation}%
with no matter about the order of the entries of $\mathbf{v}^{\left(
0\right) }$. Then perform mutations of BPS quivers which, in this
set up,
are realized by linear transformations of these vectors as follows%
\begin{equation}
\begin{tabular}{llll}
$\mathbf{v}^{\left( 0\right) }\rightarrow $ & $\mathbf{v}^{\left(
1\right) }=L_{1}\mathbf{v}^{\left( 0\right) }\rightarrow $ &
$\mathbf{v}^{\left(
2\right) }=L_{2}\mathbf{v}^{\left( 1\right) }\rightarrow $ & $\cdots $ \\
$\Theta ^{\left( 0\right) }\rightarrow $ & $\Theta ^{\left( 1\right)
}=L_{1}\Theta ^{\left( 0\right) }\rightarrow $ & $\Theta ^{\left(
2\right)
}=L_{2}\Theta ^{\left( 1\right) }\rightarrow $ & $\cdots $%
\end{tabular}%
\end{equation}
\end{itemize}

\  \  \  \newline The BPS chamber is finite if the above sequence of
mutations is finite; otherwise it is infinite. In what follows, we
will take the 4 initial BPS particles as given by the two elementary
monopoles $\mathfrak{M}_{i}$ and the two elementary dyons
$\mathfrak{D}_{i}$ with respective EM charges $b_{i} $ and $c_{i}$;
and central charges as
\begin{equation}
\begin{tabular}{llll}
$Z\left( b_{i}\right) =X_{i}$ & , & $Z\left( c_{i}\right) =Y_{i}$ &
\end{tabular}%
\end{equation}%
We also have%
\begin{equation}
\begin{tabular}{lll}
$\mathbf{v}^{\left( 0\right) }=\left(
\begin{array}{c}
b_{1} \\
b_{2} \\
b_{1} \\
b_{2}%
\end{array}%
\right) $ & , & $\Theta ^{\left( 0\right) }=\left(
\begin{array}{c}
X_{1} \\
X_{2} \\
Y_{1} \\
Y_{2}%
\end{array}%
\right) $%
\end{tabular}%
\end{equation}%
Notice that because of the ordering of the states is partial; then
one may have several BPS particles that have different masses $\left
\vert Z_{i_{1}}\right \vert \neq ...\neq \left \vert Z_{i_{4}}\right
\vert $; but with the same angle arg$Z_{i_{1}}=...=\arg Z_{i_{4}}$.
For illustration see the example of the two BPS particles $Z_{j}$
and $Z_{k}$ of fig \ref{CZ}. Obviously these BPS particles
corresponds to different nodes in the quiver as the multiplicity of
these hypermultiplets is equal to one. This feature has an
interpretation in terms of commuting basis reflections as given by
eqs \textrm{(\ref{ll}-\ref{st}), see also fig 16 reported in
appendix I}; it will be used when building the strong coupling
chamber of ADE gauge theories.

\begin{itemize}
\item BPS and anti- BPS states\newline
The angle $\arg Z_{u}$ can take values in the interval $\left[
0,2\pi \right] $, the complex $Z$ plane of the central charges at a
point $u$ is divided into two half planes:
\begin{equation}
\begin{tabular}{lll}
$\mathcal{H}_{u}^{+}=\left \{ \func{Im}Z>0\right \} $ & and & $\mathcal{H}%
_{u}^{-}=\left \{ \func{Im}Z<0\right \} $ \\
&  &
\end{tabular}%
\end{equation}%
In the upper half plane $\mathcal{H}_{u}^{+}\equiv \mathcal{H}_{u}$
lives the BPS particles and are partially ordered according to
$0<\arg Z_{u}<\pi $ while in the half plane $\mathcal{H}_{u}^{-}$
lives the anti-BPS states and we have $\pi <\arg Z_{u}<2\pi .$ The
total BPS spectrum is given by all states that live on the full
plane $\mathcal{H}_{u}$ of central charges; and so this spectrum is
CPT invariant.\newline

\item \emph{left most and right most BPS states}

By using $\arg Z_{u}$, the partial ordering of the central charges $%
Z_{u}\left( \gamma _{i}\right) \equiv Z_{i}$ in $\mathcal{H}_{u}$
allows to distinguish several sequences depending on the angles of
the central charges.\ A typical sequence of $l$ BPS particles is
given by (\ref{zbc}) where $Z_{i_{1}}$ appears at left most,
$Z_{i_{2}}$ the second left one and so on. \newline We also have
$Z_{i_{4}}$ the right most BPS, $Z_{i_{3}}$ the next right one an so
on. The knowledge of (\ref{zbc}) is important for choosing the BPS
state to begin with in performing quiver mutations.
\end{itemize}

\section{Strong coupling chambers in $SU\left( N\right) $ theories}

In this section, we study the isotropy group structure of the set $\mathcal{G%
}_{strong}^{su_{N}}$ of mutation transformations of the BPS quivers of $%
\mathcal{N}=2$ supersymmetric quantum field theories with $SU\left(
N\right) $ gauge symmetry and use this mutation group to build the
BPS spectra of the strong coupling chambers. To see how the
machinery works, we begin by studying the leading cases $N=2,3,4$;
then we give the general result for the $SU\left( N\right) $ series.
The $\mathcal{N}=2$ supersymmetric QFT's with $SO\left( 2N\right) $
and $E_{r}$ gauge symmetries will be considered in section 5.

\subsection{$SU\left( 2\right) $ theory}

\begin{figure}[tbph]
\begin{center}
\hspace{0cm} \includegraphics[width=12cm]{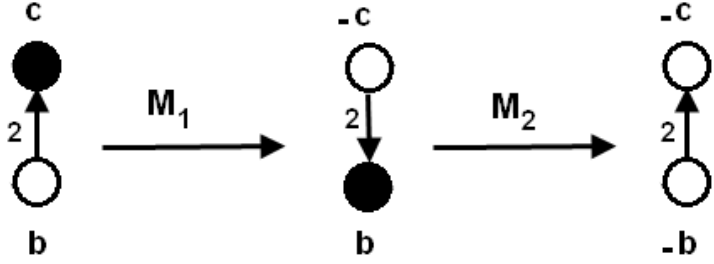}
\end{center}
\par
\vspace{-1 cm}
\caption{Strong coupling chamber using mutations of the primitive quiver $%
\mathfrak{Q}_{0}^{su_{2}}$} \label{3}
\end{figure}

This is a particular model that has been studied explicitly in \textrm{\cite%
{1A,1B}; see also \cite{02E,3A,3B,3C,31H}. Here, we }reconsider this
theory by using a groupoid approach. This study is useful as it
allows to get more insight into BPS quiver theory associated with
higher dimensional gauge symmetries.\newline The strong coupling
chamber of the effective 4D $\mathcal{N}=2$ supersymmetric pure
SU$\left( 2\right) $ gauge theory has $2$ BPS states and $2$
anti-BPS ones. These are:

\begin{itemize}
\item a monopole $\mathfrak{M}$ and a dyon $\mathfrak{D}$ with respective EM
charge $b,$ $c$; and respective complex central charges $X$ and $Y$;

\item their CPT conjugate $\mathfrak{\bar{M}}$ and $\mathfrak{\bar{D}}$
having opposite EM charge; i.e: $-b,$ $-c$.
\end{itemize}

\  \  \  \  \newline In terms of the arguments of the central
charges, the BPS states of the strong coupling chamber of this gauge
theory corresponds to
\begin{equation}
\arg Y>\arg X
\end{equation}%
The content of this chamber can be explicitly derived by applying
the quiver
mutation method of \textrm{\cite{1A,1B}} which is illustrated on fig \textrm{%
\ref{3}}

From these quiver mutations, it follows that the transformations $\mathcal{M}%
_{0,n}$ can be realized on the EM charge vector $v=\left( b,c\right)
$ as
linear mappings generated by two matrices $L_{1}$ and $L_{2}$ as follows%
\begin{equation}
\begin{tabular}{lll}
$L_{1}=\left(
\begin{array}{cc}
1 & 0 \\
0 & -1%
\end{array}%
\right) $ & , & $L_{2}=\left(
\begin{array}{cc}
-1 & 0 \\
0 & 1%
\end{array}%
\right) $%
\end{tabular}%
\end{equation}%
Notice that $L_{1}=-L_{2}=L$, this is a very particular feature that
is specific for SU$\left( 2\right) $ and has no analogue for the
case of higher dimensional gauge symmetry. We also have the
properties
\begin{equation}
\left( L_{1}\right) ^{2}=\left( L_{2}\right) ^{2}=I_{id}  \label{A}
\end{equation}%
and%
\begin{equation}
\left( L_{i}L_{j}\right) ^{m_{ij}}=I_{id}  \label{B}
\end{equation}%
with $m_{ij}$ the entries of the following integral $2\times 2$ symmetric%
\textrm{\ matrix}%
\begin{equation}
M=\left(
\begin{array}{cc}
1 & 2 \\
2 & 1%
\end{array}%
\right)  \label{C}
\end{equation}%
The relations (\ref{A}-\ref{B}-\ref{C}) turns out to play a crucial
role in the study of the \textrm{symmetry} structure of mutation
transformations in the BPS quiver theory. They encode a general
result that is valid for all ADE gauge symmetries of the BPS quiver
theory.\  \newline By using the two generators $L_{1}$ and $L_{2}$,
it is not difficult to check that a generic quiver mutation mapping
$\mathfrak{Q}_{0}^{su_{2}}$ into $\mathfrak{Q}_{n}^{su_{2}}$ depends
on the parity of the positive
integer $n$. We have%
\begin{equation}
\begin{tabular}{ll}
$\mathcal{M}_{2k}$ & $=\left( L_{2}L_{1}\right) ^{k}$ \\
$\mathcal{M}_{2k+1}$ & $=L_{1}\left( L_{2}L_{1}\right) ^{k}$%
\end{tabular}%
\end{equation}%
However, since $\left( L_{i}\right) ^{2}=I_{id}$, it follows that
the set of
mutations form a \textrm{finite dimensional group} that we denote as $%
\mathcal{G}_{strong}^{su_{2}}$. This group has \emph{4} elements\textrm{%
\footnote{$\mathcal{G}_{strong}^{su_{2}}$ is isomorphic to $\mathbb{Z}%
_{2}\times \mathbb{Z}_{2}$; see also appendix C for the link with
the finite dihedral Coxeter group Dih$_{4}.$}} namely $\left \{
I_{id},+L,-I_{id},-L\right \} $; and which, for later use, we prefer
to
write as follows%
\begin{equation}
\begin{tabular}{lll}
$\mathcal{G}_{strong}^{su_{2}}$ & $=$ & $\left \{
\begin{tabular}{llll}
$I_{id},$ & $L_{1},$ & $L_{2}L_{1},$ & $L_{1}L_{2}L_{1}$%
\end{tabular}%
\right \} $ \\
& $\equiv $ & $\left \{
\begin{tabular}{llll}
$\mathcal{M}_{0},$ & $\mathcal{M}_{1},$ & $\mathcal{M}_{2},$ & $\mathcal{M}%
_{3}$%
\end{tabular}%
\right \} $%
\end{tabular}%
\end{equation}%
or equivalently like%
\begin{equation}
\begin{tabular}{lll}
$\mathcal{G}_{strong}^{su_{2}}$ & $=$ & $\  \left \{
\begin{tabular}{llll}
$I_{id},$ & $L_{1},$ & $L_{2}L_{1},$ & $L_{2}$%
\end{tabular}%
\right \} $%
\end{tabular}%
\end{equation}%
We also have the following identities useful for generalization to
higher
dimensional gauge symmetries%
\begin{equation}
\begin{tabular}{lllll}
$\left( L_{2}L_{1}\right) ^{2}$ & $=I_{id}$ & , & $\left(
L_{1}L_{2}\right)
^{2}$ & $=I_{id}$ \\
$L_{1}L_{2}L_{1}$ & $=L_{2}$ & , & $L_{2}L_{1}L_{2}$ & $=L_{1}$%
\end{tabular}
\label{R}
\end{equation}%
and%
\begin{equation}
\begin{tabular}{ll}
$L_{2}L_{1}$ & $=L_{1}L_{2}$%
\end{tabular}%
\end{equation}%
The multiplication table of the mutation group
$\mathcal{G}_{M}^{su_{2}}$ is
as follows%
\begin{equation}
\begin{tabular}{l|llll}
$\mathcal{G}_{M}^{su_{2}}$ & $\mathcal{M}_{0}$ & $\mathcal{M}_{1}$ & $%
\mathcal{M}_{2}$ & $\mathcal{M}_{3}$ \\ \hline $\mathcal{M}_{0}$ &
$\mathcal{M}_{0}$ & $\mathcal{M}_{1}$ & $\mathcal{M}_{2}$
& $\mathcal{M}_{3}$ \\
$\mathcal{M}_{1}$ & $\mathcal{M}_{1}$ & $\mathcal{M}_{0}$ &
$\mathcal{M}_{3}$
& $\mathcal{M}_{2}$ \\
$\mathcal{M}_{2}$ & $\mathcal{M}_{2}$ & $\mathcal{M}_{3}$ &
$\mathcal{M}_{0}$
& $\mathcal{M}_{1}$ \\
$\mathcal{M}_{3}$ & $\mathcal{M}_{3}$ & $\mathcal{M}_{2}$ &
$\mathcal{M}_{1}$ & $\mathcal{M}_{0}$ \\ \hline
\end{tabular}%
\end{equation}%
Notice that the order of this discrete group is%
\begin{equation}
\text{order}\left( \mathcal{G}_{strong}^{su_{2}}\right) =\left \vert
\mathcal{G}_{strong}^{su_{2}}\right \vert =4
\end{equation}%
it is precisely the number of BPS states and anti-BPS states in the
strong coupling chamber. For later use think about this number as
\begin{equation}
\#_{bsp+antibps}=\left \vert \mathcal{G}_{strong}^{su_{2}}\right
\vert \times r
\end{equation}%
with $r$ the rank of $SU\left( 2\right) $. Below, we give the
extension of this matrix mutation group construction for the cases
of $SU\left( 3\right) $ and $SU\left( 4\right) $ models; the
generalization to the generic $SU\left( N\right) $ gauge groups will
be reported in appendix II.

\subsection{$SU\left( 3\right) $ model}

Following \textrm{\cite{1A,1B}}, the strong coupling chamber of the $%
\mathcal{N}=2$ supersymmetric pure $SU\left( 3\right) $ gauge theory
has \emph{6} BPS states and \emph{6} anti-BPS ones. These states can
be obtained by performing mutations of $\mathfrak{Q}_{0}^{su_{3}}$.
Below, we use our \textrm{algebraic} method to get the full set of
BPS states of the strong coupling chamber. This approach relies on
building the family of mutated
quivers $\mathfrak{Q}_{n}^{su_{3}}$ in the strong coupling chamber by using $%
\mathcal{G}_{strong}^{su_{3}}$.

\subsubsection{ The $\mathfrak{Q}_{n}^{su_{3}}$\emph{\ }quiver family}

First, consider the primitive quiver $\mathfrak{Q}_{0}^{su_{3}}$
given by the graph (\ref{U3}); this quiver is associated with a
point $\left( u_{1},u_{2}\right) $ of the Coulomb branch of the
moduli space of the theory; and involves only elementary BPS states.
In our approach, the quiver $\mathfrak{Q}_{0}^{su_{3}}$ is
represented by the two following:\newline \textbf{(i)} the $\left(
2r\right) ^{2}$ component vector with $r=2$; the rank of $SU\left(
3\right) $,
\begin{equation}
\mathbf{v}^{\left( 0\right) }=\left(
\begin{array}{c}
b_{1} \\
b_{2} \\
c_{1} \\
c_{2}%
\end{array}%
\right)
\end{equation}%
this vector combines the EM vector charges of the 4 elementary BPS states.%
\newline
\textbf{(ii)} the intersection matrix
\begin{equation}
\mathcal{A}_{IJ}^{\left( 0\right) }=\mathbf{v}_{I}^{\left( 0\right)
}\circ \mathbf{v}_{J}^{\left( 0\right) }.
\end{equation}%
given by eq(\ref{ps}).

\  \  \  \newline To get the expressions of the remaining BPS states
of the strong coupling
chamber, we have to perform successive mutations of $\mathfrak{Q}%
_{0}^{su_{3}}$ until reaching the quiver $\mathfrak{Q}_{0}^{su_{3}}$
again.
The mutations are by the following reflections,%
\begin{equation}
\begin{tabular}{lllll}
$\gamma _{i}^{\prime }$ & $=$ & $-\gamma _{i}$ &  &  \\
$\gamma _{j}^{\prime }$ & $=$ & $\gamma _{j}+N_{ji}\gamma _{i}$ & if & $%
N_{ji}>0$ \\
$\gamma _{j}^{\prime }$ & $=$ & $\gamma _{j}$ & if & $N_{ji}<0$%
\end{tabular}%
\end{equation}%
with the integers $N_{ji}=\gamma _{j}\circ \gamma _{i}$ given by eqs(\ref{sp}%
-\ref{ps}). \newline In what follows, we use the groupoid method to
derive the explicit BPS content of the strong coupling chamber. The
method is as follows:\newline (\textbf{1}) describe the generic BPS
quivers $\mathfrak{Q}_{n}^{su_{3}}$ of the strong coupling chamber
by the vector
\begin{equation}
\mathbf{v}^{\left( n\right) }=\left(
\begin{array}{c}
b_{1}^{\left( n\right) } \\
b_{2}^{\left( n\right) } \\
c_{1}^{\left( n\right) } \\
c_{2}^{\left( n\right) }%
\end{array}%
\right) ,\qquad n=0,1,2,...  \label{VI}
\end{equation}%
with entries describing the nodes of the quiver and intersections $\mathcal{A%
}_{IJ}^{\left( n\right) }=\mathbf{v}_{I}^{\left( n\right) }\circ \mathbf{v}%
_{J}^{\left( n\right) }$ as%
\begin{equation}
\mathcal{A}_{IJ}^{\left( n\right) }=\left(
\begin{array}{cc}
0 & b_{i}^{\left( n\right) }\circ c_{j}^{\left( n\right) } \\
c_{i}^{\left( n\right) }\circ b_{j}^{\left( n\right) } & 0%
\end{array}%
\right)  \label{AIJ}
\end{equation}%
In this view, $\mathfrak{Q}_{0}^{su_{3}}$ appears precisely as the
leading member of the family
\begin{equation}
\mathbb{B}_{strong}^{su_{3}}=\left \{ \mathfrak{Q}_{n}^{su_{3}};\text{ }%
n\geq 0\right \} .
\end{equation}%
\textrm{which should be thought of as the base of objects in
groupoid langauge}.\newline (\textbf{2}) interpret $b_{i}^{\left(
n\right) }$ and $c_{i}^{\left( n\right) }$ as the EM charge vectors
of BPS states of the strong coupling chamber. We also have
$b_{i}^{\left( 0\right) }=b_{i}$ and $c_{i}^{\left( 0\right)
}=c_{i}$. \newline The same feature holds for the central charges
$X_{i}^{\left( n\right) }$
and $Y_{i}^{\left( n\right) }$ of the BPS states%
\begin{equation}
\begin{tabular}{lll}
$X_{i}^{\left( n\right) }=Z(b_{i}^{\left( n\right) })$ & , &
$Y_{i}^{\left(
n\right) }=Z(c_{i}^{\left( n\right) })$%
\end{tabular}%
\end{equation}%
(\textbf{3}) think about the mutation from $\mathfrak{Q}_{0}^{su_{3}}$ to $%
\mathfrak{Q}_{n}^{su_{3}}$ as a linear mapping of
$\mathbf{v}^{\left( 0\right) }$ into $\mathbf{v}^{\left( n\right) }$
like,
\begin{equation}
\mathbf{v}^{\left( n\right) }=\mathcal{M}_{0}\mathbf{v}^{\left(
0\right) },\qquad n=0,1,2,...  \label{MU}
\end{equation}%
with
\begin{equation*}
\mathcal{M}_{0}=I_{id}
\end{equation*}%
and the mutations $\mathcal{M}_{0}$ given by some matrices of $GL\left( 4,%
\mathbb{Z}\right) $ that we have to determine\textrm{\footnote{%
see footnote 5}}. These matrices are obtained by using the iteration
property
\begin{equation}
\mathbf{v}^{\left( n\right) }=L_{n}\mathbf{v}^{\left( n-1\right)
},\qquad n=1,2,...
\end{equation}%
leading to the realization%
\begin{equation}
\mathcal{M}_{n}=L_{n}L_{n-1}...L_{2}L_{1}=\dprod
\limits_{k=1}^{n}L_{n+1-k} \label{RN}
\end{equation}%
and showing that the $\mathcal{M}_{n}$'s (thought of as
$\mathcal{M}_{n,0}$) are particular groupoid morphisms of
$\mathbb{G}_{su_{3}}$ acting on the BPS chambers
$\mathbb{B}_{su_{3}}$
\begin{equation}
\mathbb{G}_{su_{3}}=\left \{ \mathcal{M}_{n,m}:\mathfrak{Q}%
_{m}^{su_{3}}\rightarrow \mathfrak{Q}_{n}^{su_{3}},\qquad \mathfrak{Q}%
_{k}^{su_{3}}\in \mathbb{B}_{su_{3}}\right \}
\end{equation}%
with the binary composition \textrm{as in appendix I;
eq(\ref{br})}.\newline (\textbf{4}) To deal with (\ref{RN}), one has
to identify the appropriate ordering of the arguments of the complex
central charges $X_{i}$ and $Y_{i}$
of the two monopoles and the two dyons that make $\mathfrak{Q}_{0}^{su_{3}}$%
. It happens that the adequate\textrm{\footnote{%
The exact choice of the argument of the central charge leading to
the BPS strong coupling chamber is given\ by $\arg Y_{1}>$ $\arg
Y_{2}>$ $\arg
X_{1}> $ $\arg X_{2}$. It turns out that this choice is equivalent to (\ref%
{cd}-\ref{dc}); for an explicit proof see eqs(\ref{ll}-\ref{st}) and
discussion given there.} }choice that leads to the BPS spectrum of
the strong coupling chamber corresponds to
\begin{equation}
\begin{tabular}{ll}
$\arg Y_{1}$ & $=\arg Y_{2}$ \\
$\arg X_{1}$ & $=\arg X_{2}$%
\end{tabular}
\label{cd}
\end{equation}%
and%
\begin{equation}
\begin{tabular}{ll}
$\arg Y_{i}>\arg X_{i}$ & .%
\end{tabular}
\label{dc}
\end{equation}%
The relations (\ref{cd}) teach us that one has to treat on equal
footing the two dyons and the same thing for the two monopoles; a
helpful property which will be interpreted later and that will be
used below.

\subsubsection{Building the $\mathfrak{Q}_{n}^{su_{3}}$'s of the chamber $%
\arg Y>\arg X$}

To get the family of mutated quivers $\mathfrak{Q}_{n}^{su_{3}}$ of
the chamber $\arg Y>\arg X$, we use the property that the order of
this chamber
is finite\textrm{\footnote{%
In the BPS\ weak coupling chamber $\arg X>\arg Y$, the number of BPS
states is infinite; see sections 6 and 7.}}. Denote this order by
the positive integer as $n_{0}$; and proceeds as follows:

\begin{itemize}
\item first, mutate \emph{simultaneously} the two nodes $c_{i}^{\left(
0\right) }=c_{i}$ associated with the two dyons having central charges $%
Y_{i} $. This collective operation, which is represented by the
mutation matrix $L_{1}$, leads to the quiver
$\mathfrak{Q}_{1}^{su_{3}}$ made of two nodes $b_{i}^{\left(
1\right) }$ and two nodes $c_{i}^{\left( 1\right) }$ related by the
intersection matrix $b_{i}^{\left( 1\right) }\circ
c_{j}^{\left( 1\right) }$. Using eqs(\ref{VI}-\ref{AIJ}), we have%
\begin{equation}
\begin{tabular}{ll}
$v_{I}^{\left( 1\right) }$ & $=\left( L_{1}\right)
_{IK}v_{K}^{\left(
0\right) }$ \\
$\mathcal{A}_{IJ}^{\left( 1\right) }$ & $=\left( L_{1}\right) _{IK}\mathcal{A%
}_{KL}^{\left( 0\right) }\left( L_{1}\right) _{JL}$%
\end{tabular}%
\end{equation}%
or in a condensed manner like%
\begin{equation}
\begin{tabular}{ll}
$v^{\left( 1\right) }$ & $=L_{1}v^{\left( 0\right) }$ \\
$\mathcal{A}^{\left( 1\right) }$ & $=L_{1}\mathcal{A}^{\left(
0\right)
}L_{1}^{T}$%
\end{tabular}%
\end{equation}%
Therefore $L_{1}$ specifies completely the BPS quiver $\mathfrak{Q}%
_{1}^{su_{3}}$. Notice that due to $\left( L_{1}\right) ^{2}=I_{id}$
we have
the following inverse mutation%
\begin{equation}
\begin{tabular}{ll}
$v^{\left( 0\right) }$ & $=L_{1}v^{\left( 1\right) }$ \\
$\mathcal{A}^{\left( 0\right) }$ & $=L_{1}\mathcal{A}^{\left(
1\right)
}L_{1}^{T}$%
\end{tabular}
\label{SM}
\end{equation}

\item second, mutate simultaneously the two nodes $b_{i}^{\left( 1\right) }$
of the quiver $\mathfrak{Q}_{1}^{su_{3}}$ to end with a mutated quiver $%
\mathfrak{Q}_{2}^{su_{3}}$ with two nodes $b_{i}^{\left( 2\right) }$
and two $c_{i}^{\left( 2\right) }$ and intersection matrix
$b_{i}^{\left( 2\right)
}\circ c_{i}^{\left( 2\right) }$. Using the convention notations (\ref{VI}-%
\ref{AIJ}), we have%
\begin{equation}
\begin{tabular}{lll}
$v^{\left( 2\right) }$ & $=L_{2}v^{\left( 1\right) }$ & $=\mathcal{M}%
_{2}v^{\left( 0\right) }$ \\
$\mathcal{A}^{\left( 2\right) }$ & $=L_{2}\mathcal{A}^{\left(
1\right)
}L_{2}^{T}$ & $=\mathcal{M}_{2}\mathcal{A}^{\left( 0\right) }\mathcal{M}%
_{2}^{T}$%
\end{tabular}%
\end{equation}%
with $\mathcal{M}_{2}$ as in (\ref{RN}) and similar relations to
(\ref{SM}).

\item continue the mutation operations until reaching the quiver $\mathfrak{Q%
}_{n_{0}-1}^{su_{3}}$ made of the two nodes $b_{i}^{\left(
n_{0}-1\right) }$ and the two $c_{i}^{\left( n_{0}-1\right) }$. The
mutation relations are
then as follows%
\begin{equation}
\begin{tabular}{ll}
$v^{\left( n_{0}-1\right) }$ & $=L_{n_{0}-1}v^{\left( n_{0}-2\right) }$ \\
$\mathcal{A}^{\left( n_{0}-1\right) }$ &
$=L_{n_{0}-1}\mathcal{A}^{\left(
n_{0}-2\right) }L_{n_{0}-1}^{T}$%
\end{tabular}%
\end{equation}%
or equivalently%
\begin{equation}
\begin{tabular}{ll}
$v^{\left( n_{0}-1\right) }$ & $=\mathcal{M}_{n_{0}-1}v^{\left(
0\right) }$
\\
$\mathcal{A}^{\left( n_{0}-1\right) }$ & $=\mathcal{M}_{n_{0}-1}\mathcal{A}%
^{\left( 0\right) }\mathcal{M}_{n_{0}-1}^{T}$%
\end{tabular}%
\end{equation}

\item Because of the property of finite chamber order, the next mutation
should lead to the identity; then the quiver
$\mathfrak{Q}_{n_{0}}^{su_{3}}$
made by the two nodes $b_{i}^{\left( n_{0}\right) }$ and the two nodes $%
c_{i}^{\left( n_{0}\right) }$ has to coincide with the primitive $\mathfrak{Q%
}_{0}^{su_{3}}$. We have%
\begin{equation}
\begin{tabular}{ll}
$v^{\left( n_{0}\right) }$ & $=L_{n_{0}}v^{\left( n_{0}-1\right) }$ \\
$\mathcal{A}^{\left( n_{0}\right) }$ &
$=L_{n_{0}}\mathcal{A}^{\left(
n_{0}-1\right) }L_{n_{0}}^{T}$%
\end{tabular}%
\end{equation}%
with%
\begin{equation}
\begin{tabular}{ll}
$v^{\left( n_{0}\right) }$ & $=v^{\left( 0\right) }$ \\
$\mathcal{A}^{\left( n_{0}\right) }$ & $=\mathcal{A}^{\left( 0\right) }$%
\end{tabular}%
\end{equation}%
We also have
\begin{equation*}
\begin{tabular}{ll}
$v^{\left( n_{0}\right) }$ & $=\mathcal{M}_{n_{0}}v^{\left( 0\right) }$ \\
$\mathcal{A}^{\left( n_{0}\right) }$ & $=\mathcal{M}_{n_{0}}\mathcal{A}%
^{\left( 0\right) }\mathcal{M}_{n_{0}}^{T}$%
\end{tabular}%
\end{equation*}%
with the following constraint relation of the mutation matrices%
\begin{equation}
\mathcal{M}_{n_{0}}=L_{n_{0}}L_{n_{0}-1}...L_{2}L_{1}=I_{id}
\end{equation}
\end{itemize}

\subsection{Building the mutation set $\mathcal{G}_{strong}^{su_{3}}$}

We begin by analyzing eqs (\ref{cd}) which we use it to build the
mutation set $\mathcal{G}_{strong}^{su_{3}}$. Then we work out the
$\emph{5}$ possible mutations of the quiver
$\mathfrak{Q}_{0}^{su_{3}}$ after what we give the results.

\subsubsection{Consequence of eqs (\protect \ref{cd})}

Because of the constraint relations (\ref{cd}), the mutations of $\mathfrak{Q%
}_{0}^{su_{3}}$ obey the remarkable periodicity property,%
\begin{equation}
\begin{tabular}{lllll}
$L_{2n+1}=L_{1}$ & , & $L_{2n+2}=L_{2}$ & , & $\forall n$%
\end{tabular}
\label{L1L2}
\end{equation}%
It happens that this property is valid for the BPS quivers of any
ADE gauge symmetry; and leads to tremendous simplifications. Let us
illustrate below how this works for the case of SU$\left( 3\right)
$. Putting (\ref{L1L2}) back into (\ref{RE}), we find that the set
of matrices $\mathcal{M}_{n}$ is completely generated by $L_{1}$ and
$L_{2}$ as given by the following relations
\begin{equation}
\begin{tabular}{ll}
$\mathcal{M}_{2k}$ & $=\left( L_{2}L_{1}\right) ^{k}$ \\
$\mathcal{M}_{2k+1}$ & $=L_{1}\mathcal{M}_{2k}$%
\end{tabular}
\label{M2K}
\end{equation}%
with $L_{1}$ and $L_{2}$ obeying
\begin{equation}
\left( L_{1}\right) ^{2}=\left( L_{2}\right) ^{2}=I_{id},
\end{equation}%
as well as the identities%
\begin{equation}
\begin{tabular}{lllll}
$\left( L_{2}L_{1}\right) ^{3}$ & $=I_{id},$ &  & $\left(
L_{1}L_{2}\right) ^{3}$ & $=I_{id}\left.
\begin{array}{c}
\\
\end{array}%
\right. $ \\
$L_{1}\left( L_{2}L_{1}\right) ^{2}$ & $=L_{2},$ &  & $L_{2}\left(
L_{1}L_{2}\right) ^{2}$ & $=L_{1}\left.
\begin{array}{c}
\\
\end{array}%
\right. $%
\end{tabular}
\label{MP}
\end{equation}%
and%
\begin{equation}
L_{1}L_{2}L_{1}=L_{2}L_{1}L_{2}
\end{equation}%
By help of these relations, one can check that the set of BPS quiver
mutations (\ref{M2K}) form indeed a finite discrete group $\mathcal{G}%
_{strong}^{su_{3}}$ with order%
\begin{equation}
\left \vert \mathcal{G}_{strong}^{su_{3}}\right \vert =6
\end{equation}%
The binary multiplication table of $\mathcal{G}_{strong}^{su_{3}}$
is given
by%
\begin{equation}
\begin{tabular}{l|llllll}
$\mathcal{G}_{M}^{su_{3}}$ & $I_{id}$ & $\mathcal{M}_{1}$ &
$\mathcal{M}_{2}$ & $\mathcal{M}_{3}$ & $\mathcal{M}_{4}$ &
$\mathcal{M}_{5}$ \\ \hline
$I_{id}$ & $I_{id}$ & $\mathcal{M}_{1}$ & $\mathcal{M}_{2}$ & $\mathcal{M}%
_{3}$ & $\mathcal{M}_{4}$ & $\mathcal{M}_{5}$ \\
$\mathcal{M}_{1}$ & $\mathcal{M}_{1}$ & $I_{id}$ & $\mathcal{M}_{3}$ & $%
\mathcal{M}_{2}$ & $\mathcal{M}_{5}$ & $\mathcal{M}_{4}$ \\
$\mathcal{M}_{4}$ & $\mathcal{M}_{4}$ & $\mathcal{M}_{3}$ & $I_{id}$ & $%
\mathcal{M}_{5}$ & $\mathcal{M}_{2}$ & $\mathcal{M}_{1}$ \\
$\mathcal{M}_{3}$ & $\mathcal{M}_{3}$ & $\mathcal{M}_{4}$ &
$\mathcal{M}_{5}$
& $I_{id}$ & $\mathcal{M}_{1}$ & $\mathcal{M}_{2}$ \\
$\mathcal{M}_{2}$ & $\mathcal{M}_{2}$ & $\mathcal{M}_{5}$ &
$\mathcal{M}_{4}$
& $\mathcal{M}_{1}$ & $I_{id}$ & $\mathcal{M}_{3}$ \\
$\mathcal{M}_{5}$ & $\mathcal{M}_{5}$ & $\mathcal{M}_{2}$ &
$\mathcal{M}_{1}$ & $\mathcal{M}_{4}$ & $\mathcal{M}_{3}$ & $I_{id}$
\\ \hline
\end{tabular}
\label{TA}
\end{equation}%
To get more insight into this discrete symmetry group, it is
interesting to use Coxeter group formulation, reported in appendix
C, by considering: (a) the $2\times 2$ symmetric matrix M with
positive integer entries $m_{ij}$ as
follows%
\begin{equation}
M=\left(
\begin{array}{cc}
1 & 3 \\
3 & 1%
\end{array}%
\right)  \label{C3}
\end{equation}%
and (b) the finite set%
\begin{equation}
W\left( M\right) =\left \langle \left.
\begin{array}{c}
\\
\\
\end{array}%
\right. L_{1},L_{2}|\text{ }\left( L_{i}L_{j}\right)
^{m_{ij}}=I_{id}\right \rangle  \label{3C}
\end{equation}%
from which one recognizes that $\mathcal{G}_{strong}^{su_{3}}$ is
nothing but the matrix the Coxeter group $Dih_{6}$ with Coxeter
graph isomorphic to the Dynkin diagram of the $SU\left( 3\right) $
Lie algebra. The $Dih_{6}$ is
a discrete group having 6 elements; it is a particular group of the family $%
Dih_{2n}$ describing the 2n symmetries consisting of n rotations and
n reflections of a regular polygon with n sides.

\subsubsection{Computing the BPS spectrum and realizing $\mathcal{G}%
_{M}^{su_{3}}\simeq Dih_{6}$}

Here we\ compute explicitly the BPS states of the strong coupling
chamber of the supersymmetric spontaneously broken $SU\left(
3\right) $ gauge theory.
We also give the matrix realization of the mutation group $\mathcal{G}%
_{strong}^{su_{3}}$

\  \  \  \  \

$i)$ \emph{building the} \emph{Quiver}
$\mathfrak{Q}_{1}^{su_{3}}$\newline To get the mutated BPS quiver
$\mathfrak{Q}_{1}^{su_{3}}$ with electric magnetic vector
$\boldsymbol{v}^{\left( 1\right) }=(\boldsymbol{b}^{\left(
1\right) },\boldsymbol{c}^{\left( 1\right) })$, we have to mutate \emph{%
simultaneously} the charge vectors of the two dyons $c_{1}$ and
$c_{2}$.\ By
using the mutation rules, we get the following charge vectors%
\begin{equation}
\left(
\begin{array}{c}
b_{1}^{\left( 1\right) } \\
b_{2}^{\left( 1\right) } \\
c_{1}^{\left( 1\right) } \\
c_{2}^{\left( 1\right) }%
\end{array}%
\right) =\left(
\begin{array}{c}
b_{1}+c_{2} \\
b_{2}+c_{1} \\
-c_{1} \\
-c_{2}%
\end{array}%
\right)
\end{equation}%
from which we can learn the mutation matrix $L_{1}\equiv
\mathcal{M}_{1}$ that relates $\boldsymbol{v}^{\left( 1\right) }$
and $v^{\left( 0\right) }$,
\begin{equation}
L_{1}=\left(
\begin{array}{cccc}
I_{4} & 0 & 0 & I_{4} \\
0 & I_{4} & I_{4} & 0 \\
0 & 0 & -I_{4} & 0 \\
0 & 0 & 0 & -I_{4}%
\end{array}%
\right)  \label{22}
\end{equation}%
where $I_{4}$ stands for the $4\times 4$ identity matrix. Notice
that the rows of $L_{1}$ give precisely the EM charge vectors of the
BPS states of the quiver $\mathfrak{Q}_{1}^{su_{3}}$. Notice also
that being a triangular matrix, we have
\begin{equation}
\left( L_{1}\right) ^{2}=I_{16},\qquad \det L_{1}=1
\end{equation}%
To get the intersection matrix $\mathcal{A}^{\left( 1\right) }$, we
use the
relation%
\begin{equation}
\mathcal{A}^{\left( 1\right) }=L_{1}\mathcal{A}^{\left( 0\right)
}L_{1}^{T}
\end{equation}%
with%
\begin{equation}
\mathcal{A}^{\left( 0\right) }=\left(
\begin{array}{cccc}
0 & 0 & 2 & -1 \\
0 & 0 & -1 & 2 \\
-2 & 1 & 0 & 0 \\
1 & -2 & 0 & 0%
\end{array}%
\right)
\end{equation}%
Straightforward calculations give $\mathcal{A}^{\left( 1\right) }=-\mathcal{A%
}^{\left( 0\right) }.$ Before proceeding notice that being a
reflection, one would expect to have $\det L_{1}$ equals to $-1$;
but it is not. The point is that $L_{1}$ is not a fundamental
reflection, it is the composition of
two commuting reflections as,%
\begin{equation}
L_{1}=s_{1}t_{1}=t_{1}s_{1}  \label{ll}
\end{equation}%
with%
\begin{equation}
\begin{tabular}{llll}
$s_{1}=\left(
\begin{array}{cccc}
I_{4} & 0 & 0 & 0 \\
0 & I_{4} & I_{4} & 0 \\
0 & 0 & -I_{4} & 0 \\
0 & 0 & 0 & I_{4}%
\end{array}%
\right) $ & , & $t_{1}=\left(
\begin{array}{cccc}
I_{4} & 0 & 0 & I_{4} \\
0 & I_{4} & 0 & 0 \\
0 & 0 & I_{4} & 0 \\
0 & 0 & 0 & -I_{4}%
\end{array}%
\right) $ &
\end{tabular}
\label{st}
\end{equation}%
This feature is also valid for the generator $L_{2}$ which should be
thought as $L_{2}=s_{2}t_{2}$. This property is general; it is also
valid for higher dimensional ADE gauge symmetries to be considered
later on.

\  \  \  \

$ii)$ \emph{the} \emph{Quiver} $\mathfrak{Q}_{2}^{su_{3}}$\newline
To get the BPS quiver $\mathfrak{Q}_{2}^{su_{3}}$, we have to mutate \emph{%
simultaneously} the charge vectors $b_{1}^{\left( 1\right) }$ and $%
b_{2}^{\left( 1\right) }$ of the quiver $\mathfrak{Q}_{1}^{su_{3}}$.
We end with a new charge vector $\boldsymbol{v}^{\left( 2\right) }$
with the
following components%
\begin{equation}
\left(
\begin{array}{c}
b_{1}^{\left( 2\right) } \\
b_{2}^{\left( 2\right) } \\
c_{1}^{\left( 2\right) } \\
c_{2}^{\left( 2\right) }%
\end{array}%
\right) =\left(
\begin{array}{c}
-b_{1}-c_{2} \\
-b_{2}-c_{1} \\
b_{2} \\
b_{1}%
\end{array}%
\right)
\end{equation}%
This EM charge vector $\boldsymbol{v}^{\left( 2\right) }$ is related to $%
v^{\left( 1\right) }$ by the mutation matrix $L_{2}$ given by%
\begin{equation}
L_{2}=\left(
\begin{array}{cccc}
-I_{4} & 0 & 0 & 0 \\
0 & -I_{4} & 0 & 0 \\
0 & I_{4} & I_{4} & 0 \\
I_{4} & 0 & 0 & I_{4}%
\end{array}%
\right) ,  \label{L2}
\end{equation}%
with%
\begin{equation}
\boldsymbol{v}^{\left( 2\right) }=L_{2}v^{\left( 1\right) }
\end{equation}%
and, up on using $\boldsymbol{v}^{\left( 1\right) }=L_{1}v^{\left(
0\right)
} $, we also have%
\begin{equation}
\boldsymbol{v}^{\left( 2\right) }=L_{2}L_{1}v^{\left( 0\right)
}\equiv \mathcal{M}_{2}v^{\left( 0\right) }
\end{equation}%
Like for $L_{1}$, the matrix $L_{2}$ obeys as well $\left(
L_{2}\right) ^{2}=I_{16}$; we also have
\begin{equation}
L_{2}=s_{2}t_{2}=t_{2}s_{2}  \label{mm}
\end{equation}%
with%
\begin{equation}
\begin{tabular}{lll}
$s_{2}=\left(
\begin{array}{cccc}
-I_{4} & 0 & 0 & 0 \\
0 & I_{4} & 0 & 0 \\
0 & 0 & I_{4} & 0 \\
I_{4} & 0 & 0 & I_{4}%
\end{array}%
\right) $ & , & $t_{2}=\left(
\begin{array}{cccc}
I_{4} & 0 & 0 & 0 \\
0 & -I_{4} & 0 & 0 \\
0 & I_{4} & I_{4} & 0 \\
0 & 0 & 0 & I_{4}%
\end{array}%
\right) $%
\end{tabular}%
\end{equation}%
By combining the two successive mutations $L_{1}$ and $L_{2}$, we
get the
mutation matrix $\mathcal{M}_{2}=L_{2}L_{1}$ that maps the quiver $\mathfrak{%
Q}_{0}^{su_{3}}$ into the mutated $\mathfrak{Q}_{2}^{su_{3}}$:%
\begin{equation}
\mathcal{M}_{2}=\left(
\begin{array}{cccc}
-I_{4} & 0 & 0 & -I_{4} \\
0 & -I_{4} & -I_{4} & 0 \\
0 & I_{4} & 0 & 0 \\
I_{4} & 0 & 0 & 0%
\end{array}%
\right)
\end{equation}%
The EM charges of the BPS and anti-BPS states of this quiver are
directly
read from the rows of this matrix. We have%
\begin{equation}
\mathfrak{Q}_{2}^{su_{3}}:%
\begin{tabular}{lll}
$-\left( b_{1}+c_{2}\right) $ & , & $b_{2}$ \\
$-\left( b_{2}+c_{1}\right) $ & , & $b_{1}$%
\end{tabular}%
\end{equation}%
The intersection matrix is given by%
\begin{equation}
\mathcal{A}^{\left( 2\right) }=\mathcal{M}_{2}\mathcal{A}^{\left( 0\right) }%
\mathcal{M}_{2}^{T}=\mathcal{A}^{\left( 0\right) }
\end{equation}

\  \  \  \

$iii)$ \emph{the} \emph{quiver} $\mathfrak{Q}_{3}^{su_{3}}$ \newline
Using the property (\ref{RN}), the BPS quiver
$\mathfrak{Q}_{3}^{su_{3}}$ is
given by%
\begin{equation}
\begin{tabular}{llll}
$\boldsymbol{v}^{\left( 3\right) }=\mathcal{M}_{3}v^{\left( 0\right)
}$ & ,
&  & $\mathcal{M}_{3}=L_{1}\mathcal{M}_{2}$%
\end{tabular}%
\end{equation}%
with%
\begin{equation}
\mathcal{M}_{3}=\left(
\begin{array}{cccc}
0 & 0 & 0 & -I_{4} \\
0 & 0 & -I_{4} & 0 \\
0 & -I_{4} & 0 & 0 \\
-I_{4} & 0 & 0 & 0%
\end{array}%
\right)
\end{equation}%
and%
\begin{equation}
\left(
\begin{array}{c}
b_{1}^{\left( 3\right) } \\
b_{2}^{\left( 3\right) } \\
c_{1}^{\left( 3\right) } \\
c_{2}^{\left( 3\right) }%
\end{array}%
\right) =\left(
\begin{array}{c}
-c_{2} \\
-c_{1} \\
-b_{2} \\
-b_{1}%
\end{array}%
\right)
\end{equation}%
We also have $\mathcal{A}^{\left( 3\right) }=\mathcal{M}_{3}\mathcal{A}%
^{\left( 0\right) }\mathcal{M}_{3}^{T}$ $=-\mathcal{A}^{\left(
0\right) }$.

\  \  \  \

$iv)$ \emph{the} \emph{quiver} $\mathfrak{Q}_{4}^{su_{3}}$ \newline
the BPS quiver $\mathfrak{Q}_{4}^{su_{3}}$ is described by
\begin{equation}
\begin{tabular}{ll}
$\boldsymbol{v}^{\left( 4\right) }$ & $=\mathcal{M}_{4}v^{\left(
0\right) }$
\\
$\mathcal{A}^{\left( 4\right) }$ &
$=\mathcal{M}_{4}\mathcal{A}^{\left(
0\right) }\mathcal{M}_{4}^{T}$%
\end{tabular}%
\end{equation}%
with
\begin{equation}
\mathcal{M}_{4}=\left(
\begin{array}{cccc}
0 & 0 & 0 & I_{4} \\
0 & 0 & I_{4} & 0 \\
0 & -I_{4} & -I_{4} & 0 \\
-I_{4} & 0 & 0 & -I_{4}%
\end{array}%
\right) ,\qquad \mathcal{M}_{4}=\left( \mathcal{M}_{2}\right) ^{2}
\end{equation}%
From this mutation matrix, we learn the electric-magnetic charges of
the
quiver. We have%
\begin{equation}
\left(
\begin{array}{c}
b_{1}^{\left( 4\right) } \\
b_{2}^{\left( 4\right) } \\
c_{1}^{\left( 4\right) } \\
c_{2}^{\left( 4\right) }%
\end{array}%
\right) =\left(
\begin{array}{c}
c_{2} \\
c_{1} \\
-b_{2}-c_{1} \\
-b_{1}-c_{2}%
\end{array}%
\right)
\end{equation}

$v)$ \emph{the} \emph{quiver} $\mathfrak{Q}_{5}^{su_{3}}$ \newline
the BPS quiver $\mathfrak{Q}_{5}^{su_{3}}$ is given by%
\begin{equation}
\begin{tabular}{ll}
$\boldsymbol{v}^{\left( 5\right) }$ & $=\mathcal{M}_{5}v^{\left(
0\right) }$
\\
$\mathcal{A}^{\left( 5\right) }$ &
$=\mathcal{M}_{5}\mathcal{A}^{\left(
0\right) }\mathcal{M}_{5}^{T}=-\mathcal{A}^{\left( 0\right) }$%
\end{tabular}%
\end{equation}%
with $\mathcal{M}_{5}=L_{1}\left( \mathcal{M}_{2}\right) ^{2}$ as%
\begin{equation}
\mathcal{M}_{5}=\left(
\begin{array}{cccc}
-I_{4} & 0 & 0 & 0 \\
0 & -I_{4} & 0 & 0 \\
0 & I_{4} & I_{4} & 0 \\
I_{4} & 0 & 0 & I_{4}%
\end{array}%
\right)
\end{equation}%
The corresponding BPS states are%
\begin{equation}
\left(
\begin{array}{c}
b_{1}^{\left( 5\right) } \\
b_{2}^{\left( 5\right) } \\
c_{1}^{\left( 5\right) } \\
c_{2}^{\left( 5\right) }%
\end{array}%
\right) =\left(
\begin{array}{c}
-b_{1} \\
-b_{2} \\
b_{2}+c_{1} \\
b_{1}+c_{2}%
\end{array}%
\right)
\end{equation}

\begin{figure}[tbph]
\begin{center}
\hspace{0cm} \includegraphics[width=12cm]{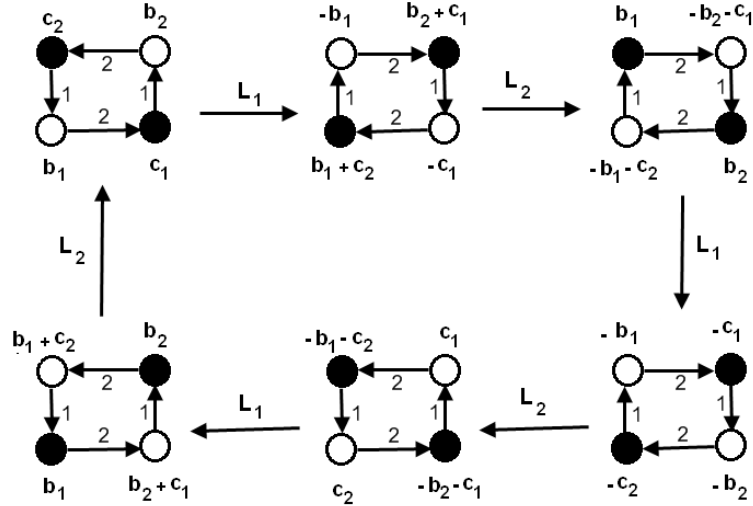}
\end{center}
\par
\vspace{-1 cm} \caption{Mutations of the quiver
$\mathfrak{Q}_{0}^{su_{3}}$ of pure supersymmetric SU$\left(
3\right) $ gauge theory} \label{SU3}
\end{figure}

$vi)$ \emph{the quiver} $\mathfrak{Q}_{6}^{su_{3}}$ \newline
the BPS quiver $\mathfrak{Q}_{6}^{su_{3}}$ is given by the charge vector $%
\boldsymbol{v}^{\left( 6\right) }$ and the intersection matrix $\mathcal{A}%
^{\left( 6\right) }$
\begin{equation}
\begin{tabular}{lll}
$\boldsymbol{v}^{\left( 6\right) }=\mathcal{M}_{6}v^{\left( 0\right)
}$ & , & $\mathcal{A}^{\left( 6\right)
}=\mathcal{M}_{6}\mathcal{A}^{\left(
0\right) }\mathcal{M}_{6}^{T}$%
\end{tabular}%
\end{equation}%
with $\mathcal{M}_{6}=\left( \mathcal{M}_{2}\right) ^{3}$ as%
\begin{equation}
\mathcal{M}_{6}=\left(
\begin{array}{cccc}
I_{4} & 0 & 0 & 0 \\
0 & I_{4} & 0 & 0 \\
0 & 0 & I_{4} & 0 \\
0 & 0 & 0 & I_{4}%
\end{array}%
\right)
\end{equation}%
The BPS states are given by
\begin{equation}
\left(
\begin{array}{c}
b_{1}^{\left( 6\right) } \\
b_{2}^{\left( 6\right) } \\
c_{1}^{\left( 6\right) } \\
c_{2}^{\left( 6\right) }%
\end{array}%
\right) =\left(
\begin{array}{c}
b_{1} \\
b_{2} \\
c_{1} \\
c_{2}%
\end{array}%
\right)  \label{44}
\end{equation}%
and are precisely the BPS states of the primitive quiver $\mathfrak{Q}%
_{0}^{su_{3}}$. The various steps of the mutations are illustrated on fig %
\ref{SU3}.
We conclude this subsection by the following summary:

1) \emph{the} $\mathfrak{Q}_{n}^{su_{3}}$ \emph{quivers}\newline
Generic quivers $\mathfrak{Q}_{n}^{su_{3}}$ describing BPS states in
the strong coupling chamber of the $\mathcal{N}=2$ QFT$_{4}$ with
spontaneously broken $SU\left( 3\right) $ gauge symmetry are
completely characterized by the $\left( v^{\left( 0\right)
},\mathcal{A}^{\left( 0\right) }\right) $ and
the mutation group $\mathcal{G}_{strong}^{su_{3}}$. The entries of $%
v^{\left( n\right) }$ and the intersection matrix are as
\begin{equation}
\begin{tabular}{ll}
$v^{\left( n\right) }$ & $=\left(
\begin{array}{c}
b^{\left( n\right) } \\
c^{\left( n\right) }%
\end{array}%
\right) $ \\
$\mathcal{A}^{\left( n\right) }$ & $=v^{\left( n\right) }\circ
v^{\left(
n\right) }$%
\end{tabular}%
,\qquad n=0,...,5  \label{N3}
\end{equation}%
with
\begin{equation}
\begin{tabular}{llll}
$b^{\left( n\right) }=\left(
\begin{array}{c}
b_{1}^{\left( n\right) } \\
b_{2}^{\left( n\right) }%
\end{array}%
\right) $ & , & $c^{\left( n\right) }=\left(
\begin{array}{c}
c_{1}^{\left( n\right) } \\
c_{2}^{\left( n\right) }%
\end{array}%
\right) $ &
\end{tabular}
\label{3N}
\end{equation}

2) \emph{BPS spectrum}\newline The BPS states of the strong coupling
chamber are given by supersymmetric
BPS states with EM charge vectors as follows%
\begin{equation}
\begin{tabular}{lllll}
$\pm b_{1}$ & $,$ & $\pm c_{1}$ & , & $\pm \left( b_{1}+c_{2}\right) $ \\
$\pm b_{2}$ & $,$ & $\pm c_{2}$ & , & $\pm \left( b_{2}+c_{1}\right) $%
\end{tabular}%
\end{equation}%
The number of the BPS and anti-BPS states is twice the order of the
group of mutation transformations $\mathcal{G}_{strong}^{su_{3}}$
which is isomorphic to the order 6 dihedral Coxeter group
\begin{equation}
\mathcal{G}_{strong}^{su_{3}}\simeq Dih_{6}
\end{equation}

3) \emph{structure of }$\mathcal{G}_{strong}^{su_{3}}$\newline
The matrix mutations $\mathcal{M}_{n}$ map the primitive $\mathfrak{Q}%
_{0}^{su_{3}}$ into $\mathfrak{Q}_{n}^{su_{3}}$. These mutations,
which can be learnt from eqs(\ref{22}-\ref{44}), are invertible and
form a discrete
group $\mathcal{G}_{strong}^{su_{3}}$ with elements%
\begin{equation}
\begin{tabular}{llllll}
$\mathcal{M}_{1},$ & $\mathcal{M}_{2},$ & $\mathcal{M}_{3},$ & $\mathcal{M}%
_{4},$ & $\mathcal{M}_{5},$ & $\mathcal{M}_{6}$%
\end{tabular}
\label{SE}
\end{equation}%
These elements obey a set of properties; in particular
\begin{equation}
\mathcal{M}_{n+6}=\mathcal{M}_{n},\qquad \left(
\mathcal{M}_{2}\right) ^{3}=I_{16}
\end{equation}%
as well as
\begin{equation}
\begin{tabular}{lll}
$\mathcal{M}_{3}=\mathcal{M}_{1}\mathcal{M}_{2}$ & , & $\mathcal{M}_{5}=%
\mathcal{M}_{1}\mathcal{M}_{4}$ \\
$\mathcal{M}_{4}=\left( \mathcal{M}_{2}\right) ^{2}$ & , & $\mathcal{M}_{6}=%
\mathcal{M}_{2}\mathcal{M}_{4}$ \\
&  &
\end{tabular}%
\end{equation}%
leading to the group multiplication table (\ref{TA}). Notice
moreover that the generators $L_{1}$ and $L_{2}$ play a symmetric
role as exhibited by the
following identities%
\begin{equation}
\begin{tabular}{llll}
$\mathcal{M}_{1}=L_{1}$ & , & $\mathcal{M}_{5}=L_{2}$ &  \\
$\mathcal{M}_{2}=L_{2}L_{1}$ & , & $\mathcal{M}_{4}=L_{1}L_{2}$ &
\end{tabular}%
\end{equation}%
and
\begin{equation}
\mathcal{M}_{3}=L_{1}L_{2}L_{1}=L_{2}L_{1}L_{2}
\end{equation}%
This property captures the fact that one needs both left and right
mutations to build the BPS chamber.

\subsection{$SU\left( 4\right) $ model}

First we give the BPS spectrum in the strong coupling chamber of
supersymmetric $SU\left( 4\right) $ gauge theory and the mutation group $%
\mathcal{G}_{strong}^{su_{4}}$. Then, we build the matrix representation of $%
\mathcal{G}_{strong}^{su_{4}}$ and derive its relation with Coxeter group $%
Dih_{8}$.

\subsubsection{BPS spectrum and $\mathcal{G}_{strong}^{su_{4}}$}

\begin{figure}[tbph]
\begin{center}
\hspace{0cm} \includegraphics[width=14cm]{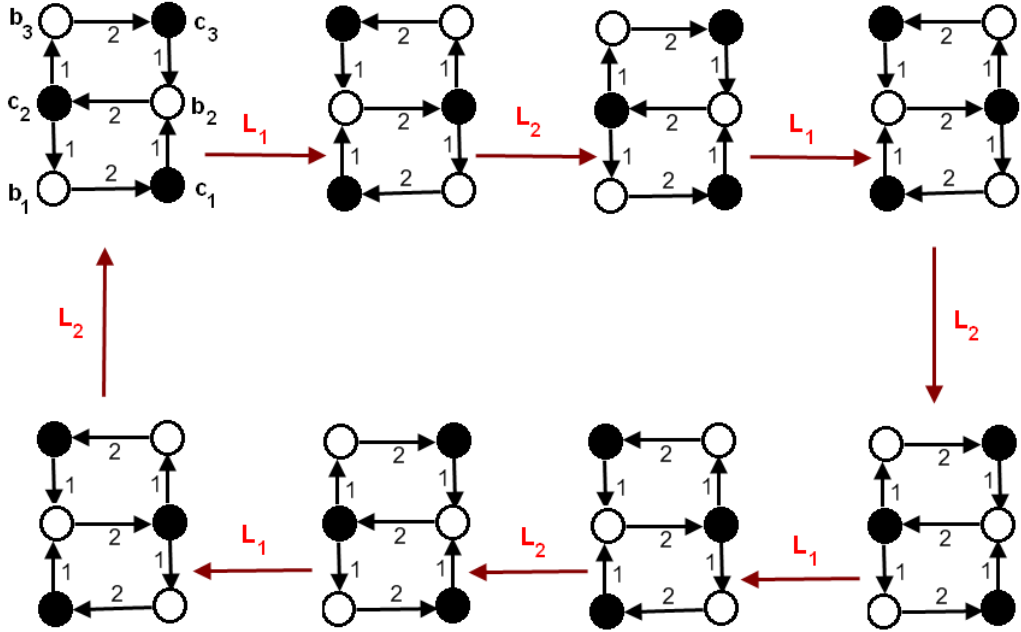}
\end{center}
\par
\vspace{-1 cm} \caption{the mutations of the elementary quiver
$\mathfrak{Q}_{0}^{su_{4}}$.} \label{SU4}
\end{figure}

These BPS states are obtained by mutating the quiver $\mathfrak{Q}%
_{0}^{su_{4}}$ as depicted in fig \ref{SU4}.
The strong coupling chamber contains \emph{12} BPS states and
\emph{12}
anti-BPS states having the electric-magnetic charges $\pm \gamma _{i}$ with $%
\gamma _{i}$ as follows:%
\begin{equation}
\begin{tabular}{llll}
$b_{1},$ & $c_{1},$ & $b_{1}+c_{2},$ & $c_{2}+b_{3},$ \\
$b_{2},$ & $c_{2},$ & $b_{2}+c_{3},$ & $b_{1}+c_{2}+b_{3}$ \\
$b_{3},$ & $c_{3},$ & $c_{1}+b_{2},$ & $c_{1}+b_{2}+c_{3}$%
\end{tabular}
\label{24}
\end{equation}%
To derive this set of BPS states, we use the relations%
\begin{equation}
\begin{tabular}{ll}
$\mathbf{v}^{\left( n\right) }$ &
$=\mathcal{M}_{n}\mathbf{v}^{\left(
0\right) }$ \\
$\mathcal{A}^{\left( n\right) }$ &
$=\mathcal{M}_{n}\mathcal{A}_{n}^{\left(
0\right) }\mathcal{M}_{n}^{T}$%
\end{tabular}%
\end{equation}%
with
\begin{equation}
\begin{tabular}{ll}
$\mathcal{M}_{2k}$ & $=\left( L_{2}L_{1}\right) ^{k}$ \\
$\mathcal{M}_{2k+1}$ & $=L_{1}\mathcal{M}_{2k}$%
\end{tabular}
\label{CH4}
\end{equation}%
Because of the constraint eqs(\ref{cd}) and (\ref{dc}), the mutation
generators $L_{1}$ and $L_{2}$ are given by
\begin{equation}
\begin{tabular}{lll}
$L_{1}=t_{3}t_{2}t_{1}$ & , & $L_{2}=s_{3}s_{2}s_{1}$%
\end{tabular}
\label{r}
\end{equation}%
with r$_{i}$ and s$_{i}$ are reflections obeying in particular%
\begin{equation}
\begin{tabular}{ll}
$\left( t_{i}\right) ^{2}=1,$ & $t_{i}t_{j}=t_{j}t_{i}$ \\
$\left( s_{i}\right) ^{2}=1,$ & $s_{i}s_{j}=s_{j}s_{i}$%
\end{tabular}%
\end{equation}%
and whose matrix representations will be given later on. From the
above identities, we show that $L_{1}$ and $L_{2}$ satisfy the
property
\begin{equation}
\left( L_{1}\right) ^{2}=\left( L_{2}\right) ^{2}=I_{id},
\end{equation}%
as well as the following identities,%
\begin{equation}
\begin{tabular}{llllll}
$\left( L_{2}L_{1}\right) ^{4}$ & $=I_{id}$ &  &  & $\left(
L_{1}L_{2}\right) ^{4}$ & $=I_{id}\left.
\begin{array}{c}
\\
\end{array}%
\right. $ \\
$L_{1}\left( L_{2}L_{1}\right) ^{3}$ & $=L_{2}$ &  &  & $L_{2}\left(
L_{1}L_{2}\right) ^{3}$ & $=L_{1}\left.
\begin{array}{c}
\\
\end{array}%
\right. $ \\
$\left( L_{2}L_{1}\right) ^{3}$ & $=L_{1}L_{2}$ &  &  & $\left(
L_{1}L_{2}\right) ^{3}$ & $=L_{2}L_{1}\left.
\begin{array}{c}
\\
\end{array}%
\right. $ \\
$L_{1}\left( L_{2}L_{1}\right) ^{2}$ & $=L_{2}L_{1}L_{2}$ &  &  & $%
L_{2}\left( L_{1}L_{2}\right) ^{2}$ & $=L_{1}L_{2}L_{1}\left.
\begin{array}{c}
\\
\end{array}%
\right. $%
\end{tabular}
\label{QE}
\end{equation}%
and%
\begin{equation}
\left( L_{2}L_{1}\right) ^{2}=\left( L_{1}L_{2}\right) ^{2}.
\end{equation}%
To get more insight into the structure of these identities, it is
interesting to introduce the $2\times 2$ symmetric matrix $M=\left(
m_{ij}\right) $ with entries as%
\begin{equation}
M=\left(
\begin{array}{cc}
1 & 4 \\
4 & 1%
\end{array}%
\right)  \label{C4}
\end{equation}%
and combine eqs(\ref{QE}) together into the set%
\begin{equation}
W\left( M\right) =\left \langle \left.
\begin{array}{c}
\\
\end{array}%
\right. L_{1},L_{2}|\text{ }\left( L_{i}L_{j}\right)
^{m_{ij}}=I_{id}\right \rangle  \label{4C}
\end{equation}%
which is nothing bat the dihedral group $Dih_{8}$. We also have,
\begin{equation}
\begin{tabular}{lllll}
$\mathcal{M}_{1}=$ & $L_{1}$ & , & $\mathcal{M}_{7}=$ & $L_{2}$ \\
$\mathcal{M}_{2}=$ & $L_{2}L_{1}$ & , & $\mathcal{M}_{6}=$ & $L_{1}L_{2}$ \\
$\mathcal{M}_{3}=$ & $L_{1}L_{2}L_{1}$ & , & $\mathcal{M}_{5}=$ & $%
L_{2}L_{1}L_{2}$%
\end{tabular}
\label{MM}
\end{equation}%
and%
\begin{equation}
\begin{tabular}{ll}
$\mathcal{M}_{0}=$ & $\mathcal{M}_{8}={\small I}_{id}$ \\
$\mathcal{M}_{4}=$ & $\left( L_{1}L_{2}\right) ^{2}=\left(
L_{2}L_{1}\right)
^{2}$%
\end{tabular}
\label{NN}
\end{equation}%
By help of these relations, we can compute the group multiplication
table;
we find:%
\begin{equation}
\begin{tabular}{l|llllllll}
$\mathcal{G}_{M}^{su_{4}}$ & ${\small I}_{id}$ & $\mathcal{M}_{1}$ & $%
\mathcal{M}_{2}$ & $\mathcal{M}_{3}$ & $\mathcal{M}_{4}$ &
$\mathcal{M}_{5}$ & $\mathcal{M}_{6}$ & $\mathcal{M}_{7}$ \\ \hline
${\small I}_{id}$ & ${\small I}_{id}$ & $\mathcal{M}_{1}$ &
$\mathcal{M}_{2}$
& $\mathcal{M}_{3}$ & $\mathcal{M}_{4}$ & $\mathcal{M}_{5}$ & $\mathcal{M}%
_{6}$ & $\mathcal{M}_{7}$ \\
$\mathcal{M}_{1}$ & $\mathcal{M}_{1}$ & ${\small I}_{id}$ &
$\mathcal{M}_{3}$
& $\mathcal{M}_{2}$ & $\mathcal{M}_{5}$ & $\mathcal{M}_{6}$ & $\mathcal{M}%
_{7}$ & $\mathcal{M}_{6}$ \\
$\mathcal{M}_{2}$ & $\mathcal{M}_{2}$ & $\mathcal{M}_{7}$ &
$\mathcal{M}_{4}$
& $\mathcal{M}_{1}$ & $\mathcal{M}_{6}$ & $\mathcal{M}_{3}$ & ${\small I}%
_{id}$ & $\mathcal{M}_{5}$ \\
$\mathcal{M}_{3}$ & $\mathcal{M}_{3}$ & $\mathcal{M}_{6}$ &
$\mathcal{M}_{5}$
& ${\small I}_{id}$ & $\mathcal{M}_{7}$ & $\mathcal{M}_{2}$ & $\mathcal{M}%
_{1}$ & $\mathcal{M}_{4}$ \\
$\mathcal{M}_{4}$ & $\mathcal{M}_{4}$ & $\mathcal{M}_{5}$ &
$\mathcal{M}_{6}$
& $\mathcal{M}_{7}$ & ${\small I}_{id}$ & $\mathcal{M}_{1}$ & $\mathcal{M}%
_{2}$ & $\mathcal{M}_{3}$ \\
$\mathcal{M}_{5}$ & $\mathcal{M}_{5}$ & $\mathcal{M}_{4}$ &
$\mathcal{M}_{7}$
& $\mathcal{M}_{6}$ & $\mathcal{M}_{1}$ & ${\small I}_{id}$ & $\mathcal{M}%
_{3}$ & $\mathcal{M}_{2}$ \\
$\mathcal{M}_{6}$ & $\mathcal{M}_{6}$ & $\mathcal{M}_{3}$ & ${\small
I}_{id}$
& $\mathcal{M}_{5}$ & $\mathcal{M}_{2}$ & $\mathcal{M}_{7}$ & $\mathcal{M}%
_{4}$ & $\mathcal{M}_{1}$ \\
$\mathcal{M}_{7}$ & $\mathcal{M}_{7}$ & $\mathcal{M}_{2}$ &
$\mathcal{M}_{1}$
& $\mathcal{M}_{4}$ & $\mathcal{M}_{3}$ & $\mathcal{M}_{6}$ & $\mathcal{M}%
_{5}$ & ${\small I}_{id}$ \\ \hline
\end{tabular}
\label{GR}
\end{equation}%
and%
\begin{equation}
\left \vert \mathcal{G}_{strong}^{su_{4}}\right \vert =8
\end{equation}%
For an explicit check of this multiplication table; use
eq(\ref{CH4}) and
the expression of the generators L$_{1}$ and L$_{2}$ given eq(\ref{L12}-\ref%
{12L}).

\subsubsection{Matrix realization of $\mathcal{G}_{strong}^{su_{4}}$}

In deriving the explicit set of the BPS\ spectrum of this theory, we
have to
use the relation (\ref{RN}) allowing to express the mutation matrices as $%
\mathcal{M}_{2k}=\left( L_{2}L_{1}\right) ^{k}$ and $\mathcal{M}_{2k+1}=L_{1}%
\mathcal{M}_{2k}$ with $0\leq k<4$ with $L_{1}$ and $L_{2}$ realized
as
\begin{equation}
L_{1}=\left(
\begin{array}{cccccc}
I_{6} & 0 & 0 & 0 & I_{6} & 0 \\
0 & I_{6} & 0 & I_{6} & 0 & I_{6} \\
0 & 0 & I_{6} & 0 & I_{6} & 0 \\
0 & 0 & 0 & -I_{6} & 0 & 0 \\
0 & 0 & 0 & 0 & -I_{6} & 0 \\
0 & 0 & 0 & 0 & 0 & -I_{6}%
\end{array}%
\right)   \label{L12}
\end{equation}%
and%
\begin{equation}
L_{2}=\left(
\begin{array}{cccccc}
-I_{6} & 0 & 0 & 0 & 0 & 0 \\
0 & -I_{6} & 0 & 0 & 0 & 0 \\
0 & 0 & -I_{6} & 0 & 0 & 0 \\
0 & I_{6} & 0 & I_{6} & 0 & 0 \\
I_{6} & 0 & I_{6} & 0 & I_{6} & 0 \\
0 & I_{6} & 0 & 0 & 0 & I_{6}%
\end{array}%
\right)   \label{12L}
\end{equation}%
where $I_{6}$ stand for 66 identity matrix; below we shall ignore
this detail by replacing $I_{6}$ by the number $1$; \textrm{see
footnote 5}.
These matrices follow from (\ref{r}) with%
\begin{equation}
\begin{tabular}{ll}
$t_{1}=\left(
\begin{array}{cccccc}
-1 & 0 & 0 & 0 & 1 & 0 \\
0 & 1 & 0 & 0 & 0 & 0 \\
0 & 0 & 1 & 0 & 0 & 0 \\
0 & 0 & 0 & 1 & 0 & 0 \\
0 & 0 & 0 & 0 & 1 & 0 \\
0 & 0 & 0 & 0 & 0 & 1%
\end{array}%
\right) ,$ & $s_{1}=\left(
\begin{array}{cccccc}
1 & 0 & 0 & 0 & 0 & 0 \\
0 & 1 & 0 & 0 & 0 & 0 \\
0 & 0 & 1 & 0 & 0 & 0 \\
0 & 1 & 0 & -1 & 0 & 0 \\
0 & 0 & 0 & 0 & 1 & 0 \\
0 & 0 & 0 & 0 & 0 & 1%
\end{array}%
\right) $%
\end{tabular}%
\end{equation}%
and%
\begin{equation}
\begin{tabular}{ll}
$t_{2}=\left(
\begin{array}{cccccc}
1 & 0 & 0 & 0 & 0 & 0 \\
0 & -1 & 0 & 1 & 0 & 1 \\
0 & 0 & 1 & 0 & 0 & 0 \\
0 & 0 & 0 & 1 & 0 & 0 \\
0 & 0 & 0 & 0 & 1 & 0 \\
0 & 0 & 0 & 0 & 0 & 1%
\end{array}%
\right) ,$ & $s_{2}=\left(
\begin{array}{cccccc}
1 & 0 & 0 & 0 & 0 & 0 \\
0 & 1 & 0 & 0 & 0 & 0 \\
0 & 0 & 1 & 0 & 0 & 0 \\
0 & 0 & 0 & 1 & 0 & 0 \\
1 & 0 & 1 & 0 & -1 & 0 \\
0 & 0 & 0 & 0 & 0 & 1%
\end{array}%
\right) $%
\end{tabular}%
\end{equation}%
as well as%
\begin{equation}
\begin{tabular}{ll}
$t_{3}=\left(
\begin{array}{cccccc}
1 & 0 & 0 & 0 & 1 & 0 \\
0 & 1 & 0 & 0 & 0 & 0 \\
0 & 0 & -1 & 0 & 0 & 0 \\
0 & 0 & 0 & 1 & 0 & 0 \\
0 & 0 & 0 & 0 & 1 & 0 \\
0 & 0 & 0 & 0 & 0 & 1%
\end{array}%
\right) ,$ & $s_{3}=\left(
\begin{array}{cccccc}
1 & 0 & 0 & 0 & 0 & 0 \\
0 & 1 & 0 & 0 & 0 & 0 \\
0 & 0 & 1 & 0 & 0 & 0 \\
0 & 0 & 0 & 1 & 0 & 0 \\
0 & 0 & 0 & 0 & 1 & 0 \\
0 & 1 & 0 & 0 & 0 & -1%
\end{array}%
\right) $%
\end{tabular}%
\end{equation}%
Moreover, using the expressions of $L_{1}$ and $L_{2}$ and eqs(\ref{MM}-\ref%
{NN}), we can compute the eight mutation matrices $\mathcal{M}_{n}$
that
generate the $\mathfrak{Q}_{n}^{su_{4}}$ quivers starting from $\mathfrak{Q}%
_{0}^{su_{4}}$. We find the following:

$\emph{i)}$\emph{\ the quiver} $\mathfrak{Q}_{1}^{su_{4}}$ \emph{from} $%
\mathfrak{Q}_{0}^{su_{4}}$\newline The mutation matrix
$\mathcal{M}_{1}$ is given by $L_{1}$; from which we
read the EM charge vector%
\begin{equation}
\mathbf{v}^{\left( 1\right) }=L_{1}\mathbf{v}^{\left( 0\right)
}=\left(
\begin{array}{c}
b_{1}+c_{2} \\
c_{1}+b_{2}+c_{3} \\
c_{2}+b_{3} \\
-c_{1} \\
-c_{2} \\
-c_{3}%
\end{array}%
\right)
\end{equation}%
and the intersection matrix%
\begin{equation}
\mathcal{A}^{\left( 1\right) }=L_{1}\mathcal{A}^{\left( 0\right) }L_{1}^{T}=-%
\mathcal{A}^{\left( 0\right) }
\end{equation}%
This mutation allows to engineer \emph{3} composite BPS states: $%
b_{1}+c_{2}, $ $c_{1}+b_{2}+c_{3},$ $c_{2}+b_{3};$ and \emph{3}
anti-BPS ones: $-c_{i}$.

$\emph{ii)}$\emph{\ the quiver} $\mathfrak{Q}_{2}^{su_{4}}$ \emph{from} $%
\mathfrak{Q}_{0}^{su_{4}}$\newline
The mutation matrix $\mathcal{M}_{2}$ leading to the quiver $\mathfrak{Q}%
_{2}^{su_{4}}$ reads as follows:%
\begin{equation}
\mathcal{M}_{2}=\left(
\begin{array}{cccccc}
-1 & 0 & 0 & 0 & -1 & 0 \\
0 & -1 & 0 & -1 & 0 & -1 \\
0 & 0 & -1 & 0 & -1 & 0 \\
0 & 1 & 0 & 0 & 0 & 1 \\
1 & 0 & 1 & 0 & 1 & 0 \\
0 & 1 & 0 & 1 & 0 & 0%
\end{array}%
\right)
\end{equation}%
From this matrix, we determine the EM charge vector%
\begin{equation}
\mathbf{v}^{\left( 2\right) }=\left(
\begin{array}{c}
-b_{1}-c_{2} \\
-c_{1}-b_{2}-c_{3} \\
-c_{2}-b_{3} \\
b_{2}+c_{3} \\
b_{1}+c_{2}+b_{3} \\
c_{1}+b_{2}%
\end{array}%
\right)
\end{equation}%
and the intersection matrix
\begin{equation}
\mathcal{A}^{\left( 2\right) }=\mathcal{M}_{2}\mathcal{A}^{\left( 0\right) }%
\mathcal{M}_{2}^{T}=\mathcal{A}^{\left( 0\right) }
\end{equation}%
This step allows also to engineer \emph{3} composite BPS states and
\emph{3} anti-BPS ones.

$\emph{iii)}$\emph{\ the quiver} $\mathfrak{Q}_{3}^{su_{4}}$ \emph{from} $%
\mathfrak{Q}_{0}^{su_{4}}$\newline The mutation matrix of
$\mathfrak{Q}_{3}^{su_{4}}$ reads as
\begin{equation}
\mathcal{M}_{3}=\left(
\begin{array}{cccccc}
0 & 0 & 1 & 0 & 0 & 0 \\
0 & 1 & 0 & 0 & 0 & 0 \\
1 & 0 & 0 & 0 & 0 & 0 \\
0 & -1 & 0 & 0 & 0 & -1 \\
-1 & 0 & -1 & 0 & -1 & 0 \\
0 & -1 & 0 & -1 & 0 & 0%
\end{array}%
\right)
\end{equation}%
and leads to%
\begin{equation}
\mathbf{v}^{\left( 3\right) }=\left(
\begin{array}{c}
b_{3} \\
b_{2} \\
b_{1} \\
-b_{2}-c_{3} \\
-b_{1}-c_{2}-b_{3} \\
-c_{1}-b_{2}%
\end{array}%
\right)
\end{equation}%
with quiver intersection matrix%
\begin{equation}
\mathcal{A}^{\left( 3\right) }=\mathcal{M}_{3}\mathcal{A}^{\left( 0\right) }%
\mathcal{M}_{3}^{T}
\end{equation}

$\emph{iv)}$\emph{\ the quiver} $\mathfrak{Q}_{4}^{su_{4}}$ \emph{from} $%
\mathfrak{Q}_{0}^{su_{4}}$\newline the matrix $\mathcal{M}_{4}$ is
given by
\begin{equation}
\mathcal{M}_{4}=\left(
\begin{array}{cccccc}
0 & 0 & -1 & 0 & 0 & 0 \\
0 & -1 & 0 & 0 & 0 & 0 \\
-1 & 0 & 0 & 0 & 0 & 0 \\
0 & 0 & 0 & 0 & 0 & -1 \\
0 & 0 & 0 & 0 & -1 & 0 \\
0 & 0 & 0 & -1 & 0 & 0%
\end{array}%
\right)
\end{equation}%
leading to%
\begin{equation}
\mathbf{v}^{\left( 4\right) }=\left(
\begin{array}{c}
-b_{3} \\
-b_{2} \\
-b_{1} \\
-c_{3} \\
-c_{2} \\
-c_{1}%
\end{array}%
\right) ,\qquad \mathcal{A}^{\left( 4\right) }=\mathcal{M}_{4}\mathcal{A}%
^{\left( 0\right) }\mathcal{M}_{4}^{T}
\end{equation}

$\emph{v)}$\emph{\ BPS quiver} $\mathfrak{Q}_{5}^{su_{4}}$ \emph{from} $%
\mathfrak{Q}_{0}^{su_{4}}$\newline In this case, we have
\begin{equation}
\mathcal{M}_{5}=\left(
\begin{array}{cccccc}
0 & 0 & -1 & 0 & -1 & 0 \\
0 & -1 & 0 & -1 & 0 & -1 \\
-1 & 0 & 0 & 0 & -1 & 0 \\
0 & 0 & 0 & 0 & 0 & 1 \\
0 & 0 & 0 & 0 & 1 & 0 \\
0 & 0 & 0 & 1 & 0 & 0%
\end{array}%
\right)
\end{equation}%
with%
\begin{equation}
\mathbf{v}^{\left( 5\right) }=\left(
\begin{array}{c}
-b_{3}-c_{2} \\
-b_{2}-c_{1}-c_{3} \\
-b_{1}-c_{2} \\
b_{3} \\
b_{2} \\
b_{1}%
\end{array}%
\right) ,\qquad \mathcal{A}^{\left( 5\right) }=\mathcal{M}_{5}\mathcal{A}%
^{\left( 0\right) }\mathcal{M}_{5}^{T}
\end{equation}

$\emph{vi)}$\emph{\ the quivers} $\mathfrak{Q}_{6/7}^{su_{4}}$ \emph{from} $%
\mathfrak{Q}_{0}^{su_{4}}$\newline The mutation matrices
$\mathcal{M}_{6}$ and $\mathcal{M}_{7}$ respectively
describing the BPS quivers $\mathfrak{Q}_{6}^{su_{4}}$ and $\mathfrak{Q}%
_{7}^{su_{4}}$ are as follows%
\begin{equation}
\begin{tabular}{lll}
$\mathcal{M}_{6}=\left(
\begin{array}{cccccc}
0 & 0 & 1 & 0 & 1 & 0 \\
0 & 1 & 0 & 1 & 0 & 1 \\
1 & 0 & 0 & 0 & 1 & 0 \\
0 & -1 & 0 & -1 & 0 & 0 \\
-1 & 0 & -1 & 0 & -1 & 0 \\
0 & -1 & 0 & 0 & 0 & -1%
\end{array}%
\right) $ & , & $\mathcal{M}_{7}=\left(
\begin{array}{cccccc}
-1 & 0 & 0 & 0 & 0 & 0 \\
0 & -1 & 0 & 0 & 0 & 0 \\
0 & 0 & -1 & 0 & 0 & 0 \\
0 & 1 & 0 & 1 & 0 & 0 \\
1 & 0 & 1 & 0 & 1 & 0 \\
0 & 1 & 0 & 0 & 0 & 1%
\end{array}%
\right) $%
\end{tabular}%
\end{equation}%
The corresponding EM charge vectors are given by%
\begin{equation}
\begin{tabular}{lll}
$\mathbf{v}^{\left( 6\right) }=\left(
\begin{array}{c}
b_{3}+c_{2} \\
b_{2}+c_{1}+c_{3} \\
b_{1}+c_{2} \\
-b_{2}-c_{1} \\
-b_{1}-b_{3}-c_{2} \\
-b_{2}-c_{2}%
\end{array}%
\right) $ & , & $\mathbf{v}^{\left( 7\right) }=\left(
\begin{array}{c}
-b_{1} \\
-b_{2} \\
-b_{3} \\
b_{2}+c_{1} \\
b_{1}+c_{2}+b_{3} \\
b_{2}+c_{3}%
\end{array}%
\right) $%
\end{tabular}%
\end{equation}%
and the intersection matrix as%
\begin{equation}
\mathcal{A}^{\left( 6\right) }=\mathcal{M}_{6}\mathcal{A}^{\left( 0\right) }%
\mathcal{M}_{6}^{T},\qquad \mathcal{A}^{\left( 7\right) }=\mathcal{M}_{7}%
\mathcal{A}^{\left( 0\right) }\mathcal{M}_{7}^{T}
\end{equation}%
Therefore the total number of BPS and anti-BPS states in the strong
coupling chamber is indeed equal to $2\left( \dim SU_{4}-3\right) $
namely
\begin{equation*}
\#_{bps+antibps}=12+12=24
\end{equation*}%
From the view of our mutation group construction, this number is
equal to the rank of $SU_{4}$\ times the order of
$\mathcal{G}_{strong}^{su_{4}}$;
this leads to%
\begin{equation}
3\times \mathcal{G}_{strong}^{su_{4}}=3\times 8=24
\end{equation}

\section{$SO\left( 2N\right) $ and $E_{r}$ models}

In this section, we extend the method developed above for $SU\left(
N\right) $ to study the BPS\ spectra of $\mathcal{N}=2$ QFTs with
$SO\left( 2r\right) $ and exceptional $E_{r}$ gauge symmetries. As
our analysis is explicit, we will focus on the examples of $SO\left(
8\right) $ and $E_{6}$ gauge groups; then we give the results for
generic $SO\left( 2r\right) $ and $E_{r}$.

\subsection{BPS states in supersymmetric $SO\left( 2r\right) $ gauge theory}

First we consider the $\mathcal{N}=2$ supersymmetric $SO\left(
8\right) $ gauge model; then we give the extension to $SO\left(
2r\right) $ for generic rank $r$.

\subsubsection{$SO\left( 8\right) $ gauge model}

\begin{figure}[tbph]
\begin{center}
\hspace{0cm} \includegraphics[width=8cm]{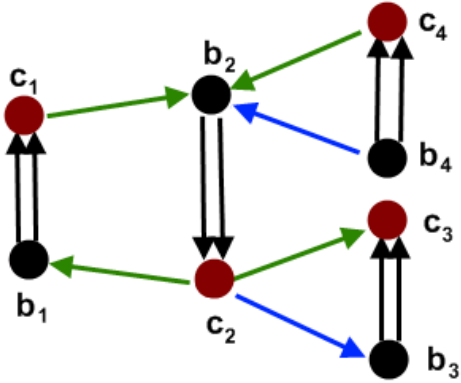}
\end{center}
\par
\vspace{-1 cm} \caption{the primitive BPS quiver
$\mathfrak{Q}_{0}^{so_{8}}$ of the supersymmetric SO$\left( 8\right)
$ gauge theory.} \label{O4}
\end{figure}

To start recall that according to results of \textrm{\cite{1A,1B}},
the strong coupling chamber of the supersymmetric $SO\left( 8\right)
$ gauge model has 24 BPS states and 24 anti-BPS ones. To work out
explicitly these states, we start by the primitive quiver
$\mathfrak{Q}_{0}^{so_{8}}$
encoding the EM charges of the 4 elementary monopoles $\mathfrak{M}_{1},...,%
\mathfrak{M}_{4}$ and the 4 elementary dyons $\mathfrak{D}_{1},...,\mathfrak{%
D}_{4}$. The EM charges of these states are respectively given by%
\begin{equation}
\begin{tabular}{llll}
$b_{i}=\left(
\begin{array}{c}
0 \\
\alpha _{i}%
\end{array}%
\right) $ & , & $c_{i}=\left(
\begin{array}{c}
\alpha _{i} \\
-\alpha _{i}%
\end{array}%
\right) $ &
\end{tabular}%
\end{equation}%
where now $\alpha _{1},...,\alpha _{2}$ are the 4 simple roots of
$SO\left( 8\right) $. The BPS quiver $\mathfrak{Q}_{0}^{so_{8}}$ of
the 4 monopoles and 4 dyons is given by fig \ref{O4}.

The intersection matrix $\mathcal{A}^{\left( 0\right) }$ of the
elementary BPS quiver $\mathfrak{Q}_{0}^{so_{8}}$ is given by
\begin{equation}
\mathcal{A}^{\left( 0\right) }=\left(
\begin{array}{cc}
0 & \alpha _{i}.\alpha _{j} \\
-\alpha _{i}.\alpha _{j} & 0%
\end{array}%
\right)
\end{equation}%
To get the remaining 16 BPS and 16 anti-BPS states, we use the
mutation group method used for $SU\left( N\right) $ gauge symmetry.
Setting the central charges $Z\left( b_{i}\right) =X_{i}$, $Z\left(
c_{i}\right) =Y_{i}$ and considering the chamber
\begin{equation}
\begin{tabular}{ll}
$\arg X_{i}$ & $=\arg X$ \\
$\arg Y_{i}$ & $=\arg Y$%
\end{tabular}%
,\qquad \forall \text{ }i=1,2,3,4
\end{equation}%
with
\begin{equation}
\arg Y>\arg X,
\end{equation}%
the mutations of the elementary quiver $\mathfrak{Q}_{0}^{so_{8}}$
are encoded in the following mutation matrix sequence
\begin{equation}
\begin{tabular}{ll}
$\mathcal{M}_{2k}$ & $=\left( L_{2}L_{1}\right) ^{k}$ \\
$\mathcal{M}_{2k+1}$ & $=L_{1}\mathcal{M}_{2k}$%
\end{tabular}%
\end{equation}%
with positive integer $k$; and where $L_{1}$ and $L_{2}$ are matrix
generators to be given later on.\ In this group theoretical method,
the EM
charge vectors of the BPS states%
\begin{equation}
v^{\left( n\right) }=\left(
\begin{array}{c}
b^{\left( n\right) } \\
c^{\left( n\right) }%
\end{array}%
\right)
\end{equation}%
appear as the rows of the mutation matrices $\mathcal{M}_{n}$. Seen
that the
BPS spectrum of this chamber is finite, it follows that the above $\mathcal{M%
}_{n}$ sequence should be periodic. It happens that the period of
the
mutations is given by $n=12$,%
\begin{equation}
\mathcal{M}_{12}=I_{id},\qquad \mathcal{M}_{n+12}=\mathcal{M}_{n},
\end{equation}%
and moreover%
\begin{equation}
\mathcal{M}_{6}=-I_{id}
\end{equation}%
This last property implies in turns that we also have%
\begin{equation}
\mathcal{M}_{n+6}=-\mathcal{M}_{n}
\end{equation}%
and so instead of determining the twelve $\mathcal{M}_{n}$'s, it is
enough to compute the first 6 elements of the mutation group namely
\begin{equation}
\begin{tabular}{lll}
$\mathcal{M}_{1},$ & $\mathcal{M}_{2},$ & $\mathcal{M}_{3}$ \\
$\mathcal{M}_{4},$ & $\mathcal{M}_{5},$ & $\mathcal{M}_{6}$%
\end{tabular}%
\end{equation}

i) \emph{the mutation matrix }$\mathcal{M}_{1}$\newline
This mutation maps $\mathfrak{Q}_{0}^{so_{8}}$ into the quiver $\mathfrak{Q}%
_{1}^{so_{8}}$; and is given by the $64\times 64$ matrix generator%
\begin{equation}
L_{1}=\left(
\begin{array}{cccccccc}
I & 0 & 0 & 0 & 0 & I & 0 & 0 \\
0 & I & 0 & 0 & I & 0 & I & I \\
0 & 0 & I & 0 & 0 & I & 0 & 0 \\
0 & 0 & 0 & I & 0 & I & 0 & 0 \\
0 & 0 & 0 & 0 & -I & 0 & 0 & 0 \\
0 & 0 & 0 & 0 & 0 & -I & 0 & 0 \\
0 & 0 & 0 & 0 & 0 & 0 & -I & 0 \\
0 & 0 & 0 & 0 & 0 & 0 & 0 & -I%
\end{array}%
\right)
\end{equation}%
where now $I=I_{8}$. This matrix acts on the EM charge vector $\mathbf{v}%
^{\left( 0\right) }=(b^{\left( 0\right) },c^{\left( 0\right) })$ of
the BPS
quiver $\mathfrak{Q}_{0}^{so_{8}}$ to give the corresponding vector $\mathbf{%
v}^{\left( 1\right) }=(b^{\left( 1\right) },c^{\left( 1\right) })$
of the
mutated Quiver $\mathfrak{Q}_{1}^{so_{8}}$. We have%
\begin{equation}
\begin{tabular}{ll}
$\mathbf{v}^{\left( 1\right) }$ &
$=\mathcal{M}_{1}\mathbf{v}^{\left(
0\right) }$ \\
$\mathcal{A}^{\left( 1\right) }$ &
$=\mathcal{M}_{1}\mathcal{A}^{\left(
0\right) }\mathcal{M}_{1}^{T}$%
\end{tabular}%
\end{equation}%
or more explicitly
\begin{equation}
\left(
\begin{array}{c}
b_{1}^{\left( 1\right) } \\
b_{2}^{\left( 1\right) } \\
b_{3}^{\left( 1\right) } \\
b_{4}^{\left( 1\right) } \\
c_{1}^{\left( 1\right) } \\
c_{2}^{\left( 1\right) } \\
c_{3}^{\left( 1\right) } \\
c_{4}^{\left( 1\right) }%
\end{array}%
\right) =\left(
\begin{array}{cccccccc}
I & 0 & 0 & 0 & 0 & I & 0 & 0 \\
0 & I & 0 & 0 & I & 0 & I & I \\
0 & 0 & I & 0 & 0 & I & 0 & 0 \\
0 & 0 & 0 & I & 0 & I & 0 & 0 \\
0 & 0 & 0 & 0 & -I & 0 & 0 & 0 \\
0 & 0 & 0 & 0 & 0 & -I & 0 & 0 \\
0 & 0 & 0 & 0 & 0 & 0 & -I & 0 \\
0 & 0 & 0 & 0 & 0 & 0 & 0 & -I%
\end{array}%
\right) \left(
\begin{array}{c}
b_{1} \\
b_{2} \\
b_{3} \\
b_{4} \\
c_{1} \\
c_{2} \\
c_{3} \\
c_{4}%
\end{array}%
\right)
\end{equation}%
leading to the following set BPS\ and anti-BPS states
\begin{equation}
\begin{tabular}{lllll}
$-c_{1},$ & $-c_{2},$ & $b_{1}+c_{2},$ & $b_{3}+c_{2},$ & $b_{4}+c_{2}$ \\
$-c_{3},$ & $-c_{4},$ & \multicolumn{3}{l}{$b_{2}+c_{1}+c_{3}+c_{4}$}%
\end{tabular}%
\end{equation}%
This mutation allows to engineer 4 composite BPS states
$b_{i}^{\left( 1\right) }$ and 4 anti-BPS ones $c_{i}^{\left(
1\right) }=-c_{i}$ with intersection matrix $\mathcal{A}^{\left(
1\right) }$.

\textbf{ii}) \emph{computing }$\mathcal{M}_{2}$\newline The matrix
$\mathcal{M}_{2}$ is obtained by performing two successive mutations
of $\mathfrak{Q}_{0}^{so_{8}}$, or equivalently a simple mutation
of $\mathfrak{Q}_{1}^{so_{8}}$. This operation gives a new quiver $\mathfrak{%
Q}_{2}^{so_{8}}$ involving other BPS and anti-BPS states. The matrix
mutation is given by $\mathcal{M}_{2}=L_{2}L_{1}$ with $L_{1}$ as above and $%
L_{2}$ like%
\begin{equation}
L_{2}=\left(
\begin{array}{cccccccc}
-I & 0 & 0 & 0 & 0 & 0 & 0 & 0 \\
0 & -I & 0 & 0 & 0 & 0 & 0 & 0 \\
0 & 0 & -I & 0 & 0 & 0 & 0 & 0 \\
0 & 0 & 0 & -I & 0 & 0 & 0 & 0 \\
0 & I & 0 & 0 & I & 0 & 0 & 0 \\
I & 0 & I & I & 0 & I & 0 & 0 \\
0 & I & 0 & 0 & 0 & 0 & I & 0 \\
0 & I & 0 & 0 & 0 & 0 & 0 & I%
\end{array}%
\right)
\end{equation}%
leading then to%
\begin{eqnarray}
\mathcal{M}_{2} &=&\left(
\begin{array}{cccccccc}
-I & 0 & 0 & 0 & 0 & -I & 0 & 0 \\
0 & -I & 0 & 0 & -I & 0 & -I & -I \\
0 & 0 & -I & 0 & 0 & -I & 0 & 0 \\
0 & 0 & 0 & -I & 0 & -I & 0 & 0 \\
0 & I & 0 & 0 & 0 & 0 & I & I \\
I & 0 & I & I & 0 & 2I & 0 & 0 \\
0 & I & 0 & 0 & I & 0 & 0 & I \\
0 & I & 0 & 0 & I & 0 & I & 0%
\end{array}%
\right) \\
&&  \notag
\end{eqnarray}%
From the rows of this matrix, we read directly the EM charge vectors $%
\mathbf{v}^{\left( 2\right) }=(b_{i}^{\left( 2\right)
},c_{i}^{\left(
2\right) })$ of the BPS and anti-BPS states. These are given by%
\begin{equation}
\begin{tabular}{lllll}
$-b_{1}-c_{2}$ &  & , & $b_{2}+c_{3}+c_{4}$ &  \\
\multicolumn{2}{l}{$-b_{2}-c_{1}-c_{3}-c_{4}$} & , & \multicolumn{2}{l}{$%
b_{1}+b_{3}+b_{4}+2c_{2}$} \\
$-b_{3}-c_{2}$ &  & , & $b_{2}+c_{1}+c_{4}$ &  \\
$-b_{4}-c_{2}$ &  & , & $b_{2}+c_{1}+c_{3}$ &
\end{tabular}%
\end{equation}%
giving 4 new BPS and 4 anti BPS states. The intersection matrix of
this BPS quiver is $\mathcal{A}^{\left( 2\right)
}=\mathcal{M}_{2}\mathcal{A}^{\left( 0\right) }\mathcal{M}_{2}^{T}$.

\textbf{iii}) \emph{computing }$\mathcal{M}_{3}$\newline
Straightforward calculations lead to
\begin{equation}
\mathcal{M}_{3}=\left(
\begin{array}{cccccccc}
0 & 0 & I & I & 0 & I & 0 & 0 \\
0 & 2I & 0 & 0 & I & 0 & I & I \\
I & 0 & 0 & I & 0 & I & 0 & 0 \\
I & 0 & I & 0 & 0 & I & 0 & 0 \\
0 & -I & 0 & 0 & 0 & 0 & -I & -I \\
-I & 0 & -I & -I & 0 & -2I & 0 & 0 \\
0 & -I & 0 & 0 & -I & 0 & 0 & -I \\
0 & -I & 0 & 0 & -I^{2} & 0 & -I & 0%
\end{array}%
\right)
\end{equation}%
giving 4 new the BPS and 4 new anti-BPS states. The EM charges
$v^{\left( 3\right) }=(b_{i}^{\left( 3\right) },c_{i}^{\left(
3\right) })$ of these
states are read from the rows of $\mathcal{M}_{3}$ and are given by%
\begin{eqnarray}
&&%
\begin{tabular}{lllll}
$b_{3}+b_{4}+c_{2}$ &  & , & $-b_{2}-c_{3}-c_{4}$ &  \\
\multicolumn{2}{l}{$2b_{2}+c_{1}+c_{3}+c_{4}$} & , & \multicolumn{2}{l}{$%
-b_{1}-b_{3}-b_{4}-2c_{2}$} \\
$b_{1}+b_{4}+c_{2}$ &  & , & $-b_{2}-c_{1}-c_{4}$ &  \\
$b_{1}+b_{3}+c_{2}$ &  & , & $-b_{2}-c_{1}-c_{3}$ &
\end{tabular}
\\
&&  \notag
\end{eqnarray}%
The intersection matrix is $\mathcal{A}^{\left( 3\right) }=\mathcal{M}_{3}%
\mathcal{A}^{\left( 0\right) }\mathcal{M}_{3}^{T}.$

iv) \emph{computing }$\mathcal{M}_{4}$\newline
the fourth order mutation matrix reads as%
\begin{equation}
\mathcal{M}_{4}=\left(
\begin{array}{cccccccc}
0 & 0 & -I & -I & 0 & -I & 0 & 0 \\
0 & -2I & 0 & 0 & -I & 0 & -I & -I \\
-I & 0 & 0 & -I & 0 & -I & 0 & 0 \\
-I & 0 & -I & 0 & 0 & -I & 0 & 0 \\
0 & I & 0 & 0 & I & 0 & 0 & 0 \\
I & 0 & I & I & 0 & I & 0 & 0 \\
0 & I & 0 & 0 & 0 & 0 & I & 0 \\
0 & I & 0 & 0 & 0 & 0 & 0 & I%
\end{array}%
\right) \
\end{equation}%
it gives other BPS states and anti-BPS ones with charge vectors
$v^{\left(
4\right) }=(b_{i}^{\left( 4\right) },c_{i}^{\left( 4\right) })$ as:%
\begin{equation}
\begin{tabular}{lllll}
$-b_{3}-b_{4}-c_{2}$ &  & , & $b_{2}+c_{1}$ &  \\
\multicolumn{2}{l}{$-2b_{2}-c_{1}-c_{3}-c_{4}$} & , & \multicolumn{2}{l}{$%
b+b_{2}+b_{3}+c_{2}$} \\
$-b_{1}-b_{4}-c_{2}$ &  & , & $b_{2}+c_{3}$ &  \\
$-b_{1}-b_{3}-c_{2}$ &  & , & $b_{2}+c_{4}$ &
\end{tabular}%
\end{equation}%
with intersection matrix given by $\mathcal{A}^{\left( 4\right) }=\mathcal{M}%
_{4}\mathcal{A}^{\left( 0\right) }\mathcal{M}_{4}^{T}.$

v) \emph{computing }$\mathcal{M}_{5}$\newline In this case we have
\begin{equation}
\mathcal{M}_{5}=\left(
\begin{array}{cccccccc}
I & 0 & 0 & 0 & 0 & 0 & 0 & 0 \\
0 & I & 0 & 0 & 0 & 0 & 0 & 0 \\
0 & 0 & I & 0 & 0 & 0 & 0 & 0 \\
0 & 0 & 0 & I & 0 & 0 & 0 & 0 \\
0 & -I & 0 & 0 & -I & 0 & 0 & 0 \\
-I & 0 & -I & -I & 0 & -I & 0 & 0 \\
0 & -I & 0 & 0 & 0 & 0 & -I & 0 \\
0 & -I & 0 & 0 & 0 & 0 & 0 & -I%
\end{array}%
\right)
\end{equation}%
giving 4 anti-BPS states with EM charges $v^{\left( 5\right)
}=(b_{i}^{\left( 5\right) },c_{i}^{\left( 5\right) })$ as reported below%
\begin{equation}
\begin{tabular}{lllll}
$b_{1}$ &  & , & $-c_{1}-b_{2}$ &  \\
\multicolumn{2}{l}{$b_{2}$} & , & \multicolumn{2}{l}{$%
-b_{2}-b_{3}-b_{4}-c_{2}$} \\
$b_{3}$ &  & , & $-b_{2}-c_{3}$ &  \\
$b_{4}$ &  & , & $-b_{2}-c_{4}$ &
\end{tabular}%
\end{equation}%
and intersection matrix as $\mathcal{A}^{\left( 5\right) }=\mathcal{M}_{5}%
\mathcal{A}^{\left( 0\right) }\mathcal{M}_{5}^{T}.$

vi) \emph{computing }$\mathcal{M}^{\left( 6\right) }$\newline This
mutation matrix reads as
\begin{equation}
\mathcal{M}^{\left( 6\right) }=\left(
\begin{array}{cccccccc}
-I & 0 & 0 & 0 & 0 & 0 & 0 & 0 \\
0 & -I & 0 & 0 & 0 & 0 & 0 & 0 \\
0 & 0 & -I & 0 & 0 & 0 & 0 & 0 \\
0 & 0 & 0 & -I & 0 & 0 & 0 & 0 \\
0 & 0 & 0 & 0 & -I & 0 & 0 & 0 \\
0 & 0 & 0 & 0 & 0 & -I & 0 & 0 \\
0 & 0 & 0 & 0 & 0 & 0 & -I & 0 \\
0 & 0 & 0 & 0 & 0 & 0 & 0 & -I%
\end{array}%
\right)
\end{equation}%
it leads to the anti- BPS image of the BPS quiver $\mathfrak{Q}_{0}^{so_{8}}$%
; the intersection matrix $\mathcal{A}^{\left( 6\right) }$ is equal to $%
\mathcal{A}^{\left( 0\right) }$.\newline Notice that the quivers
$\mathfrak{Q}_{6+n}^{so_{8}}$ are just the CPT conjugates of the
quivers $\mathfrak{Q}_{n}^{so_{8}}$; and so have the same
intersection matrix; i.e $\mathcal{A}^{\left( n+6\right) }=\mathcal{A}%
^{\left( n\right) }$.

\subsubsection{BPS spectrum and extension to $SO\left( 2r\right) $}

Combining together all BPS states making the $12$ mutated quivers $\mathfrak{%
Q}_{n}^{so_{8}}$, we get precisely the \emph{24} BPS and \emph{24}
anti-BPS states of the CPT invariant strong coupling chamber of the
$\mathcal{N}=2$ supersymmetry QFT with $SO\left( 8\right) $ gauge
symmetry. The \emph{24}
BPS states are:%
\begin{equation}
\begin{tabular}{llll}
\multicolumn{4}{l}{\  \  \  \  \  \  \  \  \  \  \  \  \  \  \  \  \
\  \  \  \  \  \  \ charge vectors of BPS states} \\ \hline
&  &  &  \\
$b_{1},$ & $b_{2},$ & $b_{3},$ & $b_{4}$ \\
$c_{1},$ & $c_{2},$ & $c_{3},$ & $c_{4}$ \\
$b_{1}+c_{2},$ & $b_{3}+c_{2},$ & $b_{4}+c_{2},$ &
$c_{1}+b_{2}+c_{3}+c_{4}$
\\
$b_{2}+c_{1},$ & $b_{2}+c_{3},$ & $b_{2}+c_{4}$ &
$b_{1}+c_{2}+b_{3}+b_{4}$
\\
$b_{2}+c_{3}+c_{4},$ & $b_{2}+c_{1}+c_{4},$ & $b_{2}+c_{1}+c_{3},$ & $%
b_{1}+b_{3}+b_{4}+2c_{2}$ \\
$b_{3}+b_{4}+c_{2},$ & $b_{1}+b_{4}+c_{2},$ & $b_{1}+b_{3}+c_{2},$ & $%
2b_{2}+c_{1}+c_{3}+c_{4}$ \\
&  &  &  \\ \hline
\end{tabular}
\label{S8}
\end{equation}%
From the $SO\left( 8\right) $ analysis, we also learn the structure of $%
\mathcal{G}_{strong}^{so_{8}}$ relating the various BPS quivers $\mathfrak{Q}%
_{n}^{so_{8}}$ of the theory. This discrete \textrm{symmetry} is
given by
the set%
\begin{equation}
\mathcal{G}_{strpng}^{so_{8}}=\left \{
\begin{tabular}{llllll}
$I_{id},$ & $\mathcal{M}_{2},$ & $\mathcal{M}_{4},$ & $\mathcal{M}_{6}$ & $%
\mathcal{M}_{8},$ & $\mathcal{M}_{10}$ \\
$\mathcal{M}_{1},$ & $\mathcal{M}_{3},$ & $\mathcal{M}_{5}$ & $\mathcal{M}%
_{7},$ & $\mathcal{M}_{9},$ & $\mathcal{M}_{11}$%
\end{tabular}%
\right \}
\end{equation}%
with
\begin{equation}
\mathcal{M}_{12}=\left( \mathcal{M}_{6}\right) ^{2}=I_{id}=I_{64}
\end{equation}%
and containing
\begin{equation}
\left \{
\begin{tabular}{llllll}
$I_{id},$ & $\mathcal{M}_{2},$ & $\mathcal{M}_{4},$ & $\mathcal{M}_{6}$ & $%
\mathcal{M}_{8},$ & $\mathcal{M}_{10}$%
\end{tabular}%
\right \}
\end{equation}%
as a $\mathbb{Z}_{6}$ abelian subsymmetry. \newline As the order
$\left \vert \mathcal{G}_{strong}^{so_{8}}\right \vert $ of this
group is \emph{12}; we have
\begin{equation}
r\times \left \vert \mathcal{G}_{strong}^{so_{8}}\right \vert
=4\times 12=48
\end{equation}%
extending the formula $r\times \left \vert \mathcal{G}_{strong}^{su_{N}}%
\right \vert $ giving the total number of the BPS and anti-BPS
states in the strong coupling chamber of the $\mathcal{N}=2$
supersymmetric $SU\left( N\right) $\ gauge theories; see
eq(\ref{S8}).\newline Moreover, if we assume that the above
$SO\left( 8\right) $ construction is valid for the full $SO\left(
2r\right) $ series, one can use the result of
\textrm{\cite{1A,1B}} to predict the structure of $\mathcal{G}%
_{strong}^{so_{2r}}$ for the full series. This discrete group $\mathcal{G}%
_{strong}^{so_{2r}}$ consists of matrix mutations $\mathcal{M}_{n}$
constrained as
\begin{equation}
\mathcal{M}_{4r-4}=I_{id}=I_{4r^{2}\times 4r^{2}}
\end{equation}%
and so we have%
\begin{equation}
\mathcal{G}_{strong}^{so_{2r}}=\left \{
\begin{tabular}{llllll}
$I_{id},$ & $\mathcal{M}_{2},$ & $\mathcal{M}_{4},$ & $\mathcal{M}_{6},$ & $%
\mathcal{\cdots },$ & $\mathcal{M}_{4r-6}$ \\
$\mathcal{M}_{1},$ & $\mathcal{M}_{3},$ & $\mathcal{M}_{5}$ & $\mathcal{M}%
_{7},$ & $\mathcal{\cdots },$ & $\mathcal{M}_{4r-5}$%
\end{tabular}%
\right \}
\end{equation}%
The order of this mutation symmetry set is
\begin{equation}
\left \vert \mathcal{G}_{strong}^{so_{2r}}\right \vert =4\left(
r-1\right)
\end{equation}%
and so the total number of BPS and anti-BPS states in the strong
coupling chamber of the $\mathcal{N}=2$ supersymmetric $SO\left(
2r\right) $\ gauge
theories, is given by%
\begin{equation}
\#_{bps+antibps}=4r\left( r-1\right)
\end{equation}%
As a check of this relation, consider the examples of the leading
elements of the $SO\left( 2r\right) $ series; in particular gauge
groups $SO\left( 4\right) \simeq SU\left( 2\right) \times SU\left(
2\right) $ and $SO\left(
6\right) \simeq SU\left( 4\right) $ considered in previous section. We have%
\begin{equation}
\begin{tabular}{l|lll}
& $SO_{4}$ \  \  \  \  \  \  \  \  \  & $SO_{6}$ \  \  \  \  \  \  \
\  \  & $SO_{8}$ \  \  \ \  \  \  \  \  \  \\ \hline $\#_{bps}$ &
$2\times 4$ & $3\times 8$ & $4\times 12\left.
\begin{array}{c}
\\
\\
\end{array}%
\right. $%
\end{tabular}%
\end{equation}%
Notice also that setting $r=\frac{N}{2}$ with N even integer, we have $%
\#_{bps+antibps}=N\left( N-2\right) $ which is precisely $2\left(
\dim SO_{N}-{\small rank}SO_{N}\right) $.

\subsection{BPS states in $\mathcal{N}=2$ supersymmetric $E_{r}$ gauge
theories}

We first study the strong coupling chamber of the BPS quiver theory in $%
\mathcal{N}=2$ supersymmetric $E_{6}$ gauge model. Then, we give the
results for the $E_{7}$ and $E_{8}$\ gauge theories by using the
mutation symmetry method; technical details are reported in appendix
III.

\subsubsection{$E_{6}$ gauge model}

The primitive quiver $\mathfrak{Q}_{0}^{E_{6}}$ of the
$\mathcal{N}=2$ supersymmetric $E_{6}$ gauge theory at some point
$u=\left(
u_{1},...,u_{6}\right) $ in the Coulomb branch is given by fig \textrm{\ref%
{E6}}. This quiver involves \emph{6} elementary monopoles $\mathfrak{M}%
_{1},...,\mathfrak{M}_{6}$ and \emph{6} elementary dyons $\mathfrak{D}%
_{1},...,\mathfrak{D}_{6}$ with respective complex central charges
as
\begin{equation}
Z\left( b_{i}\right) =X_{i},\qquad Z\left( c_{i}\right)
=Y_{i},\qquad i=1,...,6
\end{equation}%
These elementary BPS\ states have the respective EM charge vectors%
\begin{equation}
\begin{tabular}{llll}
$b_{i}=\left(
\begin{array}{c}
0 \\
\alpha _{i}%
\end{array}%
\right) $ & , & $c_{i}=\left(
\begin{array}{c}
\alpha _{i} \\
-\alpha _{i}%
\end{array}%
\right) $ &
\end{tabular}%
\end{equation}%
with intersection matrix%
\begin{equation}
\mathcal{A}^{\left( 0\right) }=\left(
\begin{array}{cc}
0 & \alpha _{i}.\alpha _{j} \\
-\alpha _{i}.\alpha _{j} & 0%
\end{array}%
\right)  \label{E}
\end{equation}%
and $\alpha _{1},...,\alpha _{6}$ the \emph{6} simple roots of
$E_{6}$.

\begin{figure}[tbph]
\begin{center}
\hspace{0cm} \includegraphics[width=8cm]{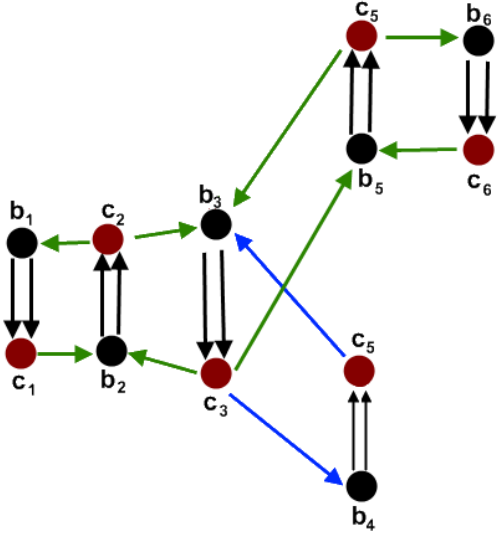}
\end{center}
\par
\vspace{-1 cm} \caption{the primitive quiver
$\mathfrak{Q}_{0}^{E_{6}}$ of the supersymmetric E$_{6}$ gauge
theory.} \label{E6}
\end{figure}

\emph{Constructing the} $\mathfrak{Q}_{n}^{E_{6}}$
\emph{quivers}\newline Following \textrm{\cite{1A,1B}}, this
supersymmetric gauge model should have \emph{72} BPS states and
\emph{72} anti-BPS ones including the \emph{12} elementary ones and
their \emph{12} CPT conjugates. To get the remaining \emph{60} BPS
and \emph{60} anti-BPS states of the strong coupling chamber of the
quiver theory, we first consider the following chamber
\begin{equation}
\begin{tabular}{l}
$\arg X_{i}=\arg X$ \\
$\arg Y_{i}=\arg Y$%
\end{tabular}%
,\qquad \forall \text{ }i=1,...,6  \label{ca}
\end{equation}%
with
\begin{equation}
\arg Y>\arg X.  \label{ac}
\end{equation}%
Then apply the mutation method used above. In this
\textrm{algebraic}
approach, mutations $\mathcal{M}_{n}$, which map the elementary quiver $%
\mathfrak{Q}_{0}^{E_{6}}$ into BPS quivers
$\mathfrak{Q}_{n}^{E_{6}}$, are realized as follows
\begin{equation}
\begin{tabular}{ll}
$\mathcal{M}_{2k}$ & $=\left( L_{2}L_{1}\right) ^{k}$ \\
$\mathcal{M}_{2k+1}$ & $=L_{1}\mathcal{M}_{2k}$%
\end{tabular}
\label{RE}
\end{equation}%
with $k$ a positive integer and $L_{1}$ and $L_{2}$ are the two
$144\times 144$ matrix generators given by eqs(\ref{L},\ref{2L}).
These matrices act on
the quiver vectors%
\begin{equation}
v^{\left( n\right) }=\left(
\begin{array}{c}
b^{\left( n\right) } \\
c^{\left( n\right) }%
\end{array}%
\right)
\end{equation}%
with
\begin{equation}
\begin{tabular}{llll}
$b^{\left( n\right) }=\left(
\begin{array}{c}
b_{1}^{\left( n\right) } \\
\vdots \\
b_{6}^{\left( n\right) }%
\end{array}%
\right) $ & , & $c^{\left( n\right) }=\left(
\begin{array}{c}
c_{1}^{\left( n\right) } \\
\vdots \\
c_{6}^{\left( n\right) }%
\end{array}%
\right) $ &
\end{tabular}%
\end{equation}%
In practice, one needs to known just the quiver
$\mathfrak{Q}_{0}^{E_{6}}$ and the mutation set; all other quivers
are completely determined by algebra. Below we describe rapidly the
derivation of the full BPS and anti-BPS spectrum of the strong
coupling chamber of this $\mathcal{N}=2$ supersymmetric gauge model;
explicit details are reported in the appendix III.

\textbf{1}) \emph{BPS} \emph{quiver
}$\mathfrak{Q}_{0}^{E_{6}}$\newline This BPS quiver is made of the
\emph{12 }elementary BPS states: $6$ monopoles and $6$ dyons with
central charges as in eqs(\ref{ca}-\ref{ac})
and EM charge vectors like%
\begin{equation}
\mathfrak{Q}_{0}^{E_{6}}:%
\begin{tabular}{ll}
$b_{i}^{\left( 0\right) }=b_{i},$ & $c_{i}^{\left( 0\right) }=c_{i}$%
\end{tabular}%
,\qquad i=1,...,6
\end{equation}%
with intersection matrix $\mathcal{A}^{\left( 0\right) }$ as in
(\ref{E}).

\textbf{2}) \emph{BPS} \emph{quiver
}$\mathfrak{Q}_{1}^{E_{6}}$\newline Performing a mutation of
$\mathfrak{Q}_{0}^{E_{6}}$ by using the
transformation $\mathcal{M}_{1}=L_{1}$, we get the quiver $\mathfrak{Q}%
_{1}^{E_{6}}$ that contains \emph{6} new BPS states with charges $%
b_{i}^{\left( 1\right) }$ and \emph{6 }anti-BPS ones with charge $%
c_{i}^{\left( 1\right) }=-c_{i}$. The BPS states are as follows:%
\begin{equation}
\mathfrak{Q}_{1}^{E_{6}}:%
\begin{tabular}{ll}
$b_{1}^{\left( 1\right) }=b_{1}+c_{2},$ & $b_{4}^{\left( 1\right)
}=b_{4}+c_{3}$ \\
$b_{2}^{\left( 1\right) }=b_{2}+c_{1}+c_{3},$ & $b_{5}^{\left(
1\right)
}=b_{5}+c_{3}+c_{6}$ \\
$b_{3}^{\left( 1\right) }=b_{3}+c_{2}+c_{4}+c_{5},$ & $b_{6}^{\left(
1\right) }=b_{6}+c_{5}$ \\
\multicolumn{2}{l}{$c_{i}^{\left( 1\right) }=-c_{i},\qquad i=1,...,6$}%
\end{tabular}%
\end{equation}%
with intersection matrix $\mathcal{A}^{\left( 1\right) }=\mathcal{M}_{1}%
\mathcal{A}^{\left( 0\right) }\mathcal{M}_{1}^{T}.$

\textbf{3}) \emph{BPS} \emph{quiver
}$\mathfrak{Q}_{2}^{E_{6}}$\newline By performing two successive
mutations on $\mathfrak{Q}_{0}^{E_{6}}$; first
by $L_{1}$ on $\mathfrak{Q}_{0}^{E_{6}}$ leading to $\mathfrak{Q}%
_{1}^{E_{6}};$ and second by $L_{2}$ on $\mathfrak{Q}_{1}^{E_{6}}$,
we end with the BPS quiver $\mathfrak{Q}_{2}^{E_{6}}$ that contains
as well 6 new BPS states with charges $(b_{i}^{\left( 2\right)
},c_{i}^{\left( 2\right) })$ given by
\begin{equation}
\mathfrak{Q}_{2}^{E_{6}}:%
\begin{tabular}{ll}
$b_{2}+c_{3},$ & $b_{1}+b_{3}+c_{2}+c_{4}+c_{5}$ \\
$b_{5}+c_{3},$ & $b_{3}+b_{6}+c_{2}+c_{4}+c_{5}$ \\
$b_{3}+c_{2}+c_{5},$ & $b_{2}+b_{4}+b_{5}+c_{1}+2c_{3}+c_{6}$ \\
\multicolumn{2}{l}{$b_{i}^{\left( 2\right) }=-b_{i}^{\left( 1\right)
},\qquad i=1,...,6$}%
\end{tabular}%
\end{equation}%
The other 6 states that complete the quiver
$\mathfrak{Q}_{2}^{E_{6}}$ are given by the anti-BPS states
$b_{i}^{\left( 2\right) }=-b_{i}^{\left(
1\right) }$. The intersection matrix $\mathcal{A}^{\left( 2\right) }=%
\mathcal{M}_{2}\mathcal{A}^{\left( 0\right) }\mathcal{M}_{2}^{T}.$

\textbf{4}) \emph{BPS} \emph{quiver
}$\mathfrak{Q}_{3}^{E_{6}}$\newline Performing three successive
mutations on $\mathfrak{Q}_{0}^{E_{6}}$; first
by $L_{1}$ on $\mathfrak{Q}_{0}^{E_{6}}$, second by $L_{2}$ on $\mathfrak{Q}%
_{1}^{E_{6}}$, and third by $L_{1}$ on $\mathfrak{Q}_{2}^{E_{6}}$,
we get the BPS content of the quiver $\mathfrak{Q}_{3}^{E_{6}}$. We
have
\begin{equation}
\mathfrak{Q}_{3}^{E_{6}}:%
\begin{tabular}{ll}
$b_{3}+c_{4}+c_{5},$ & $b_{2}+b_{4}+b_{5}+2c_{3}+c_{6}$ \\
$b_{3}+c_{2}+c_{4},$ & $b_{2}+b_{4}+b_{5}+c_{1}+2c_{3}$ \\
$b_{2}+b_{5}+c_{1}+c_{3}+c_{6},$ & $b_{1}+2b_{3}+b_{6}+2c_{2}+c_{4}+2c_{5}$%
\end{tabular}%
\end{equation}%
in addition to the anti-BPS states with charges $c_{i}^{\left(
3\right) }=-c_{i}^{\left( 2\right) }$. The intersection matrix
$\mathcal{A}^{\left( 3\right) }=\mathcal{M}_{3}\mathcal{A}^{\left(
0\right) }\mathcal{M}_{3}^{T}.$

\textbf{5}) \emph{BPS} \emph{quiver
}$\mathfrak{Q}_{4}^{E_{6}}$\newline
A mutation on the quiver $\mathfrak{Q}_{3}^{E_{6}}$ by $L_{2}$ leads to $%
\mathfrak{Q}_{4}^{E_{6}}$. The EM charge vectors of its BPS states
are as follows
\begin{equation}
\mathfrak{Q}_{4}:%
\begin{tabular}{ll}
$b_{4}+b_{5}+c_{3}+c_{6},$ & $b_{2}+b_{4}+c_{1}+c_{3},$ \\
$2b_{3}+b_{6}+c_{2}+c_{4}+2c_{5},$ & $b_{1}+2b_{3}+2c_{2}+c_{4}+c_{5},$ \\
$2b_{2}+b_{4}+2b_{5}+c_{1}+3c_{3}+c_{6}$ & $%
b_{1}+b_{3}+b_{6}+c_{2}+c_{4}+c_{5}$%
\end{tabular}%
\end{equation}%
together with $b_{i}^{\left( 4\right) }=-b_{i}^{\left( 3\right) }$.
The
intersection matrix $\mathcal{A}^{\left( 4\right) }=\mathcal{M}_{4}\mathcal{A%
}^{\left( 0\right) }\mathcal{M}_{4}^{T}.$

\textbf{6}) \emph{BPS} \emph{quiver
}$\mathfrak{Q}_{5}^{E_{6}}$\newline A further mutation by $L_{1}$ on
the quiver $\mathfrak{Q}_{4}^{E_{6}}$ gives a new quiver
$\mathfrak{Q}_{5}^{E_{6}}$ leading to 6 new BPS states with EM
charges as:%
\begin{equation}
\mathfrak{Q}_{5}^{E_{6}}:%
\begin{tabular}{ll}
$b_{3}+b_{6}+c_{2}+c_{5}\ ,$ & $b_{2}+b_{4}+2b_{5}+c_{1}+2c_{3}+c_{6}\ $ \\
$b_{2}+b_{4}+b_{5}+2c_{3}\ ,$ & $b_{1}+3b_{3}+b_{6}+2c_{2}+2c_{4}+2c_{5}$ \\
$b_{1}+b_{3}+c_{2}+c_{5}\ ,$ & $2b_{2}+b_{4}+b_{5}+c_{1}+2c_{3}+c_{6}$%
\end{tabular}%
\end{equation}%
in addition to the anti-BPS states with charges $c_{i}^{\left(
5\right) }=-c_{i}^{\left( 4\right) }$. We also have
\begin{equation}
\mathcal{A}^{\left( 5\right) }=\mathcal{M}_{5}\mathcal{A}^{\left( 0\right) }%
\mathcal{M}_{5}^{T}.
\end{equation}

\textbf{7}) \emph{BPS} \emph{quiver
}$\mathfrak{Q}_{6}^{E_{6}}$\newline
Another mutation on $\mathfrak{Q}_{5}^{E_{6}}$ by $L_{2}$ leads to $%
\mathfrak{Q}_{6}^{E_{6}}$ and induces 6 new BPS states with
electromagnetic
charges given by%
\begin{equation}
\mathfrak{Q}_{6}^{E_{6}}:%
\begin{tabular}{ll}
$b_{2}+b_{5}+c_{1}+c_{3},$ & $b_{1}+2b_{3}+b_{6}+2c_{2}+c_{5}\  \ $ \\
$2b_{3}+c_{2}+c_{4}+c_{5},$ &
$2b_{2}+2b_{4}+2b_{5}+c_{1}+c_{4}+3c_{3}+c_{6}$
\\
$b_{2}+b_{5}+c_{3}+c_{6},$ & $b_{1}+2b_{3}+b_{6}+c_{2}+c_{4}+2c_{5}$%
\end{tabular}%
\end{equation}

\textbf{8}) \emph{BPS} \emph{quiver
}$\mathfrak{Q}_{7}^{E_{6}}$\newline
Continuing the same process by mutating $\mathfrak{Q}_{6}^{E_{6}}$ by L$_{1}$%
, we obtain the quiver $\mathfrak{Q}_{7}^{E_{6}}$ that contains
other 6 new
BPS states%
\begin{equation}
\mathfrak{Q}_{7}^{E_{6}}:%
\begin{tabular}{ll}
$b_{1}+b_{3}+c_{2}+c_{4}\ ,$ & $2b_{2}+b_{4}+b_{5}+c_{1}+2c_{3}\ $ \\
$b_{2}+b_{4}+2b_{5}+2c_{3}+c_{6}\ ,$ & $%
b_{1}+3b_{3}+b_{6}+2c_{2}+c_{4}+2c_{5}$ \\
$b_{3}+b_{6}+c_{4}+c_{5}\ ,$ & $b_{2}+b_{4}+b_{5}+c_{1}+c_{3}+c_{6}$%
\end{tabular}%
\end{equation}

\textbf{9}) \emph{BPS} \emph{quiver
}$\mathfrak{Q}_{8}^{E_{6}}$\newline
This quiver has also 6 new BPS states with charges given by%
\begin{equation}
\mathfrak{Q}_{8}^{E_{6}}:%
\begin{tabular}{ll}
$b_{2}+b_{4}+c_{3},$ & $b_{1}+2b_{3}+c_{2}+c_{4}+c_{5}$ \\
$b_{1}+b_{3}+b_{6}+c_{2}+c_{5},$ &
$2b_{2}+b_{4}+2b_{5}+c_{1}+2c_{3}+c_{6}$
\\
$b_{4}+b_{5}+c_{3},$ & $2b_{3}+b_{6}+c_{2}+c_{4}+c_{5}$%
\end{tabular}%
\end{equation}

\textbf{10}) \emph{BPS} \emph{quiver }$\mathfrak{Q}_{9}^{E_{6}}$ and $%
\mathfrak{Q}_{10}^{E_{6}}$\newline These quivers involve more new
BPS states respectively given by,
\begin{equation}
\mathfrak{Q}_{9}^{E_{6}}:%
\begin{tabular}{ll}
$b_{3}+c_{5},$ & $b_{2}+b_{4}+b_{5}+c_{3}+c_{6}\ $ \\
$b_{2}+b_{5}+c_{3},$ & $b_{1}+2b_{3}+b_{6}+c_{2}+c_{4}+c_{5}$ \\
$b_{3}+c_{2},$ & $b_{2}+b_{4}+b_{5}+c_{1}+c_{3}$%
\end{tabular}%
\end{equation}%
and\emph{\ }%
\begin{equation}
\mathfrak{Q}_{10}^{E_{6}}:%
\begin{tabular}{ll}
$b_{5}+c_{6},$ & $b_{3}+b_{6}+c_{5}\ $ \\
$b_{3}+c_{4},$ & $b_{2}+b_{4}+b_{5}+c_{3}$ \\
$b_{2}+c_{1},$ & $b_{1}+b_{3}+c_{2}$%
\end{tabular}%
\end{equation}

\textbf{11}) \emph{BPS} \emph{Quiver }$\mathfrak{Q}_{11}^{E_{6}}$ and $%
\mathfrak{Q}_{12}^{E_{6}}$\newline
From the analysis reported in appendix III, the mutated quiver $\mathfrak{Q}%
_{11}^{E_{6}}$ has only anti-BPS states; and the quiver $\mathfrak{Q}%
_{12}^{E_{6}}$ is precisely the CPT conjugate of
$\mathfrak{Q}_{0}^{E_{6}}$ since it is made of the elementary
anti-BPS states.

\  \  \  \newline To conclude this analysis, we find that the total
number of BPS states in the strong coupling chamber of the
$\mathcal{N}=2$ supersymmetric E$_{6}$
gauge theory is given by%
\begin{equation}
12+10\times 6=72
\end{equation}%
which is nothing but $\dim E_{6}-6$ in agreement with the prediction
of \textrm{\cite{1A,1B}}. Along with these states, there is also
$72$ anti-BPS states. This analysis extends straightforwardly to the
E$_{7}$ and E$_{8}$ gauge theories.

\subsubsection{Building $\mathcal{G}_{strong}^{E_{r}}$}

Here, we build the structure of the discrete mutation groupoids $\mathcal{G}%
_{strong}^{E_{r}}$ associated with strong coupling chambers of the $\mathcal{%
N}=2$ supersymmetric QFT's with exceptional gauge symmetries. We
also give the corresponding BPS spectrum.

\

1)\emph{\ the mutation groupoid}
$\mathcal{G}_{strong}^{E_{6}}$\newline From the\ analysis developed
above, it follows that the discrete mutation groupoid
$\mathcal{G}_{strong}^{E_{6}}$ of the strong coupling chamber
consists of the set of mutations matrices $\mathcal{M}_{n}$ ($\equiv
\mathcal{M}_{n,0}$) given by the \textrm{realization (\ref{RE})} and
obeying the periodicity property
\begin{equation}
\mathcal{M}_{24}=I_{id}=I_{144\times 144}.
\end{equation}%
This identity tells us that the independent elements of $\mathcal{G}%
_{strong}^{E_{6}}$ are given by the following set of $144\times 144$ matrices%
\begin{equation}
\mathcal{G}_{strong}^{E_{6}}=\left \{
\begin{tabular}{llllll}
$I_{id},$ & $\mathcal{M}_{2},$ & $\mathcal{M}_{4},$ & $\mathcal{M}_{6},$ & $%
\mathcal{\cdots },$ & $\mathcal{M}_{22}$ \\
$\mathcal{M}_{1},$ & $\mathcal{M}_{3},$ & $\mathcal{M}_{5}$ & $\mathcal{M}%
_{7},$ & $\mathcal{\cdots },$ & $\mathcal{M}_{23}$%
\end{tabular}%
\right \}  \label{6E}
\end{equation}%
with $\mathcal{M}_{n}$ as in \textrm{(\ref{RE}) and (\ref{L}-\ref{2L}). }%
\newline
Eq(\ref{6E}) teaches us also that the order of this discrete groupoid $%
\mathcal{G}_{strong}^{E_{6}}$ is equal to $24$. It shows as well that $%
\mathcal{G}_{strong}^{E_{6}}$ has a finite abelian subgroup
generated by even mutations as follows
\begin{equation}
\left \{
\begin{tabular}{llllll}
$I_{id},$ & $\mathcal{M}_{2},$ & $\mathcal{M}_{4},$ & $\mathcal{M}_{6},$ & $%
\mathcal{\cdots },$ & $\mathcal{M}_{22}$%
\end{tabular}%
\right \}
\end{equation}%
This is an abelian subgroup having 12 elements and is isomorphic to $\mathbb{%
Z}_{12}$. \newline Now computing the product of the rank of the
gauge group $E_{6}$ with the order of
$\mathcal{G}_{strong}^{E_{6}}$, we find
\begin{equation}
r\times \left \vert \mathcal{G}_{M}^{E_{6}}\right \vert =6\times
24=144
\end{equation}%
This number, which splits as $72+72$, is exactly the total number of
BPS and anti-BPS states in the strong coupling chamber of the
$\mathcal{N}=2$ supersymmetric $E_{6}$\ gauge theory.

\  \  \  \  \  \newline

2)\emph{\ the mutation groupoid}
$\mathcal{G}_{strong}^{E_{7}}$\newline This set consists of the set
of mutation matrices $\mathcal{M}_{n}$ given by the
\textrm{realization (\ref{RE})} and obeying the cyclic property
\begin{equation}
\mathcal{M}_{36}=I_{id}=I_{196\times 196}.
\end{equation}%
The elements of the discrete $\mathcal{G}_{strong}^{E_{7}}$ are given by $%
196\times 196$ matrices; and are as follows%
\begin{equation}
\mathcal{G}_{strong}^{E_{7}}=\left \{
\begin{tabular}{llllll}
$I_{id},$ & $\mathcal{M}_{2},$ & $\mathcal{M}_{4},$ & $\mathcal{M}_{6},$ & $%
\mathcal{\cdots },$ & $\mathcal{M}_{24}$ \\
$\mathcal{M}_{1},$ & $\mathcal{M}_{3},$ & $\mathcal{M}_{5}$ & $\mathcal{M}%
_{7},$ & $\mathcal{\cdots },$ & $\mathcal{M}_{35}$%
\end{tabular}%
\right \}
\end{equation}%
The number of the $\mathcal{M}_{n}$'s is equal to $36$. Computing
the
product of the rank of $E_{7}$ with the order of $\mathcal{G}%
_{strong}^{E_{7}}$, we find
\begin{equation}
r\times \left \vert \mathcal{G}_{strong}^{E_{7}}\right \vert
=7\times 36=252
\end{equation}%
This number reads also as $\left( 133-7\right) +\left( 133-7\right)
$; it is precisely the number of BPS and anti-BPS states of the
strong coupling chamber of the $\mathcal{N}=2$ supersymmetric
$E_{7}$\ gauge theory.

\  \  \  \

3) \emph{the mutation groupoid}
$\mathcal{G}_{strong}^{E_{8}}$\newline
In this case, the set of mutation matrices $\mathcal{M}_{n}$ that form $%
\mathcal{G}_{strong}^{E_{8}}$ are
\begin{equation}
\mathcal{G}_{strong}^{E_{8}}=\left \{
\begin{tabular}{llllll}
$I_{id},$ & $\mathcal{M}_{2},$ & $\mathcal{M}_{4},$ & $\mathcal{M}_{6},$ & $%
\mathcal{\cdots },$ & $\mathcal{M}_{58}$ \\
$\mathcal{M}_{1},$ & $\mathcal{M}_{3},$ & $\mathcal{M}_{5}$ & $\mathcal{M}%
_{7},$ & $\mathcal{\cdots },$ & $\mathcal{M}_{59}$%
\end{tabular}%
\right \}
\end{equation}%
with the properties
\begin{equation}
\mathcal{M}_{60}=I_{id}=I_{256\times 256},\qquad \left \vert \mathcal{G}%
_{strong}^{E_{8}}\right \vert =60.
\end{equation}%
Computing the product\ $8\times \left \vert \mathcal{G}_{strong}^{E_{8}}%
\right \vert $ we find
\begin{equation}
8\times 60=480.
\end{equation}%
This number, which reads also like
\begin{equation}
\left( 248-8\right) +\left( 248-8\right)
\end{equation}%
is exactly the total number of BPS and anti-BPS states of the strong
coupling chamber of the $\mathcal{N}=2$ supersymmetric $E_{8}$\
gauge theory.

\section{Weak coupling chambers of $SU\left( 2\right) $ and $SO\left(
4\right) $}

In the reminder of this paper, we use the \textrm{algebraic} method
we have developed in previous sections to study the weak coupling
chambers of BPS quiver theories. These chambers are infinite; so we
will focus on 3 examples to illustrate the method. These examples
concern:

(\textbf{1}) the $\mathcal{N}=2$ supersymmetric $SU\left( 2\right) $
gauge theory:\newline the weak coupling chamber of this BPS quiver
theory has been considered in \textrm{\cite{1A,1B,030I}}; but here
we reconsider this chamber from the view of the mutation symmetry.
This example is also used to approach higher dimensional gauge
symmetries.

(\textbf{2}) the $\mathcal{N}=2$ supersymmetric SO$\left( 4\right) $
gauge model.\newline Because of the group homomorphism $SO\left(
4\right) \sim SU\left( 2\right) \times SU\left( 2\right) $, the BPS
quivers of this model is given by the direct generalization of the
SU$\left( 2\right) $ theory; it is made of two uncoupled SU$\left(
2\right) $ sectors.

(\textbf{3}) the $\mathcal{N}=2$ SU$\left( 3\right) $ quiver theory.
\newline The weak coupling chamber of the BPS quivers of this gauge
model may be thought of as non trivial extension of the SU$\left(
2\right) $ model. It will be considered in details in next section.\

\subsection{Weak coupling chamber of $SU\left( 2\right) $ model}

We start by building the \textrm{set of mutation symmetries} of the
weak coupling chamber of $\mathcal{N}=2$ supersymmetric $SU\left(
2\right) $ gauge model. Recall that this gauge theory has two
elementary BPS states
namely an elementary monopole and an elementary dyon.\ Denoting by $X$ and $%
Y $ the central charges of these BPS states and by $b$ and $c$ their
EM charge vectors, one distinguishes two chambers of BPS states
given by:

\begin{itemize}
\item $\arg Y>\arg X$ describing the strong coupling chamber of low energy
of $\mathcal{N}=2$ $SU\left( 2\right) $\ QFT$_{4}$. this chamber is
finite,
closed and has been considered in subsection 3.1; \textrm{see also fig} \ref%
{sw}. The corresponding mutation\textrm{\ symmetry} $\mathcal{G}%
_{strong}^{su_{2}}$ has been shown to be isomorphic to \textrm{the group}%
\begin{equation}
\mathcal{G}_{strong}^{su_{2}}=Dih_{2}\times Dih_{2}
\end{equation}

\item $\arg X>\arg Y$ associated with the infinite weak coupling chamber of
the supersymmetric theory.\ Below we study this chamber by using $\mathcal{G}%
_{weak}^{su_{2}}$.
\end{itemize}

\subsubsection{the mutation symmetry $\mathcal{G}_{weak}^{su_{2}}$}

In the case $\arg X>\arg Y$, the mutation \textrm{symmetry} $\mathcal{G}%
_{weak}^{su_{2}}$ is an infinite set with elements as in
eq(\ref{GM}). As
noticed in the discussion following (\ref{GM}) and the remark of \textrm{%
\cite{1A,1B}} concerning the need of both left and right mutations
to cover the full BPS spectrum; see also \emph{footnote 3}, the
\textrm{groupoid} structure of the set of quiver mutations requires
indeed the two infinite
sequences $\left \{ \mathcal{L}_{m}\right \} $\ and \ $\left \{ \mathcal{R}%
_{m}\right \} $; where the $\mathcal{R}_{m}$'s is the inverse of the $%
\mathcal{L}_{m}$'s. \newline Moreover, matrix representations of
mutation \textrm{symmetries} in BPS quiver theory with rank r gauge
groups indicate that mutation elements are given by $\left(
2r\right) ^{2}\times \left( 2r\right) ^{2}$ matrices with integer
entries. So in the case of SU$\left( 2\right) $, the mutation
symmetry
\begin{equation}
\mathcal{G}_{weak}^{su_{2}}=\left \{ \mathcal{L}_{m},\text{\  \ }\mathcal{R}%
_{m};\text{...}m\in Z_{+}\right \}  \label{MUT}
\end{equation}%
is an infinite subgroup of the linear group of invertible $2\times
2$ matrices with integer entries:
\begin{equation}
\mathcal{G}_{weak}^{su_{2}}\subseteq GL\left( 2,\mathbb{Z}\right)
\end{equation}%
It turns out that the derivation of the expressions of
$\mathcal{L}_{m}$\ and\ $\mathcal{R}_{m}$ depends on the parity of
the integer $m$. So we split
the $\mathcal{L}_{m}$'s and the $\mathcal{R}_{m}$'s like%
\begin{equation}
\begin{tabular}{lll}
$\mathcal{L}_{m}=\left( \mathcal{L}_{2k},\mathcal{L}_{2k+1}\right) $ & $,%
\text{\  \ }$ & $\mathcal{R}_{m}=\left( \mathcal{R}_{2k},\mathcal{R}%
_{2k+1}\right) $%
\end{tabular}
\label{LR}
\end{equation}%
with $k$ an arbitrary positive integer. Straightforward calculations
show that the explicit expressions of $\mathcal{L}_{2k}$,
$\mathcal{L}_{2k+1}$
and $\mathcal{R}_{2k}$, $\mathcal{R}_{2k+1}$ are as follows:%
\begin{equation}
\begin{tabular}{l||l}
\  \  \  \  \  \  \  \  \ left sector & \  \  \  \  \  \  \  \  \
right sector \\ \hline
&  \\
$%
\begin{tabular}{lll}
$\mathcal{L}_{2k}$ & $=$ & $\left(
\begin{array}{cc}
1+2k & 2k \\
-2k & 1-2k%
\end{array}%
\right) $ \\
&  &  \\
$\mathcal{L}_{2k+1}$ & $=$ & $\left(
\begin{array}{cc}
-1-2k & -2k \\
2k+2 & 1+2k%
\end{array}%
\right) $ \\
&  &
\end{tabular}%
$ & $%
\begin{tabular}{lll}
$\mathcal{R}_{2k}$ & $=$ & $\left(
\begin{array}{cc}
1-2k & -2k \\
2k & 1+2k%
\end{array}%
\right) $ \\
&  &  \\
$\mathcal{R}_{2k+1}$ & $=$ & $\left(
\begin{array}{cc}
2k+1 & 2k+2 \\
-2k & -2k-1%
\end{array}%
\right) $ \\
&  &
\end{tabular}%
$%
\end{tabular}
\label{REP}
\end{equation}%
These matrix mutations obey the properties,%
\begin{equation}
\left( \mathcal{L}_{2k+1}\right) ^{2}=\left(
\mathcal{R}_{2k+1}\right) ^{2}=I_{id}=I_{2\times 2},\qquad \forall k
\end{equation}%
We also have%
\begin{equation}
\begin{tabular}{ll}
$\det \mathcal{L}_{2k}$ & $=\det \mathcal{R}_{2k}=+1,$ $\  \  \forall k$ \\
$\det \mathcal{L}_{2k+1}$ & $=\det \mathcal{R}_{2k+1}=-1,$ $\  \  \forall k$%
\end{tabular}%
\end{equation}%
To establish these relations, one uses the representation%
\begin{equation}
\begin{tabular}{lll}
$\mathcal{L}_{2k}=\left( BA\right) ^{k}$ & , &
$\mathcal{R}_{2k}=\left(
AB\right) ^{k}$ \\
$\mathcal{L}_{2k+1}=A\mathcal{L}_{2k}$ & , & $\mathcal{R}_{2k+1}=B\mathcal{R}%
_{2k}$%
\end{tabular}%
\end{equation}%
Here $A$ and $B$ stand for the generators of the quiver mutations
given by the triangular matrices
\begin{equation}
\begin{tabular}{lll}
$A=\left(
\begin{array}{cc}
-1 & 0 \\
2 & 1%
\end{array}%
\right) $ & , & $B=\left(
\begin{array}{cc}
1 & 2 \\
0 & -1%
\end{array}%
\right) $%
\end{tabular}
\label{AB}
\end{equation}%
satisfying the usual reflection property $A^{2}=B^{2}=I_{id}$.

\subsubsection{BPS spectrum}

The mutation matrices $\mathcal{L}_{m}$ (resp $\mathcal{R}_{m}$) map
the quiver $\mathfrak{Q}_{0}^{su_{2}}$ made by elementary BPS states
into the
quivers $\mathfrak{Q}_{m}^{su_{2}}$ (resp $\mathfrak{Q}_{m}^{\prime su_{2}}$%
) involving composite BPS states with EM charge vectors as%
\begin{equation}
\begin{tabular}{ll||ll}
\multicolumn{2}{l||}{\  \  \  \  \  \  \  \  \  \ left sector} &
\multicolumn{2}{||l}{ \  \  \  \  \  \  \ right sector} \\ \hline
&  &  &  \\
$b^{\left( 2k\right) }$ & $=b+2kw$ & $b^{\left( 2k\right) }$ & $=b-2kw$ \\
$c^{\left( 2k\right) }$ & $=c-2kw$ & $c^{\left( 2k\right) }$ & $=c+2kw$ \\
$b^{\left( 2k+1\right) }$ & $=c-\left( 2k+1\right) w$ & $b^{\left(
2k+1\right) }$ & $=c+\left( 2k+1\right) w$ \\
$c^{\left( 2k+1\right) }$ & $=b+\left( 2k+1\right) w$ & ${\normalsize c}%
^{\left( 2k+1\right) }$ & ${\normalsize =b-}\left( 2k+1\right) {\normalsize w%
}$ \\
&  &  &  \\ \hline
\end{tabular}%
\end{equation}%
with
\begin{equation}
w=b+c  \label{w}
\end{equation}%
giving the EM charge of the W-boson vector particle. \newline From
the above results, we learn that the BPS spectrum of the weak
coupling chamber of the supersymmetric SU$\left( 2\right) $ gauge
theory is infinite; and moreover the number of BPS states grows
linearly with the positive
integer $k$.\ If defining the asymptotic limit of the mutation matrices $%
\mathcal{L}_{n}$ and by the regularized relation%
\begin{equation}
\mathcal{L}_{\infty }=\lim_{n\rightarrow \infty }\left( \frac{1}{n}\mathcal{L%
}_{n}\right) \ ,\text{ \  \  \ }\mathcal{R}_{\infty
}=\lim_{n\rightarrow \infty }\left(
\frac{1}{n}\mathcal{R}_{n}\right)  \label{lr}
\end{equation}%
we find%
\begin{equation}
\begin{tabular}{lll}
$\mathcal{L}_{\infty }^{even}=\left(
\begin{array}{cc}
1 & 1 \\
-1 & -1%
\end{array}%
\right) $ & , & $\det \mathcal{L}_{\infty }^{even}=0$ \\
&  &  \\
$\mathcal{L}_{\infty }^{odd}=\left(
\begin{array}{cc}
-1 & -1 \\
1 & 1%
\end{array}%
\right) $ & , & $\det \mathcal{L}_{\infty }^{odd}=0$%
\end{tabular}%
\end{equation}%
and
\begin{equation}
\begin{tabular}{lll}
$\mathcal{R}_{\infty }^{even}=\left(
\begin{array}{cc}
-1 & -1 \\
1 & 1%
\end{array}%
\right) $ & , & $\det \mathcal{R}_{\infty }^{even}=0$ \\
&  &  \\
$\mathcal{R}_{\infty }^{odd}=\left(
\begin{array}{cc}
1 & 1 \\
-1 & -1%
\end{array}%
\right) $ & , & $\det \mathcal{R}_{\infty }^{odd}=0$%
\end{tabular}
\label{rl}
\end{equation}%
These singular asymptotic limits describe BPS particles with charges
$\pm w=\pm \left( b+c\right) $. These vector particles have electric
charge; but no magnetic one; they are associated with
$\mathcal{N}=1$ massive $W^{\pm }$ vector multiplets.

\subsubsection{Link between $\mathcal{G}_{weak}^{su_{2}}$ and Coxeter $%
\tilde{I}_{2}\left( \infty \right) $}

In section 3, we learned that the strong coupling mutation symmetry $%
\mathcal{G}_{strong}^{su_{2}}$ is isomorphic to the dihedral group $%
I_{2}\left( 2\right) $. The latter is generated by two reflections
$r_{1}$
and $r_{2}$ obeying the group law%
\begin{equation}
\left( r_{i}r_{j}\right) ^{m_{ij}}=I_{id}
\end{equation}%
with $r_{1}$ and $r_{2}$ realized as in (\ref{A}-\ref{B}); and
$m_{ij}$ the
entries of the following integral $2\times 2$ symmetric\textrm{\ matrix}%
\begin{equation}
M=\left(
\begin{array}{cc}
1 & 2 \\
2 & 1%
\end{array}%
\right)
\end{equation}%
Here, we explore whether there exists a relation between the weak
coupling mutation group $\mathcal{G}_{weak}^{su_{2}}$ and infinite
Coxeter
symmetries. Indeed, there is a close relation; the mutation group $\mathcal{G%
}_{weak}^{su_{2}}$ is linked to the infinite limit $m\rightarrow
\infty $\ of the Coxeter group $I_{2}\left( m\right) $. More
precisely, we have
\begin{equation}
\mathcal{G}_{weak}^{su_{2}}\simeq \tilde{I}_{2}\left( \infty \right)
\end{equation}%
To get this relation, it is helpful to first set the two reflections (\ref%
{AB}) as $s_{1}=A$ and $s_{2}=B$ satisfying
\begin{equation}
\left( s_{i}s_{j}\right) ^{m_{ij}}=I_{id}
\end{equation}%
and then look for the Coxeter group $W\left( M\right) $ with matrix $%
M=\left( m_{ij}\right) $ that lead to the mutation group $\mathcal{G}%
_{weak}^{su_{2}}$. From the matrix realization (\ref{AB}), it
follows that
we have%
\begin{equation}
M=\left(
\begin{array}{cc}
1 & \infty \\
\infty & 1%
\end{array}%
\right)
\end{equation}%
Moreover, using the link between the matrix $M$ of Coxeter groups
and the Cartan matrix K of Lie algebras namely
\begin{equation}
K_{ij}=-2\cos \frac{\pi }{m_{ij}}  \label{KM}
\end{equation}%
it results that the the Coxeter group of the weak coupling chamber
is given by the Dynkin graph of the twisted affine $SU\left(
2\right) $ Kac Moody algebra with generalized Cartan matrix as
\begin{equation}
K=\left(
\begin{array}{cc}
2 & -2 \\
-2 & 2%
\end{array}%
\right) ,\qquad \det K=0
\end{equation}

\subsection{Extension to SO$\left( 4\right) $ model}

Here we study the extension of the weak mutation symmetry $\mathcal{G}%
_{weak}^{su_{2}}$ to $\mathcal{N}=2$ supersymmetric QFTs with
SO$\left( 4\right) $ gauge model.

\subsubsection{mutation symmetry $\mathcal{G}_{weak}^{so_{4}}$}

\begin{figure}[tbph]
\begin{center}
\hspace{0cm} \includegraphics[width=12cm]{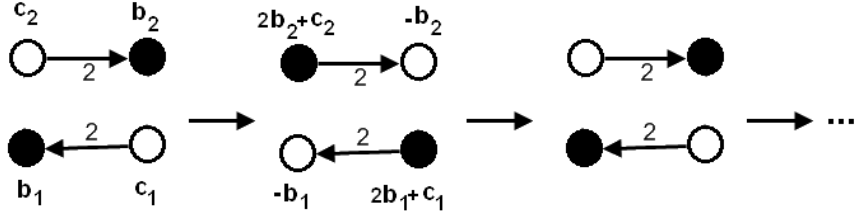}
\end{center}
\par
\vspace{-1 cm} \caption{mutations of BPS quivers of SO$\left(
4\right) $ theory} \label{A11}
\end{figure}

The primitive quiver of this model is given by two disconnected
SU$\left( 2\right) $ BPS quivers as depicted in fig(\ref{A11}).
To get the structure of $\mathcal{G}_{weak}^{so_{4}}$, we use the
group homomorphism $SO\left( 4\right) \sim SU\left( 2\right) \times
SU\left(
2\right) $. This property suggests that the weak mutation symmetry $\mathcal{%
G}_{weak}^{so_{4}}$ should be given by the direct product of two $\mathcal{G}%
_{weak}^{su_{2}}$ copies%
\begin{equation}
\mathcal{G}_{weak}^{so_{4}}\simeq \mathcal{G}_{weak}^{su_{2}}\times \mathcal{%
G}_{weak}^{\prime su_{2}}
\end{equation}%
This group is contained in the linear group of $4\times 4$
invertible
integral matrices%
\begin{equation}
\mathcal{G}_{weak}^{so_{4}}\subseteq GL\left( 2,\mathbb{Z}\right)
\times GL\left( 2,\mathbb{Z}\right) \subseteq GL\left(
4,\mathbb{Z}\right)
\end{equation}%
The elements of this discrete group are given by the following
$4\times 4$
matrices%
\begin{equation}
\begin{tabular}{llll}
$\left(
\begin{array}{cc}
\mathcal{L}_{n}^{su_{2}} & 0 \\
0 & I_{2}%
\end{array}%
\right) $, &  & $\left(
\begin{array}{cc}
\mathcal{R}_{n}^{su_{2}} & 0 \\
0 & I_{2}%
\end{array}%
\right) ,$ & $n\in \mathbb{Z}_{+}$ \\
&  &  &
\end{tabular}%
\end{equation}%
and%
\begin{equation}
\begin{tabular}{llll}
$\left(
\begin{array}{cc}
I_{2} & 0 \\
0 & \mathcal{L}_{m}^{\prime su_{2}}%
\end{array}%
\right) $, &  & $\left(
\begin{array}{cc}
I_{2} & 0 \\
0 & \mathcal{R}_{m}^{\prime su_{2}}%
\end{array}%
\right) ,$ & $m\in \mathbb{Z}_{+}$ \\
&  &  &
\end{tabular}%
\end{equation}%
with $\mathcal{L}_{n}^{su_{2}}$ and $\mathcal{R}_{n}^{su_{2}}$\ as in (\ref%
{LR}).\newline
By help of the results on $\mathcal{G}_{weak}^{su_{2}}$, it is clear that $%
\mathcal{G}_{weak}^{so_{4}}$ is given by the direct product of two
copies of $\tilde{I}_{2}\left( \infty \right) $ leading to the
infinite dimensional
Coxeter group%
\begin{equation}
\mathcal{G}_{weak}^{so_{4}}\simeq \tilde{I}_{2}\left( \infty \right)
\times \tilde{I}_{2}\left( \infty \right)
\end{equation}%
In this case, the Coxeter matrix M reads as%
\begin{equation}
M=\left(
\begin{array}{cccc}
1 & \infty & 0 & 0 \\
\infty & 1 & 0 & 0 \\
0 & 0 & 1 & \infty \\
0 & 0 & \infty & 1%
\end{array}%
\right)
\end{equation}%
which, by help of (\ref{KM}), leads to the following twisted affine SO$%
\left( 4\right) $ Kac-Moody with Cartan matrix%
\begin{equation}
K^{so_{4}}=\left(
\begin{array}{cccc}
2 & -2 & 0 & 0 \\
-2 & 2 & 0 & 0 \\
0 & 0 & 2 & -2 \\
0 & 0 & -2 & 2%
\end{array}%
\right)
\end{equation}%
with $\det K^{so_{4}}=0$.

\subsubsection{BPS spectrum}

Using results on the BPS quiver of the weak coupling chamber of the $%
\mathcal{N}=2$ supersymmetric QFT with SU$\left( 2\right) $ gauge
symmetry, we can deduce the corresponding one for the SO$\left(
4\right) $ gauge
model. We have:%
\begin{equation}
\begin{tabular}{ll||ll}
\multicolumn{2}{l||}{\  \  \  \  \  \  \  \  \  \ left sector of
SU$_{1}\left(
2\right) $} & \multicolumn{2}{||l}{\  \  \  \  \  \  \ right sector of SU$%
_{1}\left( 2\right) $} \\ \hline
&  &  &  \\
$b_{1}^{\left( 2k\right) }$ & $=b_{1}+2kw_{1}$ & $b_{1}^{\left(
2k\right) }$
& $=b_{1}-2kw_{1}$ \\
$c_{1}^{\left( 2k\right) }$ & $=c_{1}-2kw_{1}$ & $c_{1}^{\left(
2k\right) }$
& $=c_{1}+2kw_{1}$ \\
$b_{1}^{\left( 2k+1\right) }$ & $=c_{1}-\left( 2k+1\right) w_{1}$ & $%
b_{1}^{\left( 2k+1\right) }$ & $=c_{1}+\left( 2k+1\right) w_{1}$ \\
$c_{1}^{\left( 2k+1\right) }$ & $=b_{1}+\left( 2k+1\right) w_{1}$ & $%
{\normalsize c}_{1}^{\left( 2k+1\right) }$ & ${\normalsize
=b_{1}-}\left(
2k+1\right) {\normalsize w}_{1}$ \\
&  &  &  \\ \hline
\end{tabular}%
\end{equation}%
and%
\begin{equation}
\begin{tabular}{ll||ll}
\multicolumn{2}{l||}{\  \  \  \  \  \  \  \  \  \ left sector of
SU$_{2}\left(
2\right) $} & \multicolumn{2}{||l}{\  \  \  \  \  \  \ right sector of SU$%
_{2}\left( 2\right) $} \\ \hline
&  &  &  \\
$b_{2}^{\left( 2k\right) }$ & $=b_{2}+2kw_{2}$ & $b_{2}^{\left(
2k\right) }$
& $=b_{2}-2kw_{2}$ \\
$c_{2}^{\left( 2k\right) }$ & $=c_{2}-2kw_{2}$ & $c_{2}^{\left(
2k\right) }$
& $=c_{2}+2kw_{2}$ \\
$b_{2}^{\left( 2k+1\right) }$ & $=c_{2}-\left( 2k+1\right) w_{2}$ & $%
b_{2}^{\left( 2k+1\right) }$ & $=c_{2}+\left( 2k+1\right) w_{2}$ \\
$c_{2}^{\left( 2k+1\right) }$ & $=b_{2}+\left( 2k+1\right) w_{2}$ & $%
{\normalsize c}_{2}^{\left( 2k+1\right) }$ & ${\normalsize
=b_{2}-}\left(
2k+1\right) {\normalsize w}_{2}$ \\
&  &  &  \\ \hline
\end{tabular}%
\end{equation}%
with
\begin{equation}
w_{i}=b_{i}+c_{i}
\end{equation}%
associated with the two BPS W-bosons. Moreover the asymptotic limits
of the mutation matrices $\mathcal{L}_{n}^{su_{2}}$ and
$\mathcal{R}_{n}^{su_{2}}$
leads to%
\begin{equation}
\begin{tabular}{lll}
$\mathcal{L}_{\infty }^{su_{2}}=\left(
\begin{array}{cccc}
1 & 1 & 0 & 0 \\
-1 & -1 & 0 & 0 \\
0 & 0 & 1 & 0 \\
0 & 0 & 0 & 1%
\end{array}%
\right) $ & , & $\mathcal{L}_{\infty }^{\prime su_{2}}=\left(
\begin{array}{cccc}
1 & 0 & 0 & 0 \\
0 & 1 & 0 & 0 \\
0 & 0 & 1 & 1 \\
0 & 0 & -1 & -1%
\end{array}%
\right) $%
\end{tabular}%
\end{equation}%
and%
\begin{equation}
\begin{tabular}{lll}
$\mathcal{R}_{\infty }^{su_{2}}=\left(
\begin{array}{cccc}
-1 & -1 & 0 & 0 \\
1 & 1 & 0 & 0 \\
0 & 0 & 1 & 0 \\
0 & 0 & 0 & 1%
\end{array}%
\right) $ & , & $\mathcal{R}_{\infty }^{\prime su_{2}}=\left(
\begin{array}{cccc}
1 & 0 & 0 & 0 \\
0 & 1 & 0 & 0 \\
0 & 0 & -1 & -1 \\
0 & 0 & 1 & 1%
\end{array}%
\right) $%
\end{tabular}%
\end{equation}%
These are singular asymptotic limits describing the four $W_{i}^{\pm
}$ vector BPS particles of charges $\pm w_{i}=\pm \left(
b_{i}+c_{i}\right) $.

\section{Weak coupling chamber of $SU\left( 3\right) $ theory}

First we build the weak mutation symmetry
$\mathcal{G}_{weak}^{su_{3}}$; then we give the BPS spectrum of its
weak coupling chamber; or more precisely the BPS chamber that
follows from the extension of the
construction used in deriving the weak coupling BPS states in the case of SU$%
\left( 2\right) $ theory.

\subsection{the mutation symmetry $\mathcal{G}_{weak}^{su_{3}}$}

We start by recalling that quiver mutations for rank r gauge
symmetries may
realized in terms of invertible matrices forming a particular subset $%
\mathcal{G}_{weak}^{su_{3}}$ of GL$\left( 4r^{2},\mathbb{Z}\right)
$. In the case of $\mathcal{N}=2$ supersymmetric QFT with $SU\left(
3\right) $ gauge group, the mutation symmetry
$\mathcal{G}_{weak}^{su_{3}}$ of the weak
coupling chamber is therefore given by an infinite subset of $GL\left( 16,%
\mathbb{Z}\right) $\textrm{\ }containing\textrm{\ }$\mathcal{G}%
_{weak}^{su_{2}}$ as a subset. However, because of the fact that
quiver mutations are only sensitive to the $2r$ building blocks of
(\ref{rep}); \emph{see also footnote 5}, corresponding to the
subgroup
\begin{equation}
GL\left( 4,\mathbb{Z}\right) \otimes I_{4}\subset GL\left( 16,\mathbb{Z}%
\right)
\end{equation}%
with $I_{4}$\ standing for the $4\times 4$ identity matrix, we have
\begin{equation}
\mathcal{G}_{weak}^{su_{3}}\subset GL\left( 4,\mathbb{Z}\right)
\otimes I_{4}
\end{equation}%
So the groupoid $\mathcal{G}_{weak}^{su_{3}}$ is given by an
infinite set of $16\times 16$ matrices that factorize into matrix
blocks as $N_{ij}\otimes
I_{4}$ of the form%
\begin{equation}
\left(
\begin{array}{cccc}
N_{11}I_{4} & N_{12}I_{4} & N_{13}I_{4} & N_{14}I_{4} \\
N_{21}I_{4} & N_{22}I_{4} & N_{23}I_{4} & N_{24}I_{4} \\
N_{31}I_{4} & N_{32}I_{4} & N_{22}I_{4} & N_{34}I_{4} \\
N_{41}I_{4} & N_{42}I_{4} & N_{43}I_{4} & N_{44}I_{4}%
\end{array}%
\right)  \label{NI}
\end{equation}%
with $N_{ij}$ integers. Moreover, seen that the BPS weak chamber of the $%
SU\left( 3\right) $ theory should contain as singular limits the 6
massive gauge particles $W^{\pm \alpha _{i}}$ with EM charges as
\begin{equation}
w_{i}^{\pm }=\pm \left( b_{i}+c_{i}\right) ,\qquad i=1,2,3
\label{WA}
\end{equation}%
and mimicking the construction of the weak coupling chamber of the
SU$\left( 2\right) $ theory that lead to eqs(\ref{lr}-\ref{rl}), it
follows that the weak coupling chamber of the $SU\left( 3\right) $
theory\textrm{\ }should be
generated by \emph{6} involutions denoted as%
\begin{equation}
\begin{tabular}{lllll}
$\mathcal{A}_{1},$ $\mathcal{B}_{1}$ & $;$ & $\mathcal{A}_{2},$ $\mathcal{B}%
_{2}$ & $;$ & $\mathcal{A}_{3},$ $\mathcal{B}_{3}$%
\end{tabular}
\label{ABC}
\end{equation}%
with $\mathcal{A}_{i}^{2}=\mathcal{B}_{i}^{2}=I_{16}$. Each one of
these
involutions is associated with a root of SU$\left( 3\right) $; and each $%
\left( \mathcal{A}_{i},\mathcal{B}_{i}\right) $ pair, which is in
1:1 correspondence with the pair of roots $\left( \alpha
_{i},-\alpha _{i}\right) $, generates an SU$\left( 2\right) $ type
BPS weak coupling chamber. Therefore, generic elements of the set
$\mathcal{G}_{weak}^{su_{3}}$
are given by the typical monomials%
\begin{equation}
\dprod_{n_{i},m_{j}}\mathcal{A}_{3}^{m_{j_{3}}}\mathcal{B}_{3}^{n_{i_{3}}}%
\mathcal{A}_{2}^{m_{j_{2}}}\mathcal{B}_{2}^{n_{i_{2}}}\mathcal{A}%
_{1}^{m_{j_{1}}}\mathcal{B}_{1}^{n_{i_{1}}}  \label{GE}
\end{equation}%
with integers $n_{i},m_{j}=0,1$ and $i,j$ two arbitrary positive
integers.

\subsubsection{Identifying sub-symmetries $G_{i}$ of $\mathcal{G}%
_{weak}^{su_{3}}$}

A way to approach $\mathcal{G}_{weak}^{su_{3}}$ is to use special
properties of the weak chamber of the SU$\left( 3\right) $ gauge
theory; in particular eqs (\ref{ABC},\ref{GE},\ref{WA}) which we
develop below:

\  \  \  \  \

1) $\mathcal{G}_{weak}^{su_{3}}$ \emph{has 3} $\mathcal{G}_{weak}^{su_{2}}$%
\emph{s as proper subsymmetries} \newline
The weak coupling chamber of the BPS quiver theory of the supersymmetric SU$%
\left( 3\right) $ gauge theory must contains as a singular limit the
6 massive vector bosons $W^{\pm \alpha _{i}}$ with EM charges
\ref{WA}
following from the gauge symmetry breaking%
\begin{equation}
SU\left( 3\right) \rightarrow U\left( 1\right) \times U\left(
1\right)
\end{equation}%
So the $\mathcal{G}_{weak}^{su_{3}}$ should contain three basic
infinite series of type eqs(\ref{REP}-\ref{AB}) as sub-chambers
associated with the
subsets%
\begin{equation}
\begin{tabular}{lll}
$G_{\alpha _{1}},$ & $G_{\alpha _{2}},$ & $G_{\alpha _{3}}$%
\end{tabular}%
\end{equation}%
These three groupoids $G_{\alpha _{i}}$ are in one to one with the
positive roots of $SU\left( 3\right) $; each $G_{\alpha _{i}}$ is
generated by the pair $\mathcal{A}_{i},\mathcal{B}_{i}$,
respectively describing left mutations and right ones, and so is
isomorphic to the weak coupling chamber
of the $\mathcal{N}=2$ supersymmetric SU$\left( 2\right) $ gauge theory,%
\begin{equation}
G_{\alpha _{i}}\simeq \mathcal{G}_{weak}^{su_{2}}  \label{GS}
\end{equation}

2) \emph{six} \emph{generators} \newline As mentioned earlier, the
set $\mathcal{G}_{weak}^{su_{3}}$ has 6 basic
generators in one to one correspondence with the roots $\pm \alpha _{1},$ $%
\pm \alpha _{2},$ $\pm \alpha _{3}$ of SU$\left( 3\right) $. These
are given
by eqs(\ref{ABC},\ref{NI}) with the convention notation $\mathcal{A}_{i},$ $%
\mathcal{B}_{i}$ for exhibiting the correspondence with the roots
$\alpha _{i}$. Moreover seen that $\alpha _{3}=\alpha _{1}+\alpha
_{2}$, one expects
that $\mathcal{A}_{3},$ $\mathcal{B}_{3}$ to be also expressed in terms of $%
\mathcal{A}_{1},$ $\mathcal{B}_{1}$ and $\mathcal{A}_{2},$
$\mathcal{B}_{2}$.

\  \  \  \

\begin{figure}[tbph]
\begin{center}
\hspace{0cm} \includegraphics[width=12cm]{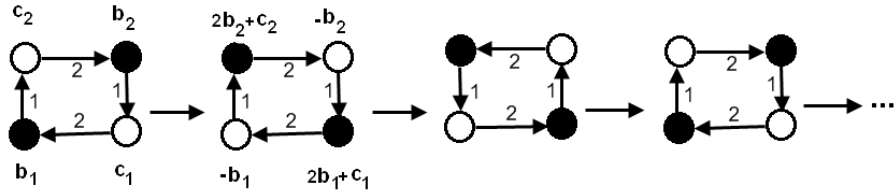}
\end{center}
\par
\vspace{-1 cm}
\caption{particular quiver mutations of the elementary BPS quiver of SU$%
\left( 3\right) $ gauge theory} \label{B11}
\end{figure}

\emph{Deriving }$G_{\alpha _{1}}$\emph{\ and }$G_{\alpha
_{2}}$\newline Thinking for a while about $G_{\alpha _{1}}$\emph{\
}and\emph{\ }$G_{\alpha _{2}}$ as two uncoupled subsets and using
the homomorphism (\ref{GS}), we can write down the generators of
$G_{\alpha _{1}}$ as follows:
\begin{equation}
\begin{tabular}{lll}
$\mathcal{A}_{1}=\left(
\begin{array}{cccc}
-I_{4} & 0 & 0 & 0 \\
2I_{4} & I_{4} & 0 & 0 \\
0 & 0 & I_{4} & 0 \\
0 & 0 & 0 & I_{4}%
\end{array}%
\right) $ & $,$ & $\mathcal{B}_{1}=\left(
\begin{array}{cccc}
I_{4} & 2I_{4} & 0 & 0 \\
0 & -I_{4} & 0 & 0 \\
0 & 0 & I_{4} & 0 \\
0 & 0 & 0 & I_{4}%
\end{array}%
\right) $%
\end{tabular}%
\end{equation}%
satisfying $\mathcal{A}_{1}^{2}=\mathcal{B}_{1}^{2}=I_{id}$.
Similarly, the
subset $G_{\alpha _{2}}$ of $\mathcal{G}_{weak}^{su_{3}}$ is generated by%
\begin{equation}
\begin{tabular}{lll}
$\mathcal{A}_{2}=\left(
\begin{array}{cccc}
I_{4} & 0 & 0 & 0 \\
0 & I_{4} & 0 & 0 \\
0 & 0 & I_{4} & 2I_{4} \\
0 & 0 & 0 & -I_{4}%
\end{array}%
\right) $ & $,$ & $\mathcal{B}_{2}=\left(
\begin{array}{cccc}
I_{4} & 0 & 0 & 0 \\
0 & I_{4} & 0 & 0 \\
0 & 0 & -I_{4} & 0 \\
0 & 0 & 2I_{4} & I_{4}%
\end{array}%
\right) ,$%
\end{tabular}%
\end{equation}%
with $\mathcal{A}_{2}^{2}=\mathcal{B}_{2}^{2}=I_{id}$ ($\equiv
I_{16}$). The $G_{\alpha _{1}}$\emph{\ }and\emph{\ }$G_{\alpha
_{2}}$ describe two classes of particular mutations (two SU$\left(
2\right) $ type sub-chambers) on the elementary BPS quiver
$\mathfrak{Q}_{0}^{su_{3}}$; see fig(\ref{B11}).
Recall that this quiver is represented by the EM charge vector%
\begin{equation}
v_{i}=\left(
\begin{array}{c}
b_{1} \\
c_{1} \\
b_{2} \\
c_{2}%
\end{array}%
\right)  \label{OC}
\end{equation}%
and the intersection matrix $\mathcal{I}_{ij}=v_{i}\circ v_{j}$.
\newline Generic elements of $G_{\alpha _{1}}$ and $G_{\alpha _{2}}$
that are uncoupled are respectively given by the following sets
\begin{equation}
G_{\alpha _{1}}=\left \{ \left(
\begin{array}{cc}
\mathcal{L}_{n}^{\left( \alpha _{1}\right) } & 0 \\
0 & I_{8}%
\end{array}%
\right) ,\text{ \ }\left(
\begin{array}{cc}
\mathcal{R}_{n}^{\left( \alpha _{1}\right) } & 0 \\
0 & I_{8}%
\end{array}%
\right) ,\text{ \ }n\in \mathbb{Z}_{+}\left.
\begin{array}{c}
\\
\\
\end{array}%
\right. \right \}
\end{equation}%
leaving invariant the components $\left( b_{2},c_{2}\right) $ of
(\ref{OC}); and
\begin{equation}
G_{\alpha _{2}}=\left \{ \left(
\begin{array}{cc}
I_{8} & 0 \\
0 & \mathcal{L}_{m}^{\left( \alpha _{2}\right) }%
\end{array}%
\right) ,\text{ \ }\left(
\begin{array}{cc}
I_{8} & 0 \\
0 & \mathcal{R}_{m}^{\left( \alpha _{2}\right) }%
\end{array}%
\right) ,\text{ \ }m\in \mathbb{Z}_{+}\left.
\begin{array}{c}
\\
\\
\end{array}%
\right. \right \}
\end{equation}%
leaving invariant $\left( b_{1},c_{1}\right) $ of (\ref{OC}). In the
above
relations, $\mathcal{L}_{n}^{\left( \alpha _{1}\right) },\mathcal{R}%
_{n}^{\left( \alpha _{1}\right) }$ and $\mathcal{L}_{n}^{\left(
\alpha _{2}\right) },\mathcal{R}_{n}^{\left( \alpha _{2}\right) }$
are as follows:
\begin{equation}
\begin{tabular}{lll}
$\mathcal{L}_{2k}^{\left( \alpha _{i}\right) }=\left( \mathcal{B}_{i}%
\mathcal{A}_{i}\right) ^{k}$ & , & $\mathcal{L}_{2k+1}^{\left(
\alpha _{i}\right) }=\mathcal{A}_{i}\left(
\mathcal{B}_{i}\mathcal{A}_{i}\right)
^{k}$ \\
$\mathcal{R}_{2k}^{\left( \alpha _{i}\right) }=\left( \mathcal{A}_{i}%
\mathcal{B}_{i}\right) ^{k}$ & , & $\mathcal{R}_{2k+1}^{\left(
\alpha _{i}\right) }=\mathcal{B}_{i}\left(
\mathcal{A}_{i}\mathcal{B}_{i}\right)
^{k}$%
\end{tabular}%
\end{equation}%
Clearly $G_{\alpha _{1}}$ and $G_{\alpha _{2}}$ are isomorphic to $\mathcal{G%
}_{weak}^{su_{2}}$,
\begin{equation}
G_{\alpha _{1}}\simeq \mathcal{G}_{weak}^{su_{2}}\simeq G_{\alpha
_{2}}
\end{equation}%
Notice also that using the identity matrix $I_{8}$ and the matrix
$J$ given below, the generators of $G_{\alpha _{1}}$ and $G_{\alpha
_{2}}$ can be also put into the remarkable form
\begin{equation}
\begin{tabular}{ll}
$\mathcal{L}_{2k}^{\left( \beta \right) }$ & $=I_{8}+2kJ$ \\
$\mathcal{R}_{2k}^{\left( \beta \right) }$ & $=I_{8}-2kJ$%
\end{tabular}%
\end{equation}%
and%
\begin{equation}
\begin{tabular}{ll}
$\mathcal{L}_{2k+1}^{\left( \beta \right) }$ & $=A-2kJ$ \\
$\mathcal{R}_{2k+1}^{\left( \beta \right) }$ & $=B+2kJ$%
\end{tabular}%
\end{equation}%
with $\beta $ standing for $\alpha _{1},$ $\alpha _{2}$. We also
have
\begin{equation}
\begin{tabular}{lll}
$J=\left(
\begin{array}{cc}
I_{4} & I_{4} \\
-I_{4} & -I_{4}%
\end{array}%
\right) $ & $,$ &
\end{tabular}
\label{J}
\end{equation}%
and the useful property
\begin{equation}
\left( J\right) ^{2}=0
\end{equation}%
allowing tremendous simplifications in doing explicit computations.

\subsubsection{Building $\mathcal{G}_{weak}^{su_{3}}$}

Here we complete the above study by constructing the subsymmetry
$G_{\alpha _{3}}$ of $\mathcal{G}_{weak}^{su_{3}}$; then we
\textrm{give the content of} $\mathcal{G}_{weak}^{su_{3}}$. An
inspection of the quiver mutations that link the $G_{\alpha _{1}}$
and $G_{\alpha _{2}}$ chambers suggests that the
elements of $G_{\alpha _{3}}$ are generated by the mutation matrices,%
\begin{equation}
\begin{tabular}{lll}
$\mathcal{A}_{3}=\left(
\begin{array}{cccc}
-I_{4} & 0 & -I_{4} & -I_{4} \\
2I_{4} & I_{4} & I_{4} & I_{4} \\
I_{4} & I_{4} & I_{4} & 2I_{4} \\
-I_{4} & -I_{4} & 0 & -I_{4}%
\end{array}%
\right) $ & $,$ & $\mathcal{B}_{3}=\left(
\begin{array}{cccc}
I_{4} & 2I_{4} & I_{4} & I_{4} \\
0 & -I_{4} & -I_{4} & -I_{4} \\
-I_{4} & -I_{4} & -I_{4} & 0 \\
I_{4} & I_{4} & 2I_{4} & I_{4}%
\end{array}%
\right) $%
\end{tabular}
\label{GAB}
\end{equation}%
these matrices couple the two subsets (sub-groupoids) $G_{\alpha _{1}}$ and $%
G_{\alpha _{2}}$ of the groupoid $\mathcal{G}_{weak}^{su_{2}}$.
Moreover, using the expression of the $8\times 8$ matrices $A$, $B$,
$J$ given previously and which we recall below
\begin{equation}
\begin{tabular}{llll}
$A=\left(
\begin{array}{cc}
-I_{4} & 0 \\
2I_{4} & I_{4}%
\end{array}%
\right) ,$ & $B=\left(
\begin{array}{cc}
I_{4} & 2I_{4} \\
0 & -I_{4}%
\end{array}%
\right) ,$ & $J=\left(
\begin{array}{cc}
I_{4} & I_{4} \\
-I_{4} & -I_{4}%
\end{array}%
\right) $ &
\end{tabular}%
\end{equation}%
we can rewrite the generators (\ref{GAB}) into $2\times 2$ block
matrices as
follows%
\begin{equation}
\begin{tabular}{lll}
$\mathcal{A}_{3}=\left(
\begin{array}{cc}
A & -J \\
J & B%
\end{array}%
\right) $ & $,$ & $\mathcal{B}_{3}=\left(
\begin{array}{cc}
B & J \\
-J & A%
\end{array}%
\right) ,$%
\end{tabular}%
\end{equation}%
Furthermore, using the properties
\begin{equation}
\begin{tabular}{lll}
$A^{2}=I_{8},$ & $B^{2}=I_{8},$ & $J^{2}=0,$%
\end{tabular}%
\end{equation}%
we have
\begin{equation}
\begin{tabular}{lll}
$\mathcal{A}_{3}^{2}=I_{16}$ & $,$ & $\mathcal{B}_{3}^{2}=I_{16}$%
\end{tabular}
\label{KA}
\end{equation}%
showing that $\mathcal{A}_{3}$ and $\mathcal{B}_{3}$ are also
involutions of
the BPS quivers. We also have the useful identities%
\begin{equation}
\begin{tabular}{lll}
$AB=I_{8}-2J,$ & $AJ=-J,$ & $JA=J$ \\
$BA=I_{8}+2J,$ & $BJ=-J,$ & $JB=J$%
\end{tabular}
\label{BA}
\end{equation}%
and
\begin{equation}
\begin{tabular}{lll}
$AB+BA=2I_{8},$ & $AJ+JA=0,$ & $AJ-JA=-2J$ \\
$AB-BA=-4J,$ & $BJ+JB=0,$ & $BJ-JB=-2J$%
\end{tabular}%
\end{equation}%
as well as%
\begin{equation}
\begin{tabular}{l}
$\left( AB\right) ^{k}=I_{8}-2kJ$ \\
$\left( BA\right) ^{k}=I_{8}+2kJ$%
\end{tabular}%
\end{equation}%
To get the elements of $G_{\alpha _{3}}$ $\subset
\mathcal{G}_{weak}^{su_{3}}
$, we take advantage for the isomorphism%
\begin{equation}
G_{\alpha _{3}}\simeq \mathcal{G}_{weak}^{su_{2}}
\end{equation}%
to build $\mathcal{L}_{n}^{\left( \alpha _{3}\right) }$ and $\mathcal{R}%
_{n}^{\left( \alpha _{3}\right) }$ that read as follows:%
\begin{equation}
G_{\alpha _{3}}=\left \{
\begin{tabular}{llll}
$\mathcal{L}_{2k}^{\left( \alpha _{3}\right) },$ & $\mathcal{L}%
_{2k+1}^{\left( \alpha _{3}\right) },$ & $\mathcal{R}_{2k}^{\left(
\alpha
_{3}\right) },$ & $\mathcal{R}_{2k+1}^{\left( \alpha _{3}\right) }$%
\end{tabular}%
;\text{ }k\in \mathbb{Z}^{+}\right \}
\end{equation}%
with%
\begin{equation}
\begin{tabular}{lllll}
$\mathcal{L}_{2k}^{\left( \alpha _{3}\right) }$ & $=\left( \mathcal{B}_{3}%
\mathcal{A}_{3}\right) ^{k}$ & , & $\mathcal{R}_{2k}^{\left( \alpha
_{3}\right) }$ & $=\left( \mathcal{A}_{3}\mathcal{B}_{3}\right) ^{k}$ \\
$\mathcal{L}_{2k+1}^{\left( \alpha _{3}\right) }$ &
$=\mathcal{A}_{3}\left(
\mathcal{B}_{3}\mathcal{A}_{3}\right) ^{k}$ & , & $\mathcal{R}%
_{2k+1}^{\left( \alpha _{3}\right) }$ & $=\mathcal{B}_{3}\left( \mathcal{A}%
_{3}\mathcal{B}_{3}\right) ^{k}$%
\end{tabular}%
\end{equation}%
and
\begin{equation}
\mathcal{L}_{2k}^{\left( \alpha _{3}\right) }=\left(
\begin{array}{cc}
I_{8}+2kJ & 2kJ \\
-2kJ & I_{8}-2kJ%
\end{array}%
\right)
\end{equation}%
as well as%
\begin{equation}
\mathcal{R}_{2k}^{\left( \alpha _{3}\right) }=\left(
\begin{array}{cc}
I_{8}-2kJ & -2kJ \\
2kJ & I_{8}+2kJ%
\end{array}%
\right)
\end{equation}%
We also have
\begin{equation}
\mathcal{L}_{2k+1}^{\left( \alpha _{3}\right) }=\left(
\begin{array}{cc}
A-2kJ & -\left( 2k+1\right) J \\
\left( 2k+1\right) J & B+2kJ%
\end{array}%
\right)
\end{equation}%
and%
\begin{equation}
\mathcal{R}_{2k+1}^{\left( \alpha _{3}\right) }=\left(
\begin{array}{cc}
B+2kJ & \left( 2k+1\right) J \\
-\left( 2k+1\right) J & A-2kJ%
\end{array}%
\right)
\end{equation}%
Knowing the explicit expressions of the generators $\mathcal{A}_{i},$ $%
\mathcal{B}_{i}$, one can then compute the elements of $\mathcal{G}%
_{weak}^{su_{3}}$; these are given by the monomials (\ref{GE}).

\subsection{BPS spectrum}

The weak coupling BPS chamber of the $\mathcal{N}=2$ supersymmetric
QFT with SU$\left( 3\right) $ gauge symmetry is given by the
following $\left(
k,l,p\right) $-positive integer series%
\begin{equation}
\begin{array}{cc}
\left[ 1+2\left( k+p\right) \right] b_{1}+2\left( k+p\right)
c_{1}+2pb_{2}+2pc_{2} &  \\
-2\left( k+p\right) b_{1}-\left[ 2\left( k+p\right) -1\right]
c_{1}-2pb_{2}-2pc_{2} &  \\
2pb_{1}+2pc_{1}+\left[ 1+2\left( l+p\right) \right] b_{2}+2\left(
l+p\right)
c_{2} &  \\
-2pb_{1}-2pc_{1}-2\left( l+p\right) b_{2}+\left[ 2\left( l+p\right)
-1\right] c_{2} &
\end{array}%
\end{equation}%
and
\begin{equation}
\begin{array}{cc}
\left[ 1-2\left( k+p\right) \right] b_{1}-2\left( k+p\right)
c_{1}-2pb_{2}-2pc_{2} &  \\
2\left( k+p\right) b_{1}+\left[ 1+2\left( k+p\right) \right]
c_{1}+2pb_{2}+2pc_{2} &  \\
-2pb_{1}-2pc_{1}-\left[ 2\left( l+p\right) -1\right] b_{2}-2\left(
l+p\right) c_{2} &  \\
2pb_{1}+2pc_{1}+2\left( l+p\right) b_{2}+\left[ 1+2\left( l+p\right)
\right] c_{2} &
\end{array}%
\end{equation}%
These EM charge vectors c\textrm{an be also put into blocks of
matrices as}

\begin{eqnarray*}
&&\left(
\begin{array}{cccc}
\left[ 1+2\left( k+p\right) \right] I_{4}\text{ \  \  \  \ } &
2\left(
k+p\right) I_{4} & 2pI_{4} & 2pI_{4} \\
-2\left( k+p\right) I_{4} & \left[ 1-2\left( k+p\right) \right]
I_{4}\text{
\  \  \  \ } & -2pI_{4} & -2pI_{4} \\
2pI_{4} & 2pI_{4} & \left[ 1+2\left( l+p\right) \right] I_{4}\text{
\  \  \  \ }
& 2\left( l+p\right) I_{4} \\
-2pI_{4} & -2pI_{4} & -2\left( l+p\right) I_{4} & \left[ 1-2\left(
l+p\right) \right] I_{4}%
\end{array}%
\right) \\
&&
\end{eqnarray*}%
and%
\begin{eqnarray*}
&&\left(
\begin{array}{cccc}
\left[ 1-2\left( k+p\right) \right] I_{4}\text{ \  \  \  \ } &
-2\left(
k+p\right) I_{4} & -2pI_{4} & -2pI_{4} \\
2\left( k+p\right) I_{4} & \left[ 1+2\left( k+p\right) \right]
I_{4}\text{ \
\  \  \ } & 2pI_{4} & 2pI_{4} \\
-2pI_{4} & -2pI_{4} & \left[ 1-2\left( l+p\right) \right]
I_{4}\text{ \  \  \
\ } & -2\left( l+p\right) I_{4} \\
2pI_{4} & 2pI_{4} & 2\left( l+p\right) I_{4} & \left[ 1+2\left( l+p\right) %
\right] I_{4}%
\end{array}%
\right) \\
&&
\end{eqnarray*}%
This BPS chamber contains the weak sub-chambers associated of the
mutations sets $G_{\alpha _{1}},G_{\alpha _{2}},G_{\alpha _{3}}$.
They appear as one-integer series; we have:

\begin{itemize}
\item the $\left( k,0,0\right) $- series

The mutation matrices reads as
\begin{equation}
\mathcal{L}_{2k}^{\left( \alpha _{1}\right) }=\left(
\begin{array}{cccc}
\left( 1+2k\right) I_{4} & 2kI_{4} & 0 & 0 \\
-2kI_{4} & \left( 1-2k\right) I_{4} & 0 & 0 \\
0 & 0 & I_{4}\  & 0 \\
0 & 0 & 0 & I_{4}%
\end{array}%
\right)
\end{equation}%
and%
\begin{equation}
\mathcal{R}_{2k}^{\left( \alpha _{1}\right) }=\left(
\begin{array}{cccc}
\left( 1-2k\right) I_{4} & -2kI_{4} & 0 & 0 \\
2kI_{4} & \left( 1+2k\right) I_{4} & 0 & 0 \\
0 & 0 & I_{4}\  & 0 \\
0 & 0 & 0 & I_{4}\
\end{array}%
\right)
\end{equation}%
with respective large k limit given by%
\begin{equation}
\begin{tabular}{ll}
$2k\left(
\begin{array}{cccc}
I_{4} & I_{4} & 0 & 0 \\
-I_{4} & -I_{4} & 0 & 0 \\
0 & 0 & I_{4}\  & 0 \\
0 & 0 & 0 & I_{4}%
\end{array}%
\right) ,$ & $2k\left(
\begin{array}{cccc}
-I_{4} & -I_{4} & 0 & 0 \\
I_{4} & I_{4} & 0 & 0 \\
0 & 0 & I_{4}\  & 0 \\
0 & 0 & 0 & I_{4}%
\end{array}%
\right) $%
\end{tabular}%
\end{equation}%
describing the EM charge vectors $\pm w_{1}=\pm \left(
b_{1}+c_{1}\right) $ of the W$_{\pm \alpha _{1}}$-bosons together
with a monopole and a dyon. There are also two more matrices
$\mathcal{L}_{2k+1}^{\left( \alpha _{1}\right) }$ and
$\mathcal{R}_{2k+1}^{\left( \alpha _{1}\right) }$ that we have not
reported here and which are given by (\ref{REP}).

\item the $\left( 0,l,0\right) $- series

this series is given by
\begin{equation}
\left(
\begin{array}{cccc}
I_{4} & 0 & 0 & 0 \\
0 & I_{4} & 0 & 0 \\
0 & 0 & \left( 1+2l\right) I_{4} & 2lI_{4} \\
0 & 0 & -2lI_{4} & \left( 1-2l\right) I_{4}%
\end{array}%
\right)
\end{equation}%
and%
\begin{equation}
\left(
\begin{array}{cccc}
I_{4} & 0 & 0 & 0 \\
0 & I_{4} & 0 & 0 \\
0 & 0 & \left( 1-2l\right) I_{4} & -2lI_{4} \\
0 & 0 & 2lI_{4} & \left( 1+2l\right) I_{4}%
\end{array}%
\right)
\end{equation}%
with respective large $l$ limit as follows%
\begin{equation}
\begin{tabular}{ll}
$2l\left(
\begin{array}{cccc}
I_{4} & 0 & 0 & 0 \\
0 & I_{4} & 0 & 0 \\
0 & 0 & I_{4} & I_{4} \\
0 & 0 & -I_{4} & -I_{4}%
\end{array}%
\right) ,$ & $2l\left(
\begin{array}{cccc}
I_{4} & 0 & 0 & 0 \\
0 & I_{4} & 0 & 0 \\
0 & 0 & -I_{4} & -I_{4} \\
0 & 0 & I_{4} & I_{4}%
\end{array}%
\right) $%
\end{tabular}%
\end{equation}%
describing the EM charge vectors $\pm w_{2}=\pm \left(
b_{2}+c_{2}\right) $ of the $W_{\pm \alpha _{2}}$-bosons together
with a monopole and a dyon.

\item the $\left( 0,0,p\right) $- series

We have
\begin{eqnarray}
&&\left(
\begin{array}{cccc}
\left( 1+2p\right) I_{4} & 2pI_{4} & 2pI_{4} & 2pI_{4} \\
-2pI_{4} & \left( 1-2p\right) I_{4} & -2pI_{4} & -2pI_{4} \\
2pI_{4} & 2pI_{4} & \left( 1+2p\right) I_{4} & 2pI_{4} \\
-2pI_{4} & -2pI_{4} & -2pI_{4} & \left( 1-2p\right) I_{4}%
\end{array}%
\right) \\
&&  \notag
\end{eqnarray}%
and%
\begin{eqnarray}
&&\left(
\begin{array}{cccc}
\left( 1-2p\right) I_{4} & -2pI_{4} & -2pI_{4} & -2pI_{4} \\
2pI_{4} & \left( 1+2p\right) I_{4} & 2pI_{4} & 2pI_{4} \\
-2pI_{4} & -2pI_{4} & \left( 1-2p\right) I_{4} & -2pI_{4} \\
2pI_{4} & 2pI_{4} & 2pI_{4} & \left( 1+2p\right) I_{4}%
\end{array}%
\right) \\
&&  \notag
\end{eqnarray}%
with respective large $p$ limit given by%
\begin{equation}
\begin{tabular}{ll}
$2p\left(
\begin{array}{cccc}
1 & 1 & 1 & 1 \\
-1 & -1 & -1 & -1 \\
1 & 1 & 1 & 1 \\
-1 & -1 & -1 & -1%
\end{array}%
\right) \otimes I_{4},$ & $2p\left(
\begin{array}{cccc}
-1 & -1 & -1 & -1 \\
1 & 1 & 1 & 1 \\
-1 & -1 & -1 & -1 \\
1 & 1 & 1 & 1%
\end{array}%
\right) \otimes I_{4}$%
\end{tabular}%
\end{equation}
\end{itemize}

describing the electric-magnetic charge vectors $\pm w_{3}=\pm
\left( b_{1}+c_{1}+b_{2}+c_{2}\right) $ of the massive W$_{\pm
\left( \alpha _{1}+\alpha _{2}\right) }$- bosons.\newline Using the
$w_{i}$ charge of W-bosons, the BPS spectrum of the weak chamber,
following from our construction based on borrowing specific features
from
the Lie algebra of the SU$\left( 3\right) $ gauge symmetry, reads as%
\begin{equation}
\begin{tabular}{ll||ll}
\multicolumn{2}{l||}{\  \  \  \  \  \  \  \  \  \ left sector of
SU$\left( 3\right) $} & \multicolumn{2}{||l}{\  \  \  \  \  \  \
right sector of SU$\left( 3\right) $}
\\ \hline
&  &  &  \\
$b_{1}^{\left( 2k,p\right) }$ & $=b_{1}+2kw_{1}+2pw_{3}$ &
$b_{1}^{\left(
2k,p\right) }$ & $=b_{1}-2kw_{1}-2pw_{3}$ \\
$c_{1}^{\left( 2k,p\right) }$ & $=c_{1}-2kw_{1}-2pw_{3}$ &
$c_{1}^{\left(
2k,p\right) }$ & $=c_{1}+2kw_{1}+2pw_{3}$ \\
$b_{1}^{\left( 2k+1,p\right) }$ & $=c_{1}-\left( 2k+1\right)
w_{1}+2pw_{3}$ & $b_{1}^{\left( 2k+1,p\right) }$ & $=c_{1}+\left(
2k+1\right) w_{1}-2pw_{3}$
\\
$c_{1}^{\left( 2k+1,p\right) }$ & $=b_{1}+\left( 2k+1\right)
w_{1}-2pw_{3}$
& ${\normalsize c}_{1}^{\left( 2k+1,p\right) }$ & ${\normalsize =b_{1}-}%
\left( 2k+1\right) {\normalsize w}_{1}+2pw_{3}$ \\
&  &  &  \\
$b_{2}^{\left( 2l,p\right) }$ & $=b_{2}+2lw_{2}+2pw_{3}$ &
$b_{2}^{\left(
2k,p\right) }$ & $=b_{2}-2lw_{2}-2pw_{3}$ \\
$c_{2}^{\left( 2l,p\right) }$ & $=c_{2}-2lw_{2}-2pw_{3}$ &
$c_{2}^{\left(
2k,p\right) }$ & $=c_{2}+2lw_{2}+2pw_{3}$ \\
$b_{2}^{\left( 2l+1,p\right) }$ & $=c_{2}-\left( 2l+1\right)
w_{2}+2pw_{3}$ & $b_{2}^{\left( 2k+1,p\right) }$ & $=c_{2}+\left(
2l+1\right) w_{2}-2pw_{3}$
\\
$c_{2}^{\left( 2l+1,p\right) }$ & $=b_{2}+\left( 2l+1\right)
w_{2}-2pw_{3}$
& ${\normalsize c}_{2}^{\left( 2k+1,p\right) }$ & ${\normalsize =b_{2}-}%
\left( 2l+1\right) {\normalsize w}_{2}+2pw_{3}$ \\
&  &  &  \\ \hline
\end{tabular}%
\end{equation}

\section{Conclusion and comment}

Motivated by recent results on BPS quiver theory, we have developed
in this paper an algebraic method to study the BPS\ spectra of
$\mathcal{N}=2$ supersymmetric QFT$_{4}$ with ADE gauge symmetries.
After describing useful tools on BPS quiver theory along the line
of\  \textrm{\cite{1A,1B}}, we have shown that BPS spectra of
$\mathcal{N}=2$ QFT$_{4}$ is completely determined
by the primitive quiver $\mathfrak{Q}_{0}^{G}$, described by the pair $%
\left( \mathbf{v}^{\left( 0\right) },\mathcal{A}^{\left( 0\right) }\right) $%
; and the set $\mathcal{G}_{Mut}^{G}=\left \{ \mathcal{M}_{m,n}:\mathfrak{Q}%
_{n}^{G}\rightarrow \mathfrak{Q}_{m}^{G};\text{ }m,n\geq 0\right \}
$ of quiver mutations which correspond precisely to arrows in the
groupoid language; see appendix I.\newline The mutation symmetries
$\mathcal{G}_{Mut}^{G}$, with G standing for any ADE gauge group,
have been explicitly worked out; and were shown to be either finite
and closed or infinite depending on the BPS\ chambers that
correspond to groupoid orbits. In the finite case describing strong
coupling BPS chambers, we have shown that BPS chambers are given by
finite and closed groupoid orbits with total number of BPS and
anti-BPS states given by the number of $\mathcal{M}_{m,0}$ mutations
in $\mathcal{G}_{strong}^{G}$ as
reported in eq(\ref{DR}). We have also derived the relation between\ $%
\mathcal{G}_{strong}^{G}$ and the discrete groups $Dih_{2h}$ with
$h$ the Coxeter number of the ADE gauge symmetries. \newline Then,
we have used the same method to study the infinite weak coupling
chambers of $\mathcal{N}=2$ supersymmetric QFT with $SU\left( 2\right) $, $%
SO\left( 4\right) $ and $SU\left( 3\right) $ gauge symmetries. We
have derived the corresponding mutation symmetries and the
associated BPS spectra. We believe that this construction applies as
well to get the weak coupling chambers for all ADE gauge symmetries
as well as supersymmetric gauge symmetries with flavor matters.

\  \  \  \  \newline \textbf{Acknowledgements}: This research work
is supported by URAC09/CNRS.

\section{Appendix I: Groupoids}

Physical symmetries are generally thought of in terms of groups and
their
representations. However there are plenty of objects, such as the set $%
\mathcal{G}_{Mut}$ considered in this study, which exhibit what we
clearly recognize as symmetries, but which strictly speaking do not
form groups. The structure of such kind of exotic symmetries are
described by groupoids first
introduced by H. Brandt \textrm{\cite{030J,030K}}, see also \textrm{\cite%
{030L,31II,32II,33II,AW,3I} and refs therein}. In this appendix, we
first review some basic properties of groupoids and then illustrate
the construction on the set $\mathcal{G}_{Mut}$ of the quiver
mutations of the strong and weak BPS chambers of $\mathcal{N}=2$
QFT$_{4}$ we have been studying in this paper.

\subsection{Set up of the structure}

Roughly, one can think about groupoids as a generalization of groups
in the sense that their algebraic structures are almost the same
except that in groupoids not all elements can be composed. Following
\textrm{\cite{AW,MW}},
the elements of groupoids can be often presented as collections of triples $%
\left( x,g,y\right) $ obeying the groupoid relations (i)-(iv) given
below.

\subsubsection{Definition}

Given a set $\mathbb{B}$ of objects $x$, a groupoid $\mathbb{G}$ over $%
\mathbb{B}$ (sometimes denoted like $\mathbb{G}\rightrightarrows \mathbb{B}$%
) is given by the set
\begin{equation}
\mathbb{G}=\left \{ \mathrm{g}\equiv \left( x,g,y\right) \text{ \
}|\text{ \ }x,y\in \mathbb{B}\text{ and }x=gy\right \}  \label{47}
\end{equation}%
together with the following: \newline (a) A partially defined binary
operation $\left( \mathrm{g,h}\right) \rightarrow \mathrm{gh}$,
which by using the triples $\left( x,g,y\right) $,
reads explicitly like%
\begin{equation}
\left( x,g,y\right) .\left( y,h,z\right) =\left( x,gh,z\right)
\label{br}
\end{equation}%
(b) Two maps $S,$ $T$ from $\mathbb{G}$ to $\mathbb{B}$, known as
source and
target maps, and acting on the groupoid elements like%
\begin{equation}
\begin{tabular}{lll}
$S:$ & $\left( x,g,y\right) \rightarrow x$ &  \\
$T:$ & $\left( x,g,y\right) \rightarrow y$ &
\end{tabular}
\label{ss}
\end{equation}%
The binary operation (\ref{br}) obeys the 4 properties: \newline
$i)$ It is defined only for certain pairs of elements; the product \textrm{gh%
} is defined only when source and target maps are constrained as
\begin{equation}
T\left( \mathrm{g}\right) =S\left( \mathrm{h}\right)  \label{50}
\end{equation}%
see also fig \ref{MO} to fix the ideas\newline $ii)$ It is
associative; if one of the products $\left( \mathrm{gh}\right)
\mathrm{h}$ or $\mathrm{g}\left( \mathrm{hk}\right) $ is defined,
then so is the other and they are equal
\begin{equation}
\left( \mathrm{gh}\right) \mathrm{k}=\mathrm{g}\left(
\mathrm{hk}\right) \label{51}
\end{equation}%
$iii)$ For each element \textrm{g} in $\mathbb{G}$, there are left
and right identity elements $\mathrm{\lambda }_{g}$ and
$\mathrm{\varrho }_{g}$ such
that%
\begin{equation}
\mathrm{\lambda }_{g}\mathrm{g}=\mathrm{g}=\mathrm{g\varrho }_{g}
\label{52}
\end{equation}%
$iv)$ Each \textrm{g }$\in \mathbb{G}$ has an inverse
$\mathrm{g}^{-1}\in \mathbb{G}$ for which
\begin{equation}
\mathrm{gg}^{-1}=\mathrm{\lambda }_{g}\mathrm{,\qquad g}^{-1}\mathrm{g}=%
\mathrm{\varrho }_{g}  \label{53}
\end{equation}%
where $\mathrm{\lambda }_{g}$ and $\mathrm{\varrho }_{g}$ are as in
(iii).

\subsubsection{Some particular properties}

Here we give some special and useful features on groupoids and their
orbits.

\emph{a) groupoids and groups}\newline
From the above definition, one learns that in case where the basis set $%
\mathbb{B}$ contains a single element $x$, all pairs of groupoid
elements can be composed in any order; and so the groupoid is a
group.
\begin{figure}[tbph]
\begin{center}
\hspace{0cm} \includegraphics[width=12cm]{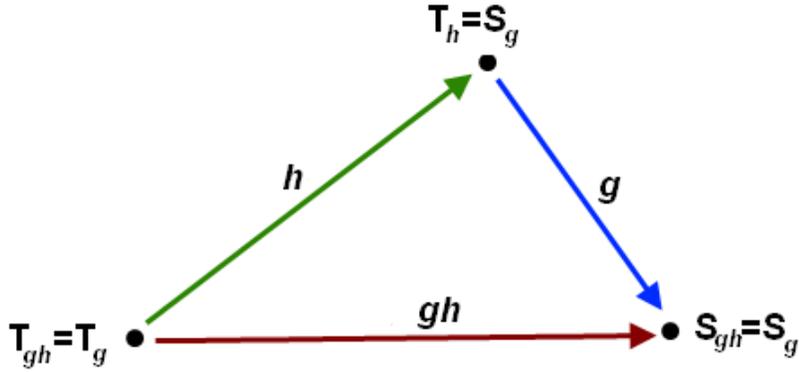}
\end{center}
\par
\vspace{-1 cm} \caption{Composition of arrows in groupoids}
\label{MO}
\end{figure}
Inversly, any group can be considered as a groupoid over a point.
More generally, given a collection of points, then the collection of
groups over those points is a groupoid.

\emph{b) arrow representation}\newline From eqs(\ref{47}-\ref{53})
and the illustrating figure one also learns that each element
$\mathrm{g}$ of the groupoid $\mathbb{G}$ can be represented by
an \emph{arrow} pointing from $T\left( g\right) $ to $S\left( g\right) $ in $%
\mathbb{B}$ as follows%
\begin{equation}
T\left( g\right) \text{\  \ }\overset{g}{\longrightarrow }\text{ \  \  \ }%
S\left( g\right)
\end{equation}%
where $S$ and $T$ are the source and target maps of eq(\ref{ss}). In
this graphic representation, to be used later on, the \emph{groupoid
multiplication} \emph{law} corresponds then to the \emph{composition
of the arrows} placing the head of a given arrow $h$ to the tail of
the previous one $g$ as shown on fig \ref{MO}.

\emph{c) other properties}\newline Using the arrow representation,
one can read directly the basic groupoid properties; in particular
the following,
\begin{equation}
\begin{tabular}{lllll}
$S_{g^{-1}}$ & $=T_{g}$ & , & $T_{g^{-1}}$ & $=S_{g}$ \\
$S_{gh}$ & $=S_{g}$ & , & $T_{gh}$ & $=T_{h}$%
\end{tabular}
\label{ob}
\end{equation}%
and%
\begin{equation}
\begin{tabular}{lllll}
$\left( \lambda _{g}g\right) h$ & $=\lambda _{g}\left( gh\right) $ & , & $%
g\left( h\varrho _{g}\right) $ & $=\left( gh\right) \varrho _{g}$%
\end{tabular}%
\end{equation}%
Moreover, since for each object $x\in \mathbb{B}$, there is an
identity arrow $id_{x}\in \mathbb{G}$, one can embed the set
$\mathbb{B}$ of objects in the groupoid;
\begin{equation}
id:\mathbb{B\hookrightarrow G}
\end{equation}%
allowing to think about $\mathbb{B}$ as a subset of $\mathbb{G}$.

\emph{d) orbit of the groupoid}\newline An orbit of the groupoid
$\mathbb{G}$ over $\mathbb{B}$ is an equivalence
class for the relation%
\begin{equation}
\begin{tabular}{llll}
$x\sim _{\mathbb{G}}y$ & , & $x,y\in \mathbb{B}$ &
\end{tabular}%
\end{equation}%
if and only if there is a groupoid element $g\in \mathbb{G}$ such that%
\begin{equation}
\begin{tabular}{lll}
$S\left( g\right) =x$ & , & $T\left( g\right) =y$%
\end{tabular}%
\end{equation}

\begin{figure}[tbph]
\begin{center}
\hspace{0cm} \includegraphics[width=14cm]{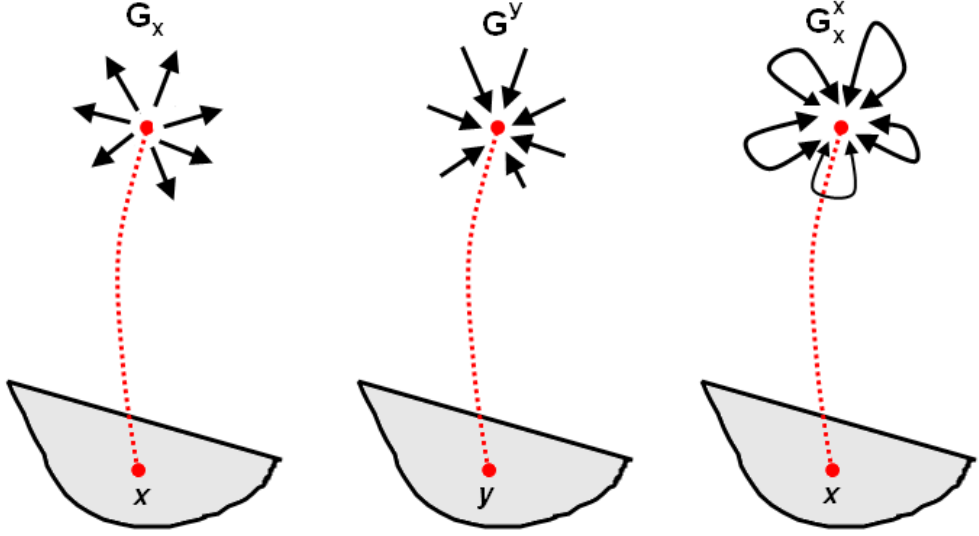}
\end{center}
\par
\vspace{-1 cm} \caption{on left the source fiber $S^{-1}\left(
x\right) $, in middle the
target fiber $T^{-1}\left( y\right) $ and on right the isotropy group $%
S^{-1}\left( x\right) \cap T^{-1}\left( x\right) $} \label{star}
\end{figure}

\emph{e)\ fibers and isotropy group}\newline If
$\mathbb{G}\rightrightarrows \mathbb{B}$ is a groupoid and $x,y\in
\mathbb{B}$, then we have the following notions:\newline i) the
\emph{source-fiber} at $x$ is the set $\mathbb{G}_{x}$ of all arrows
from $x$, namely%
\begin{equation}
\mathbb{G}_{x}=\mathbb{G}\left( x,.\right) =S^{-1}\left( x\right)
=\left \{ g\in G\text{ }|\text{ }S\left( g\right) =x\right \}
\label{O}
\end{equation}%
ii) the \emph{target-fiber} at $y$ is the set of all arrows to y, namely%
\begin{equation}
\mathbb{G}^{y}=\mathbb{G}\left( .,y\right) =T^{-1}\left( y\right)
=\left \{ g\in G\text{ }|\text{ }T\left( g\right) =y\right \}
\label{T}
\end{equation}%
iii) the set of arrows from $x$ to y is%
\begin{equation}
\mathbb{G}_{x}^{y}=\mathbb{G}\left( x,y\right) =S^{-1}\left(
x\right) \cap
T^{-1}\left( y\right) =\left \{ g\in G\text{ }|\text{ }x\overset{g}{%
\rightarrow }y\right \}  \label{TT}
\end{equation}%
\textbf{iv}) the isotropy group at $x$ is the set of \emph{self-arrows} of $%
x $, namely%
\begin{equation}
\mathbb{G}_{x}^{x}=\mathbb{G}\left( x,x\right) =S^{-1}\left(
x\right) \cap
T^{-1}\left( x\right) =\left \{ g\in G\text{ }|\text{ }x\overset{g}{%
\rightarrow }x\right \}  \label{S}
\end{equation}

\  \  \  \newline Notice that the set $\mathbb{G}_{x}^{x}$ has a
group structure; indeed the multiplication in $\mathbb{G}_{x}^{x}$
is inherited from $\mathbb{G}$; it is associative and is defined for
each $g\in $ $\mathbb{G}_{x}^{x}$ since
\begin{equation}
S\left( x\right) =x=T\left( x\right) .
\end{equation}
The identity $id_{x}$ is in $\mathbb{G}_{x}^{x}$, and if $g\in $ $\mathbb{G}%
_{x}^{x}$, then $g^{-1}\in $ $\mathbb{G}_{x}^{x}$ since
\begin{equation}
S\left( g^{-1}\right) =T\left( g\right) =x=S\left( g\right) =T\left(
g^{-1}\right) .
\end{equation}%
Notice also that one might picture the sets
\begin{equation}
\begin{tabular}{lllll}
$\mathbb{G}_{x}$ & $,$ & $\mathbb{G}^{y}$ & $,$ & $\mathbb{G}_{x}^{x}$%
\end{tabular}%
\end{equation}%
as \emph{dandelions} above each point in the base as in fig
\ref{star}. These sets are sometimes called the \emph{star}, the
\emph{costar} and the the \emph{vertex group} respectively.
Along with special features given above; one can consider many other
aspects of groupoids by mimicking the theory of groups; for details
on some of these aspects and also for applications of groupoids in
physical literature see \textrm{\cite{3I}} and refs therein for the
important role they play in topological field theory. With these
general tools at hand, we turn to the study of
$\mathcal{G}_{Mut}^{G}$.

\subsection{Quiver mutations}

In the case of the quiver mutations considered in this paper, the base set $%
\mathbb{B}$ of the groupoid $\mathcal{G}_{Mut}^{G}$ is given by the
chambers
of the BPS quivers; in other words the objects $x$ are the quivers $%
\mathfrak{Q}_{n}^{G}$ with $n$ integer.

\subsubsection{Mutation groupoid}

In the arrow representation introduced above, the groupoid $\mathcal{G}%
_{Mut}^{G}$ can be pictured as the collection of the $\mathfrak{Q}_{n}^{G}$%
's with the various arrows connecting them as follows:%
\begin{equation}
\mathcal{G}_{Mut}^{G}=\left \{ \mathcal{M}_{_{n,m}}:\text{ }\mathfrak{Q}%
_{n}^{G}\text{\  \ }\rightarrow \text{\  \  \ }\mathfrak{Q}_{m}^{G}\text{; \  \ }%
n,m\in \mathbb{Z}\text{ }\right \}  \label{df}
\end{equation}%
with $\mathcal{M}_{_{n,m}}$ invertible maps whose basic properties
are given below. Using this arrow notation, we immediately figure
out the groupoid structure of the set of mutations of BPS quivers as
well as their useful properties.

\  \  \  \  \

$\left( a\right) $ \emph{Composition law in }$\mathcal{G}_{Mut}^{G}$\emph{:}%
\newline
The binary operation in the set of quiver mutations
$\mathcal{G}_{Mut}^{G}$ is given by the following arrow
multiplication
\begin{equation}
\mathfrak{Q}_{n}^{G}\overset{\mathcal{M}_{_{n,m}}}{\longrightarrow }%
\mathfrak{Q}_{m}^{G}\overset{\mathcal{M}_{_{m,k}}}{\longrightarrow }%
\mathfrak{Q}_{k}^{G}\text{ \ }=\text{ \ }\mathfrak{Q}_{n}^{G}\overset{%
\mathcal{M}_{_{n,k}}}{\longrightarrow }\mathfrak{Q}_{k}^{G}
\end{equation}%
from which we learn that not all mutations can be composed since we
should have
\begin{equation}
\mathcal{M}_{_{n,m}}\mathcal{M}_{_{l,k}}=\delta
_{ml}\mathcal{M}_{_{n,k}} \label{21}
\end{equation}%
showing that mutation pairings are defined only if $m=l$; here
$\delta _{ml}$ is the usual Kronecker symbol.

\  \  \

$\left( b\right) $ \emph{S and T maps in
}$\mathcal{G}_{Mut}^{G}$\newline
The source and target mappings $S$ and $T$ in the mutation groupoid $%
\mathcal{G}_{Mut}^{G}$ are given by the analogue of eqs(\ref{ss});
they are
defined as follows%
\begin{equation}
\begin{tabular}{lll}
$S\left( \mathfrak{Q}_{n}^{G}\overset{\mathcal{M}_{_{n,m}}}{\longrightarrow }%
\mathfrak{Q}_{m}^{G}\overset{\mathcal{M}_{_{m,k}}}{\longrightarrow }%
\mathfrak{Q}_{k}^{G}\right) $ & $=$ & $S\left( \mathfrak{Q}_{n}^{G}\overset{%
\mathcal{M}_{_{n,m}}}{\longrightarrow }\mathfrak{Q}_{m}^{G}\right) $ \\
&  &  \\
$T\left( \mathfrak{Q}_{n}^{G}\overset{\mathcal{M}_{_{n,m}}}{\longrightarrow }%
\mathfrak{Q}_{m}^{G}\overset{\mathcal{M}_{_{m,k}}}{\longrightarrow }%
\mathfrak{Q}_{k}^{G}\right) $ & $=$ & $T\left( \mathfrak{Q}_{m}^{G}\overset{%
\mathcal{M}_{_{m,k}}}{\longrightarrow }\mathfrak{Q}_{k}^{G}\right) $%
\end{tabular}%
\end{equation}%
and obey eq(\ref{ob}).

\  \  \

$\left( c\right) $ \emph{Inverse in }$\mathcal{G}_{Mut}^{G}$\newline
The inverse of a generic quiver mutation of $\mathcal{G}_{Mut}^{G}$
is given
by the map%
\begin{equation}
\begin{tabular}{lll}
$\iota \left( \mathfrak{Q}_{n}^{G}\overset{\mathcal{M}_{_{n,m}}}{%
\longrightarrow }\mathfrak{Q}_{m}^{G}\right) $ & $=$ & $\mathfrak{Q}_{n}^{G}%
\overset{\mathcal{M}_{_{n,m}}^{-1}}{\longrightarrow }\mathfrak{Q}_{m}^{G}$%
\end{tabular}%
\end{equation}%
which, by using the definition (\ref{df}), leads to
\begin{equation}
\mathcal{M}_{_{n,m}}^{-1}=\mathcal{M}_{_{m,n}}
\end{equation}%
From this relation, we also learn that in the case where $m=n$, we
have
\begin{equation}
\mathcal{M}_{_{n,n}}^{-1}=\mathcal{M}_{_{n,n}}  \label{id}
\end{equation}%
showing that the $\mathcal{M}_{_{n,n}}$'s are nothing but the identities $%
id_{\mathfrak{Q}_{n}^{G}}$ of the mutation groupoid.

\  \  \

$\left( d\right) $ \emph{Left and right identities in }$\mathcal{G}%
_{Mut}^{G} $\newline The left and right identities\emph{\ }in
$\mathcal{G}_{Mut}^{G}$ are as in
previous relation (\ref{id}); they correspond to%
\begin{equation}
\begin{tabular}{lll}
$\mathfrak{Q}_{n}^{G}\overset{\mathcal{M}_{_{n,m}}}{\longrightarrow }%
\mathfrak{Q}_{m}^{G}\overset{\mathcal{M}_{_{m,m}}}{\longrightarrow }%
\mathfrak{Q}_{m}^{G}$ & $=$ & $\mathfrak{Q}_{n}^{G}\overset{\mathcal{M}%
_{_{n,m}}}{\longrightarrow }\mathfrak{Q}_{m}^{G}$%
\end{tabular}%
\end{equation}%
which can be also written like%
\begin{equation}
\mathfrak{Q}_{n}^{G}\overset{\mathcal{M}_{_{n,n}}}{\longrightarrow }%
\mathfrak{Q}_{n}^{G}\overset{\mathcal{M}_{_{n,m}}}{\longrightarrow }%
\mathfrak{Q}_{m}^{G}
\end{equation}%
From these two graphic equations, we read the left and right
identities; i.e
the analogue of eq(\ref{53}). We also have%
\begin{equation}
\begin{tabular}{lll}
$\mathcal{M}_{_{n,m}}\mathcal{M}_{_{m,m}}$ & $=\mathcal{M}_{_{n,m}}=$ & $%
\mathcal{M}_{_{n,n}}\mathcal{M}_{_{n,m}}$%
\end{tabular}%
\end{equation}

\  \  \  \  \  \  \  \  \

$\left( e\right) $ \emph{Left and right multiplication in }$\mathcal{G}%
_{Mut}^{G}$ \newline
The analogue of $\mathrm{gg}^{-1}=\mathrm{\lambda }_{g}$ is given by%
\begin{equation}
\begin{tabular}{lll}
$\mathfrak{Q}_{n}^{G}\overset{\mathcal{M}_{_{n,m}}}{\longrightarrow }%
\mathfrak{Q}_{m}^{G}\overset{\mathcal{M}_{_{n,m}}^{-1}}{\longrightarrow }%
\mathfrak{Q}_{n}^{G}$ & $=$ & $\mathfrak{Q}_{n}^{G}\overset{\mathcal{M}%
_{_{n,n}}}{\longrightarrow }\mathfrak{Q}_{n}^{G}$%
\end{tabular}%
\end{equation}%
Similarly, the analogue of
$\mathrm{g}^{-1}\mathrm{g}=\mathrm{\varrho }_{g}$
is as follows%
\begin{equation}
\begin{tabular}{lll}
$\mathfrak{Q}_{m}^{G}\overset{\mathcal{M}_{_{n,m}}^{-1}}{\longrightarrow }%
\mathfrak{Q}_{n}^{G}\overset{\mathcal{M}_{_{n,m}}}{\longrightarrow }%
\mathfrak{Q}_{m}^{G}$ & $=$ & $\left( \mathfrak{Q}_{m}^{G}\overset{\mathcal{M%
}_{_{m,m}}}{\longrightarrow }\mathfrak{Q}_{m}^{G}\right) $%
\end{tabular}%
\end{equation}%
These relations lead respectively to%
\begin{equation}
\begin{tabular}{ll}
$\mathcal{M}_{_{n,m}}\mathcal{M}_{_{n,m}}^{-1}$ & $=\mathcal{M}_{_{n,m}}%
\mathcal{M}_{_{m,n}}=\mathcal{M}_{_{n,n}}$ \\
&  \\
$\mathcal{M}_{_{n,m}}^{-1}\mathcal{M}_{_{n,m}}$ & $=\mathcal{M}_{_{m,n}}%
\mathcal{M}_{_{n,m}}=\mathcal{M}_{_{m,m}}$%
\end{tabular}%
\end{equation}%
The properties (a)-(e) show that the set of BPS quiver mutations $\mathcal{G}%
_{Mut}^{G}$ form indeed a groupoid.

\subsubsection{Examples}

We end this study by giving the arrow representation of strong and
weak
orbits of the quiver mutations for two examples: (\textbf{i}) \emph{SU}$%
\left( 2\right) $ BPS theory and (\textbf{ii}) the \emph{SU(3)}
quiver theory.

\  \

\begin{figure}[tbph]
\begin{center}
\hspace{0cm} \includegraphics[width=10cm]{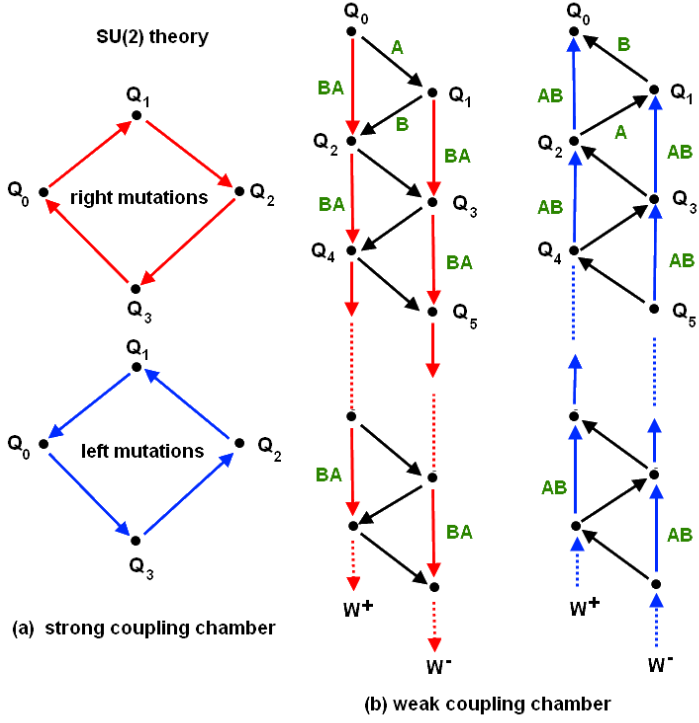}
\end{center}
\par
\vspace{-1 cm} \caption{groupoid orbits of quiver mutations in
SU$\left( 2\right) $ BPS quiver theory. (a) We have represented both
left and right mutations in the finite strong coupling BPS chamber
from which we can learn directly the inverse of mutations, the orbit
is closed and so corresponds to a periodic cycle. (b) left and right
mutations of the infinite weak coupling BPS chamber. Right mutations
are the inverse of left ones. $W^{\pm }$ refer to the quiver of
gauge bosons corresponding to alternate singular limits of the
mutations.} \label{sw}
\end{figure}

\emph{SU}$\left( 2\right) $\emph{\ case: strong and weak coupling chambers}%
\newline
In the arrow representation, the successive quiver mutations of the $%
\mathfrak{Q}_{k}^{su_{2}}$'s for both the strong and weak coupling
chambers of SU$\left( 2\right) $ are as depicted in fig \ref{sw}.
the strong coupling chamber is represented by a closed loop while
the weak one is given by open sequences that culminate at two
singular limits describing the two gauge particles $W^{\pm }$ as
depicted in the figure. The divergence of the limit $k\rightarrow
\infty $ is due to the alternate
property of the sequences (\ref{LR}) namely%
\begin{equation}
\mathcal{L}_{m}=\left( \mathcal{L}_{2k},\mathcal{L}_{2k+1}\right)
,\qquad \mathcal{R}_{m}=\left(
\mathcal{R}_{2k},\mathcal{R}_{2k+1}\right)
\end{equation}%
obeying%
\begin{equation}
\begin{tabular}{ll}
$\det \mathcal{L}_{2k}$ & $=\det \mathcal{R}_{2k}=+1\ $ \\
$\det \mathcal{L}_{2k+1}$ & $=\det \mathcal{R}_{2k+1}=-1\ $%
\end{tabular}%
\end{equation}%
and%
\begin{equation}
\begin{tabular}{llll}
$A:$ & $\mathcal{L}_{2k}$ & $\rightarrow $ & $\mathcal{L}_{2k+1}$ \\
$B:$ & $\mathcal{R}_{2k}$ & $\rightarrow $ & $\mathcal{R}_{2k+1}$%
\end{tabular}%
\end{equation}%
with $\det A=\det B=-1.$

\  \  \  \  \

\begin{figure}[tbph]
\begin{center}
\hspace{0cm} \includegraphics[width=14cm]{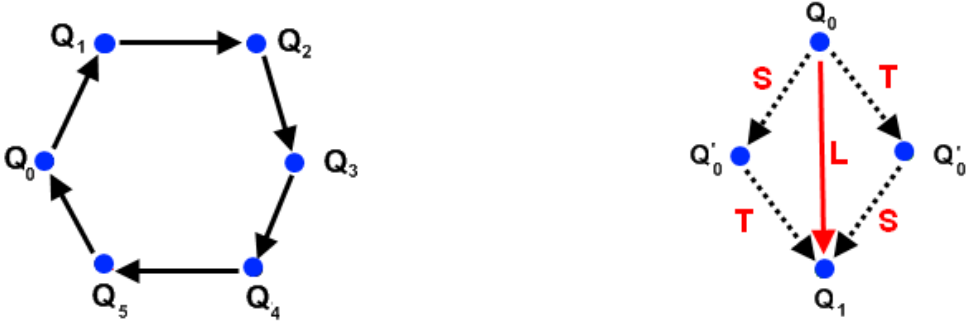}
\end{center}
\par
\vspace{-1 cm} \caption{(a) groupoid orbit of mutations in the
strong coupling chamber with isotropy group
$\mathcal{G}_{strong}^{su_{3}}$. (b) arrow representation of
eqs(\protect \ref{ll}-\protect \ref{mm}) describing the commuting reflections $%
s_{i}t_{i}=t_{i}s_{i}=L_{i}.$} \label{orb}
\end{figure}

\emph{SU}$\left( 3\right) $\emph{\ case: strong coupling
chamber}\newline The arrow representation of the $SU\left( 3\right)
$ BPS quiver theory is very rich; it involves 4 basic reflections
$s_{1},s_{2}$ and $t_{1},t_{2}$
whose matrix realisations are as in eq(\ref{st}). Because of the property $%
s_{i}t_{i}=t_{i}s_{i}$, we have considered their composition
\begin{equation}
L_{i}=s_{i}t_{i}
\end{equation}%
to build the strong coupling BPS chamber whose sequence of arrows is
given by fig \ref{orb}.
This closure of the arrow flows extends to the strong coupling BPS
chambers of all $\mathcal{N}=2$ supersymmetric QFTs with ADE gauge
symmetries.

\  \  \  \

\begin{figure}[tbph]
\begin{center}
\hspace{0cm} \includegraphics[width=12cm]{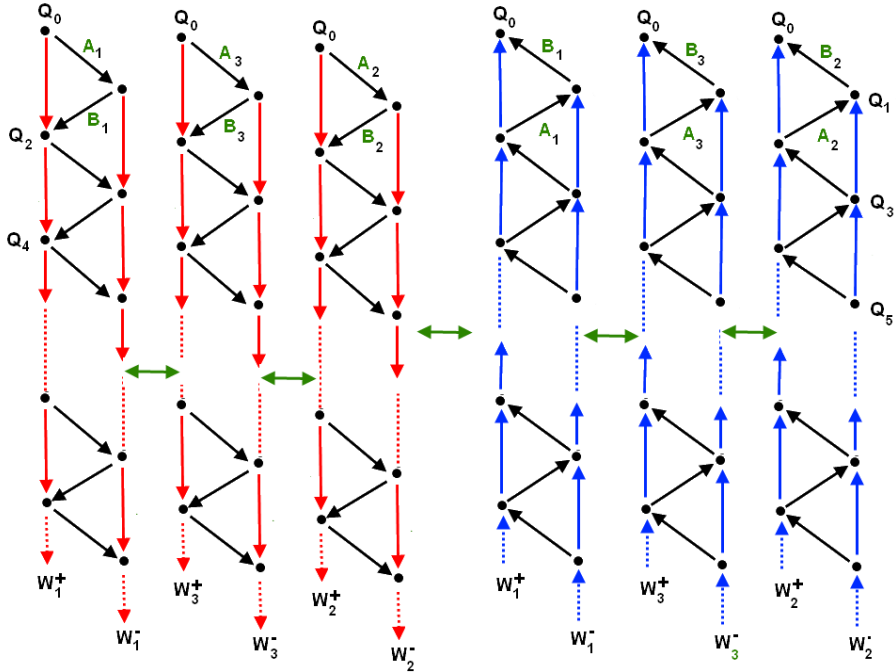}
\end{center}
\par
\vspace{-1 cm} \caption{Weak coupling chamber of SU$\left( 3\right)
$ model: 6 sequences of arrows in $\mathcal{G}_{weak}^{su_{3}}$
representing left and right mutations in SU$\left( 3\right) $ BPS
quiver theory; each sub-chamber, made of the sequence
$A_{i}B_{i}...A_{i}B_{i}$ and inverses, is isomorphic to the SU(2)
chamber of fig \protect \ref{sw}. These sequences tend towards 6
singular limits corresponding to the 6 massive gauge bosons of the
SU$\left( 3\right) $ BPS quiver theory. Dahed cross arrows between
the linear sequences are to indicate the passage between
subchambers.} \label{ro}
\end{figure}

\emph{SU}$\left( 3\right) $\emph{\ case: weak coupling
chambers}\newline In the weak coupling chamber of the SU$\left(
3\right) $ theory, the sequences of arrows are open and are much
more complex than the ones of the SU$\left( 2\right) $ theory. These
sequences involve infinite sub-chambers
that culminate at \emph{6} singular limits%
\begin{equation}
\begin{tabular}{lllll}
$W_{\alpha _{1}}^{\pm }$ & , & $W_{\alpha _{2}}^{\pm }$ & , &
$W_{\alpha
_{2}}^{\pm }$%
\end{tabular}
\label{ww}
\end{equation}%
describing the 6 massive gauge bosons of the supersymmetric
$SU\left( 3\right) $ gauge model. In fig \ref{ro}, we give those
sequences of arrows describing mutations in BPS quiver theory that
lead to (\ref{ww}).
The alternate sequences of this figure are given by%
\begin{equation}
\mathcal{L}_{m}^{\left( \alpha \right) }=\left(
\mathcal{L}_{2k}^{\left( \alpha \right) },\mathcal{L}_{2k+1}^{\left(
\alpha \right) }\right) ,\qquad \mathcal{R}_{m}^{\left( \alpha
\right) }=\left( \mathcal{R}_{2k}^{\left( \alpha \right)
},\mathcal{R}_{2k+1}^{\left( \alpha \right) }\right)
\end{equation}%
obeying%
\begin{equation}
\begin{tabular}{ll}
$\det \mathcal{L}_{2k}^{\left( \alpha \right) }$ & $=\det \mathcal{R}%
_{2k}^{\left( \alpha \right) }=+1\ $ \\
$\det \mathcal{L}_{2k+1}^{\left( \alpha \right) }$ & $=\det \mathcal{R}%
_{2k+1}^{\left( \alpha \right) }=-1\ $%
\end{tabular}%
\end{equation}%
with%
\begin{equation}
\begin{tabular}{llll}
$A_{\alpha }:$ & $\mathcal{L}_{2k}^{\left( \alpha \right) }$ &
$\rightarrow $
& $\mathcal{L}_{2k+1}^{\left( \alpha \right) }$ \\
$B_{\alpha }:$ & $\mathcal{R}_{2k}$ & $\rightarrow $ & $\mathcal{R}%
_{2k+1}^{\left( \alpha \right) }$%
\end{tabular}%
\end{equation}%
and $\det A_{\alpha }=\det B_{\alpha }=-1$ and where $\alpha $ a
positive root of $SU\left( 3\right) .$

\section{Appendix II: Extension to generic $SU\left( N\right) $ models}

In this appendix, we extend the analysis performed in section 4 to $\mathcal{%
N}=2$ supersymmetric QFTs with spontaneously broken $SU\left(
N\right) $ gauge invariance down to $U^{N-1}\left( 1\right) $.\

\subsection{Strong coupling chamber in $SU\left( N\right) $ models}

We start by recalling that the number $\#_{bps}$ of BPS states in
the strong coupling chamber of the $\mathcal{N}=2$ supersymmetric
pure $SU\left( N\right) $ quantum field theory is given by the
dimension of $SU\left(
N\right) $ minus its rank $r$%
\begin{equation}
\#_{bps}=\left( N^{2}-1\right) -N-1
\end{equation}%
This finite chamber includes the $\left( N-1\right) $ elementary monopoles $%
\mathfrak{M}_{i}$; and the $\left( N-1\right) $ elementary dyons $\mathfrak{D%
}_{i}$ as well as composite ones. The $\mathfrak{M}_{i}$ and $\mathfrak{D}%
_{i}$\ particles have respective complex central charges $X_{i}$, $Y_{i}\ $%
and EM charge vectors $b_{i},$ $c_{i}$. The latters are
2r-dimensional vectors which read in terms of the simple roots
$\alpha _{i}$ of $SU\left(
N\right) $ as follows%
\begin{equation}
\begin{tabular}{lll}
$b_{i}=\left(
\begin{array}{c}
0 \\
\alpha _{i}%
\end{array}%
\right) $, & $c_{i}=\left(
\begin{array}{c}
\alpha _{i} \\
-\alpha _{i}%
\end{array}%
\right) ,$ & $i=1,...,r$%
\end{tabular}%
\end{equation}%
The strong coupling chamber contains also an equal number of
anti-BPS states with negative EM charge vectors $-b_{i}$ and
$-c_{i}$. So the total number
of BPS and anti-BPS states is $2N\left( N-1\right) $. By using the rank $%
r=N-1$ of the $SU\left( N\right) $ gauge symmetry, the total number
of BPS states can be put in the form
\begin{equation}
\#_{bps}+\#_{anti-bps}=2r\left( r+1\right)
\end{equation}%
In what follows, we want to show that this number can be also
interpreted in terms of the order of the mutation group
$\mathcal{G}_{strong}^{su_{r+1}}$
like%
\begin{equation}
\#_{bps}+\#_{anti-bps}=r\times \left \vert \mathcal{G}_{strong}^{su_{r+1}}%
\right \vert  \label{RO}
\end{equation}%
with
\begin{equation}
\left \vert \mathcal{G}_{strong}^{su_{r+1}}\right \vert =2\left(
r+1\right) \equiv \mathrm{g}  \label{OR}
\end{equation}%
Recall that in the cases of leading $SU\left( r+1\right) $ gauge
symmetries considered in previous sections, we have obtained the
results
\begin{equation}
\begin{tabular}{l|llllll}
& \  \  \  & $SU_{2}$ &  & $SU_{3}$ \  \  \  &  & $SU_{4}$ \\ \hline
$\  \ r\left.
\begin{array}{c}
\\
\end{array}%
\right. $ & $\  \ $ & $\  \ 1$ &  & $\  \ 2$ &  & $\  \ 3$ \\
$\  \  \mathrm{g}\left.
\begin{array}{c}
\\
\end{array}%
\right. $ & $\  \ $ & $\  \ 4$ &  & $\  \ 6$ &  & $\  \ 8$ \\
$r\times \mathrm{g}$ \  \  \ $\left.
\begin{array}{c}
\\
\end{array}%
\right. $ &  & $\  \ 4$ &  & $\ 12$ &  & $\ 24$%
\end{tabular}%
\end{equation}%
which agree with eqs(\ref{OR}-\ref{RO}). Recall also that these
chambers correspond to the central charges of the elementary BPS
states with arguments $\arg X_{i},$ $\arg Y_{i}$ constrained as
follows
\begin{equation}
\begin{tabular}{llll}
$\arg Y_{i}$ & $=\arg Y$ & , &  \\
$\arg X_{i}$ & $=\arg X$ & , & $\forall i=1,...,r$%
\end{tabular}%
\end{equation}%
and%
\begin{equation}
\begin{tabular}{ll}
$\arg Y>\arg X$ & .%
\end{tabular}%
\end{equation}%
The derivation of the BPS states is obtained by performing quiver
mutations of the quiver $\mathfrak{Q}_{0}^{su_{N}}$ made by the
elementary BPS states. Extending the method we have used for
$SU\left( 2\right) $, $SU\left( 3\right) $ and $SU\left( 4\right) $
to generic N, the BPS and anti-BPS states are read from the rows of
the matrix mutations $\mathcal{M}_{n}$
mapping the quiver $\mathfrak{Q}_{0}^{su_{N}}$ into $\mathfrak{Q}%
_{n}^{su_{N}}$ with $0<n\leq 2N$. The mutation matrices
$\mathcal{M}_{n}$ are constructed below.

\subsection{Building $\mathcal{G}_{strong}^{su_{N}}$}

Here we build the discrete symmetry $\mathcal{G}_{strong}^{su_{N}}$
of mutation transformations interchanging the BPS quivers of the
strong coupling chamber of the $\mathcal{N}=2$ supersymmetric pure
$SU\left( N\right) $ quantum field theory.\newline
To that purpose, we use special properties of the two matrix generators $%
L_{1}$ and $L_{2}$ of the mutations. These generators are triangular
matrices of the form%
\begin{equation}
L_{1}=\left(
\begin{array}{cccccc}
I_{r} & \cdots & 0 &  &  &  \\
\vdots & \ddots & 0 &  & {\Large S} &  \\
0 & \cdots & I_{r} &  &  &  \\
&  &  & -I_{r} & \cdots & 0 \\
& {\Large 0} &  & \vdots & \ddots & \vdots \\
&  &  & 0 & \cdots & -I_{r}%
\end{array}%
\right)
\end{equation}%
and%
\begin{equation}
L_{2}=\left(
\begin{array}{cccccc}
-I_{r} & \cdots & 0 &  &  &  \\
\vdots & \ddots & 0 &  & {\Large 0} &  \\
0 & \cdots & -I_{r} &  &  &  \\
&  &  & I_{r} & \cdots & 0 \\
& {\Large S} &  & \vdots & \ddots & \vdots \\
&  &  & 0 & \cdots & I_{r}%
\end{array}%
\right)
\end{equation}%
with $I_{r}$ standing for the $r\times r$ identity matrix and the $%
r^{2}\times r^{2}$ matrix $S$ as follows%
\begin{equation}
S=\left(
\begin{array}{cccccc}
0 & I_{r} & 0 & \cdots &  & 0 \\
I_{r} & 0 & I_{r} &  &  & 0 \\
0 & I_{r} & 0 &  &  &  \\
\vdots &  &  & \ddots &  &  \\
0 &  &  &  &  & I_{r} \\
0 &  & \cdots &  & I_{r} & 0%
\end{array}%
\right)
\end{equation}%
These are non commuting matrices, $L_{1}L_{2}\neq L_{2}L_{1}$, obeying%
\begin{equation}
\left( L_{1}\right) ^{2}=\left( L_{2}\right) ^{2}=I_{id}
\end{equation}%
as well as identities extending eqs(\ref{B}-\ref{C}),
(\ref{C3}-\ref{3C})
and (\ref{C4}-\ref{4C})\ that can be put in the form%
\begin{equation}
\left \langle L_{1},L_{2}|\text{ }\left( L_{i}L_{j}\right)
^{m_{ij}}=I_{id}\left.
\begin{array}{c}
\\
\\
\end{array}%
\right. \right \rangle
\end{equation}%
with $m_{ij}$ the entries of the matrix%
\begin{equation}
M=\left(
\begin{array}{cc}
1 & N \\
N & 1%
\end{array}%
\right)
\end{equation}%
Using the expression of the $L_{1}$ and $L_{2}$ matrices, one can
explicitly
check, after a lengthy but straightforward calculations, that the set $%
\mathcal{G}_{strong}^{su_{N}}$ is a non abelian symmetry with $2N$
elements.
In the case $N=2K$ for instance, the group elements are given by%
\begin{equation}
\mathcal{G}_{strong}^{su_{N}}=\left \{
\begin{tabular}{lll}
$I_{id};$ &  &  \\
$\mathcal{M}_{2},\mathcal{M}_{4},...,\mathcal{M}_{2N-2}$ & $;$ & $\mathcal{M}%
_{1},\mathcal{M}_{3},...,\mathcal{M}_{2N-1}$%
\end{tabular}%
\right \}
\end{equation}%
or equivalently%
\begin{equation}
\mathcal{G}_{strong}^{su_{N}}=\left \{
\begin{tabular}{lll}
$I_{id};$ &  &  \\
$\mathcal{N}_{2},\mathcal{N}_{4},...,\mathcal{N}_{2N-2}$ & $;$ & $\mathcal{N}%
_{1},\mathcal{N}_{3},...,\mathcal{N}_{2N-1}$%
\end{tabular}%
\right \}
\end{equation}%
where we have set%
\begin{equation}
\begin{tabular}{ll}
$\mathcal{M}_{2k}$ & $=\left( L_{2}L_{1}\right) ^{k}$ \\
$\mathcal{N}_{2k}$ & $=\left( L_{1}L_{2}\right) ^{k}$%
\end{tabular}%
\end{equation}%
and%
\begin{equation}
\begin{tabular}{ll}
$\mathcal{M}_{2k+1}$ & $=L_{1}\mathcal{M}_{2k}$ \\
$\mathcal{N}_{2k+1}$ & $=L_{2}\mathcal{N}_{2k}$%
\end{tabular}%
\end{equation}%
We also have the identities%
\begin{equation}
\begin{tabular}{lll}
$\mathcal{M}_{K}=\mathcal{N}_{K}$ & , & $\mathcal{M}_{2N\pm n}=\mathcal{N}%
_{2N\mp n}$%
\end{tabular}%
\end{equation}%
and the periodic properties%
\begin{equation}
\begin{tabular}{ll}
$\mathcal{M}_{0}=\mathcal{M}_{2N}$ & $=I_{id}$ \\
$\mathcal{N}_{0}=\mathcal{N}_{2N}$ & $=I_{id}$%
\end{tabular}%
\end{equation}%
Notice that the $\mathcal{M}_{n}$ and $\mathcal{N}_{m}$ mutation
matrices satisfy as well as
\begin{equation}
\begin{tabular}{ll}
$\mathcal{M}_{n}\mathcal{M}_{m}$ & $=\mathcal{M}_{n+m}$ \\
$\mathcal{N}_{n}\mathcal{N}_{m}$ & $=\mathcal{N}_{n+m}$%
\end{tabular}%
\end{equation}%
and%
\begin{equation}
\mathcal{M}_{n}\mathcal{N}_{m}=\left \{
\begin{array}{ccc}
\mathcal{M}_{n-m} & \text{if } & n>m \\
I_{id} & \text{if } & n=m \\
\mathcal{N}_{m-n} & \text{if } & n<m%
\end{array}%
\right.
\end{equation}%
From these relations, we learn that the subset $\left \{ \mathcal{M}_{2n},%
\mathcal{N}_{2n}\right \} $ with $0\leq n\leq N$ of $\mathcal{G}%
_{strong}^{su_{N}}$ form a discrete abelian subsymmetry isomorphic to $%
\mathbb{Z}_{N}$.

\section{Appendix III: Quiver mutations in $E_{6}$ model}

In this appendix, we give the explicit expressions of the mutation matrices $%
\mathcal{M}_{n}$ relating the various BPS quivers
$\mathfrak{Q}_{n}^{E_{6}}$ of the strong coupling chamber of the 4D
space time $\mathcal{N}=2$
supersymmetric E$_{6}$ gauge theory spontaneously broken down to $%
U^{6}\left( 1\right) $. From these matrices $\mathcal{M}_{n}$, one
learns
directly the EM charge vectors%
\begin{equation}
\begin{tabular}{lll}
$b_{1}^{\left( n\right) },...,b_{6}^{\left( n\right) }$ & , &
$c_{1}^{\left(
n\right) },...,c_{6}^{\left( n\right) }$%
\end{tabular}%
\end{equation}%
of the BPS states and their anti-BPS partners as the rows of the $\mathcal{M}%
_{n}$'s. One also learns the intersection matrices by help of the formula%
\begin{equation}
\begin{tabular}{ll}
$\mathcal{A}^{\left( n\right) }$ &
$=\mathcal{M}_{n}\mathcal{A}^{\left(
0\right) }\mathcal{M}_{n}^{T}$%
\end{tabular}%
\end{equation}%
From the results of section 5, it follows that the $\mathcal{M}_{n}$
mutations are $144\times 144$ matrices generated by $L_{1}$ and
$L_{2}$
given by%
\begin{equation}
L_{1}=\left(
\begin{array}{cccccccccccc}
{\small I} & {\small 0} & {\small 0} & {\small 0} & {\small 0} &
{\small 0} & {\small 0} & {\small I} & {\small 0} & {\small 0} &
{\small 0} & {\small 0}
\\
{\small 0} & {\small I} & {\small 0} & {\small 0} & {\small 0} &
{\small 0} & {\small I} & {\small 0} & {\small I} & {\small 0} &
{\small 0} & {\small 0}
\\
{\small 0} & {\small 0} & {\small I} & {\small 0} & {\small 0} &
{\small 0} & {\small 0} & {\small I} & {\small 0} & {\small I} &
{\small I} & {\small 0}
\\
{\small 0} & {\small 0} & {\small 0} & {\small I} & {\small 0} &
{\small 0} & {\small 0} & {\small 0} & {\small I} & {\small 0} &
{\small 0} & {\small 0}
\\
{\small 0} & {\small 0} & {\small 0} & {\small 0} & {\small I} &
{\small 0} & {\small 0} & {\small 0} & {\small I} & {\small 0} &
{\small 0} & {\small I}
\\
{\small 0} & {\small 0} & {\small 0} & {\small 0} & {\small 0} &
{\small I} & {\small 0} & {\small 0} & {\small 0} & {\small 0} &
{\small I} & {\small 0}
\\
{\small 0} & {\small 0} & {\small 0} & {\small 0} & {\small 0} &
{\small 0}
& {\small -I} & {\small 0} & {\small 0} & {\small 0} & {\small 0} & {\small 0%
} \\
{\small 0} & {\small 0} & {\small 0} & {\small 0} & {\small 0} &
{\small 0}
& {\small 0} & {\small -I} & {\small 0} & {\small 0} & {\small 0} & {\small 0%
} \\
{\small 0} & {\small 0} & {\small 0} & {\small 0} & {\small 0} &
{\small 0}
& {\small 0} & {\small 0} & {\small -I} & {\small 0} & {\small 0} & {\small 0%
} \\
{\small 0} & {\small 0} & {\small 0} & {\small 0} & {\small 0} &
{\small 0}
& {\small 0} & {\small 0} & {\small 0} & {\small -I} & {\small 0} & {\small 0%
} \\
{\small 0} & {\small 0} & {\small 0} & {\small 0} & {\small 0} &
{\small 0}
& {\small 0} & {\small 0} & {\small 0} & {\small 0} & {\small -I} & {\small 0%
} \\
{\small 0} & {\small 0} & {\small 0} & {\small 0} & {\small 0} &
{\small 0}
& {\small 0} & {\small 0} & {\small 0} & {\small 0} & {\small 0} & {\small -I%
}%
\end{array}%
\right)  \label{L}
\end{equation}%
and%
\begin{eqnarray}
L_{2} &=&\left(
\begin{array}{cccccccccccc}
{\small -I} & {\small 0} & {\small 0} & {\small 0} & {\small 0} &
{\small 0} & {\small 0} & {\small 0} & {\small 0} & {\small 0} &
{\small 0} & {\small 0}
\\
{\small 0} & {\small -I} & {\small 0} & {\small 0} & {\small 0} &
{\small 0} & {\small 0} & {\small 0} & {\small 0} & {\small 0} &
{\small 0} & {\small 0}
\\
{\small 0} & {\small 0} & {\small -I} & {\small 0} & {\small 0} &
{\small 0} & {\small 0} & {\small 0} & {\small 0} & {\small 0} &
{\small 0} & {\small 0}
\\
{\small 0} & {\small 0} & {\small 0} & {\small -I} & {\small 0} &
{\small 0} & {\small 0} & {\small 0} & {\small 0} & {\small 0} &
{\small 0} & {\small 0}
\\
{\small 0} & {\small 0} & {\small 0} & {\small 0} & {\small -I} &
{\small 0} & {\small 0} & {\small 0} & {\small 0} & {\small 0} &
{\small 0} & {\small 0}
\\
{\small 0} & {\small 0} & {\small 0} & {\small 0} & {\small 0} &
{\small -I} & {\small 0} & {\small 0} & {\small 0} & {\small 0} &
{\small 0} & {\small 0}
\\
{\small 0} & {\small I} & {\small 0} & {\small 0} & {\small 0} &
{\small 0} & {\small I} & {\small 0} & {\small 0} & {\small 0} &
{\small 0} & {\small 0}
\\
{\small I} & {\small 0} & {\small I} & {\small 0} & {\small 0} &
{\small 0} & {\small 0} & {\small I} & {\small 0} & {\small 0} &
{\small 0} & {\small 0}
\\
{\small 0} & {\small I} & {\small 0} & {\small I} & {\small I} &
{\small 0} & {\small 0} & {\small 0} & {\small I} & {\small 0} &
{\small 0} & {\small 0}
\\
{\small 0} & {\small 0} & {\small I} & {\small 0} & {\small 0} &
{\small 0} & {\small 0} & {\small 0} & {\small 0} & {\small I} &
{\small 0} & {\small 0}
\\
{\small 0} & {\small 0} & {\small I} & {\small 0} & {\small 0} &
{\small I} & {\small 0} & {\small 0} & {\small 0} & {\small 0} &
{\small I} & {\small 0}
\\
{\small 0} & {\small 0} & {\small 0} & {\small 0} & {\small I} &
{\small 0}
& {\small 0} & {\small 0} & {\small 0} & {\small 0} & {\small 0} & {\small I}%
\end{array}%
\right)  \label{2L} \\
&&  \notag
\end{eqnarray}%
where each entry stands for a diagonal $12\times 12$ matrix: that is $%
0=0_{12\times 12}$ and $I=I_{12\times 12}$. From these matrices, we
can compute the set of mutations
\begin{equation}
\mathcal{G}_{M}^{E_{6}}=\left \{
\begin{tabular}{lll}
$\mathcal{M}_{2k}$ & $=\left( L_{2}L_{1}\right) ^{k}$ &  \\
$\mathcal{M}_{2k+1}$ & $=L_{1}\mathcal{M}_{2k}$ &
\end{tabular}%
;\text{ \ }0\leq k\leq 11\right \}
\end{equation}%
We construct below the explicit expressions of these matrices.

\begin{itemize}
\item \emph{\ matrix} $\mathcal{M}_{1}$ and \emph{quiver} $\mathfrak{Q}%
_{1}^{E_{6}}$

This matrix is equal to the generator $L_{1}$; then the EM charge
vectors of the BPS states of the quiver $\mathfrak{Q}_{1}^{E_{6}}$
are as follows:
\begin{equation}
\mathfrak{Q}_{1}^{E_{6}}:\left \{
\begin{tabular}{lll}
$b_{1}+c_{2},$ & $b_{2}+c_{1}+c_{3},$ & $b_{3}+c_{2}+c_{4}+c_{5},$ \\
$b_{4}+c_{3},$ & $b_{5}+c_{3}+c_{6},$ & $b_{6}+c_{5},$%
\end{tabular}%
\right \}
\end{equation}

\item \emph{\ the quiver} $\mathfrak{Q}_{2}^{E_{6}}$

the matrix $\mathcal{M}_{2}$ is given by $L_{2}L_{1}$ and reads
explicitly as
\begin{equation}
\mathcal{M}_{2}=\left(
\begin{array}{cccccccccccc}
-1 & 0 & 0 & 0 & 0 & 0 & 0 & -1 & 0 & 0 & 0 & 0 \\
0 & -1 & 0 & 0 & 0 & 0 & -1 & 0 & -1 & 0 & 0 & 0 \\
0 & 0 & -1 & 0 & 0 & 0 & 0 & -1 & 0 & -1 & -1 & 0 \\
0 & 0 & 0 & -1 & 0 & 0 & 0 & 0 & -1 & 0 & 0 & 0 \\
0 & 0 & 0 & 0 & -1 & 0 & 0 & 0 & -1 & 0 & 0 & -1 \\
0 & 0 & 0 & 0 & 0 & -1 & 0 & 0 & 0 & 0 & -1 & 0 \\
0 & 1 & 0 & 0 & 0 & 0 & 0 & 0 & 1 & 0 & 0 & 0 \\
1 & 0 & 1 & 0 & 0 & 0 & 0 & 1 & 0 & 1 & 1 & 0 \\
0 & 1 & 0 & 1 & 1 & 0 & 1 & 0 & 2 & 0 & 0 & 1 \\
0 & 0 & 1 & 0 & 0 & 0 & 0 & 1 & 0 & 0 & 1 & 0 \\
0 & 0 & 1 & 0 & 0 & 1 & 0 & 1 & 0 & 1 & 1 & 0 \\
0 & 0 & 0 & 0 & 1 & 0 & 0 & 0 & 1 & 0 & 0 & 0%
\end{array}%
\right)
\end{equation}%
The rows of this matrix give the EM $b_{i}^{\left( 2\right) }$ and $%
c_{i}^{\left( 2\right) }$ charge vectors of the BPS states of the quiver $%
\mathfrak{Q}_{2}^{E_{6}}$:%
\begin{equation}
\mathfrak{Q}_{2}^{E_{6}}:\left \{
\begin{tabular}{lll}
$b_{2}+c_{3},$ & $b_{1}+b_{3}+c_{2}+c_{4}+c_{5},$ & $%
b_{2}+b_{4}+b_{5}+c_{1}+2c_{3}+c_{6}$ \\
$b_{5}+c_{3},$ & $b_{3}+b_{6}+c_{2}+c_{4}+c_{5},$ & $b_{3}+c_{2}+c_{5}$%
\end{tabular}%
\right \}
\end{equation}%
This quiver has also 6 BPS states.

\item \emph{\ the quiver }$\mathfrak{Q}_{3}^{E_{6}}$

The mutation matrix $\mathcal{M}_{3}=L_{1}\mathcal{M}_{2}$ reads as:%
\begin{equation}
\mathcal{M}_{3}=\left(
\begin{array}{cccccccccccc}
0 & 0 & 1 & 0 & 0 & 0 & 0 & 0 & 0 & 1 & 1 & 0 \\
0 & 1 & 0 & 1 & 1 & 0 & 0 & 0 & 2 & 0 & 0 & 1 \\
1 & 0 & 2 & 0 & 0 & 1 & 0 & 2 & 0 & 1 & 2 & 0 \\
0 & 1 & 0 & 0 & 1 & 0 & 1 & 0 & 1 & 0 & 0 & 1 \\
0 & 1 & 0 & 1 & 1 & 0 & 1 & 0 & 2 & 0 & 0 & 0 \\
0 & 0 & 1 & 0 & 0 & 0 & 0 & 1 & 0 & 1 & 0 & 0 \\
0 & -1 & 0 & 0 & 0 & 0 & 0 & 0 & -1 & 0 & 0 & 0 \\
-1 & 0 & -1 & 0 & 0 & 0 & 0 & -1 & 0 & -1 & -1 & 0 \\
0 & -1 & 0 & -1 & -1 & 0 & -1 & 0 & -2 & 0 & 0 & -1 \\
0 & 0 & -1 & 0 & 0 & 0 & 0 & -1 & 0 & 0 & -1 & 0 \\
0 & 0 & -1 & 0 & 0 & -1 & 0 & -1 & 0 & -1 & -1 & 0 \\
0 & 0 & 0 & 0 & -1 & 0 & 0 & 0 & -1 & 0 & 0 & 0%
\end{array}%
\right)
\end{equation}%
The BPS states are as follows%
\begin{equation}
\mathfrak{Q}_{3}^{E_{6}}:\left \{
\begin{tabular}{lll}
$b_{3}+c_{4}+c_{5},$ & $b_{2}+b_{4}+b_{5}+2c_{3}+c_{6},$ & $%
b_{1}+2b_{3}+b_{6}+2c_{2}+c_{4}+2c_{5}$ \\
$b_{3}+c_{2}+c_{4},$ & $b_{2}+b_{4}+b_{5}+c_{1}+2c_{3},$ & $%
b_{2}+b_{5}+c_{1}+c_{3}+c_{6}$%
\end{tabular}%
\right \}
\end{equation}%
and there are also 6 new BPS states.

\item \emph{The matrix} $\mathcal{M}_{4}$

This matrix is given by
\begin{equation}
\mathcal{M}_{4}=\left(
\begin{array}{cccccccccccc}
0 & 0 & -1 & 0 & 0 & 0 & 0 & 0 & 0 & -1 & -1 & 0 \\
0 & -1 & 0 & -1 & -1 & 0 & 0 & 0 & -2 & 0 & 0 & -1 \\
-1 & 0 & -2 & 0 & 0 & -1 & 0 & -2 & 0 & -1 & -2 & 0 \\
0 & -1 & 0 & 0 & -1 & 0 & -1 & 0 & -1 & 0 & 0 & -1 \\
0 & -1 & 0 & -1 & -1 & 0 & -1 & 0 & -2 & 0 & 0 & 0 \\
0 & 0 & -1 & 0 & 0 & 0 & 0 & -1 & 0 & -1 & 0 & 0 \\
0 & 0 & 0 & 1 & 1 & 0 & 0 & 0 & 1 & 0 & 0 & 1 \\
0 & 0 & 2 & 0 & 0 & 1 & 0 & 1 & 0 & 1 & 2 & 0 \\
0 & 2 & 0 & 1 & 2 & 0 & 1 & 0 & 3 & 0 & 0 & 1 \\
1 & 0 & 1 & 0 & 0 & 1 & 0 & 1 & 0 & 1 & 1 & 0 \\
1 & 0 & 2 & 0 & 0 & 0 & 0 & 2 & 0 & 1 & 1 & 0 \\
0 & 1 & 0 & 1 & 0 & 0 & 1 & 0 & 1 & 0 & 0 & 0%
\end{array}%
\right)
\end{equation}%
from which we learn the following 6 BPS states%
\begin{equation}
\mathfrak{Q}_{4}^{E_{6}}:\left \{
\begin{tabular}{ll}
$b_{4}+b_{5}+c_{3}+c_{6},$ & $b_{2}+b_{4}+c_{1}+c_{3},$ \\
$2b_{3}+b_{6}+c_{2}+c_{4}+2c_{5},$ & $b_{1}+2b_{3}+2c_{2}+c_{4}+c_{5},$ \\
$2b_{2}+b_{4}+2b_{5}+c_{1}+3c_{3}+c_{6}$ & $%
b_{1}+b_{3}+b_{6}+c_{2}+c_{4}+c_{5}$%
\end{tabular}%
\right \}
\end{equation}

\item \emph{the matrix }$\mathcal{M}_{5}$

We have
\begin{equation}
\mathcal{M}_{5}=\left(
\begin{array}{cccccccccccc}
0 & 0 & 1 & 0 & 0 & 1 & 0 & 1 & 0 & 0 & 1 & 0 \\
0 & 1 & 0 & 1 & 2 & 0 & 1 & 0 & 2 & 0 & 0 & 1 \\
1 & 0 & 3 & 0 & 0 & 1 & 0 & 2 & 0 & 2 & 2 & 0 \\
0 & 1 & 0 & 1 & 1 & 0 & 0 & 0 & 2 & 0 & 0 & 0 \\
0 & 2 & 0 & 1 & 1 & 0 & 1 & 0 & 2 & 0 & 0 & 1 \\
1 & 0 & 1 & 0 & 0 & 0 & 0 & 1 & 0 & 0 & 1 & 0 \\
0 & 0 & 0 & -1 & -1 & 0 & 0 & 0 & -1 & 0 & 0 & -1 \\
0 & 0 & -2 & 0 & 0 & -1 & 0 & -1 & 0 & -1 & -2 & 0 \\
0 & -2 & 0 & -1 & -2 & 0 & -1 & 0 & -3 & 0 & 0 & -1 \\
-1 & 0 & -1 & 0 & 0 & -1 & 0 & -1 & 0 & -1 & -1 & 0 \\
-1 & 0 & -2 & 0 & 0 & 0 & 0 & -2 & 0 & -1 & -1 & 0 \\
0 & -1 & 0 & -1 & 0 & 0 & -1 & 0 & -1 & 0 & 0 & 0%
\end{array}%
\right)
\end{equation}%
leading to the following 6 new BPS states%
\begin{equation}
\mathfrak{Q}_{5}^{E_{6}}:\left \{
\begin{tabular}{ll}
$b_{3}+b_{6}+c_{2}+c_{5}\ ,$ & $b_{2}+b_{4}+2b_{5}+c_{1}+2c_{3}+c_{6}\ $ \\
$b_{2}+b_{4}+b_{5}+2c_{3}\ ,$ & $b_{1}+3b_{3}+b_{6}+2c_{2}+2c_{4}+2c_{5}$ \\
$b_{1}+b_{3}+c_{2}+c_{5}\ ,$ & $2b_{2}+b_{4}+b_{5}+c_{1}+2c_{3}+c_{6}$%
\end{tabular}%
\right \}
\end{equation}

\item \emph{the matrix }$\mathcal{M}_{6}$

This matrix reads as follows%
\begin{equation}
\mathcal{M}_{6}=\left(
\begin{array}{cccccccccccc}
0 & 0 & -1 & 0 & 0 & -1 & 0 & -1 & 0 & 0 & -1 & 0 \\
0 & -1 & 0 & -1 & -2 & 0 & -1 & 0 & -2 & 0 & 0 & -1 \\
-1 & 0 & -3 & 0 & 0 & -1 & 0 & -2 & 0 & -2 & -2 & 0 \\
0 & -1 & 0 & -1 & -1 & 0 & 0 & 0 & -2 & 0 & 0 & 0 \\
0 & -2 & 0 & -1 & -1 & 0 & -1 & 0 & -2 & 0 & 0 & -1 \\
-1 & 0 & -1 & 0 & 0 & 0 & 0 & -1 & 0 & 0 & -1 & 0 \\
0 & 1 & 0 & 0 & 1 & 0 & 1 & 0 & 1 & 0 & 0 & 0 \\
1 & 0 & 2 & 0 & 0 & 1 & 0 & 2 & 0 & 1 & 1 & 0 \\
0 & 2 & 0 & 2 & 2 & 0 & 1 & 0 & 3 & 0 & 0 & 1 \\
0 & 0 & 2 & 0 & 0 & 0 & 0 & 1 & 0 & 1 & 1 & 0 \\
1 & 0 & 2 & 0 & 0 & 1 & 0 & 1 & 0 & 1 & 2 & 0 \\
0 & 1 & 0 & 0 & 1 & 0 & 0 & 0 & 1 & 0 & 0 & 1%
\end{array}%
\right)
\end{equation}%
leading to the BPS spectrum%
\begin{equation}
\mathfrak{Q}_{6}^{E_{6}}:\left \{
\begin{tabular}{ll}
$b_{2}+b_{5}+c_{1}+c_{3},$ & $b_{1}+2b_{3}+b_{6}+2c_{2}+c_{5}\  \ $ \\
$2b_{3}+c_{2}+c_{4}+c_{5},$ &
$2b_{2}+2b_{4}+2b_{5}+c_{1}+c_{4}+3c_{3}+c_{6}$
\\
$b_{2}+b_{5}+c_{3}+c_{6},$ & $b_{1}+2b_{3}+b_{6}+c_{2}+c_{4}+2c_{5}$%
\end{tabular}%
\right \}
\end{equation}

\item \emph{the matrix }$\mathcal{M}_{7}$

This matrix is given by%
\begin{equation}
\mathcal{M}_{7}=\left(
\begin{array}{cccccccccccc}
1 & 0 & 1 & 0 & 0 & 0 & 0 & 1 & 0 & 1 & 0 & 0 \\
0 & 2 & 0 & 1 & 1 & 0 & 1 & 0 & 2 & 0 & 0 & 0 \\
1 & 0 & 3 & 0 & 0 & 1 & 0 & 2 & 0 & 1 & 2 & 0 \\
0 & 1 & 0 & 1 & 1 & 0 & 1 & 0 & 1 & 0 & 0 & 1 \\
0 & 1 & 0 & 1 & 2 & 0 & 0 & 0 & 2 & 0 & 0 & 1 \\
0 & 0 & 1 & 0 & 0 & 1 & 0 & 0 & 0 & 1 & 1 & 0 \\
0 & -1 & 0 & 0 & -1 & 0 & -1 & 0 & -1 & 0 & 0 & 0 \\
-1 & 0 & -2 & 0 & 0 & -1 & 0 & -2 & 0 & -1 & -1 & 0 \\
0 & -2 & 0 & -2 & -2 & 0 & -1 & 0 & -3 & 0 & 0 & -1 \\
0 & 0 & -2 & 0 & 0 & 0 & 0 & -1 & 0 & -1 & -1 & 0 \\
-1 & 0 & -2 & 0 & 0 & -1 & 0 & -1 & 0 & -1 & -2 & 0 \\
0 & -1 & 0 & 0 & -1 & 0 & 0 & 0 & -1 & 0 & 0 & -1%
\end{array}%
\right)
\end{equation}%
The corresponding BPS states are%
\begin{equation}
\mathfrak{Q}_{7}^{E_{6}}:\left \{
\begin{tabular}{ll}
$b_{1}+b_{3}+c_{2}+c_{4}\ ,$ & $2b_{2}+b_{4}+b_{5}+c_{1}+2c_{3}\ $ \\
$b_{2}+b_{4}+2b_{5}+2c_{3}+c_{6}\ ,$ & $%
b_{1}+3b_{3}+b_{6}+2c_{2}+c_{4}+2c_{5}$ \\
$b_{3}+b_{6}+c_{4}+c_{5}\ ,$ & $b_{2}+b_{4}+b_{5}+c_{1}+c_{3}+c_{6}$%
\end{tabular}%
\right \}
\end{equation}

\item \emph{the matrix }$\mathcal{M}_{8}$

We have%
\begin{equation}
\mathcal{M}_{8}=\left(
\begin{array}{cccccccccccc}
-1 & 0 & -1 & 0 & 0 & 0 & 0 & -1 & 0 & -1 & 0 & 0 \\
0 & -2 & 0 & -1 & -1 & 0 & -1 & 0 & -2 & 0 & 0 & 0 \\
-1 & 0 & -3 & 0 & 0 & -1 & 0 & -2 & 0 & -1 & -2 & 0 \\
0 & -1 & 0 & -1 & -1 & 0 & -1 & 0 & -1 & 0 & 0 & -1 \\
0 & -1 & 0 & -1 & -2 & 0 & 0 & 0 & -2 & 0 & 0 & -1 \\
0 & 0 & -1 & 0 & 0 & -1 & 0 & 0 & 0 & -1 & -1 & 0 \\
0 & 1 & 0 & 1 & 0 & 0 & 0 & 0 & 1 & 0 & 0 & 0 \\
1 & 0 & 2 & 0 & 0 & 0 & 0 & 1 & 0 & 1 & 1 & 0 \\
0 & 2 & 0 & 1 & 2 & 0 & 1 & 0 & 2 & 0 & 0 & 1 \\
1 & 0 & 1 & 0 & 0 & 1 & 0 & 1 & 0 & 0 & 1 & 0 \\
0 & 0 & 2 & 0 & 0 & 1 & 0 & 1 & 0 & 1 & 1 & 0 \\
0 & 0 & 0 & 1 & 1 & 0 & 0 & 0 & 1 & 0 & 0 & 0%
\end{array}%
\right)
\end{equation}%
leading to%
\begin{equation}
\mathfrak{Q}_{8}^{E_{6}}:\left \{
\begin{tabular}{ll}
$b_{2}+b_{4}+c_{3},$ & $b_{1}+2b_{3}+c_{2}+c_{4}+c_{5}$ \\
$b_{1}+b_{3}+b_{6}+c_{2}+c_{5},$ &
$2b_{2}+b_{4}+2b_{5}+c_{1}+2c_{3}+c_{6}$
\\
$b_{4}+b_{5}+c_{3},$ & $2b_{3}+b_{6}+c_{2}+c_{4}+c_{5}$%
\end{tabular}%
\right \}
\end{equation}

\item \emph{the matrix }$\mathcal{M}_{9}$

Here we have:

\begin{equation}
\mathcal{M}_{9}=\left(
\begin{array}{cccccccccccc}
{\small 0} & {\small 0} & {\small 1} & {\small 0} & {\small 0} &
{\small 0} & {\small 0} & {\small 0} & {\small 0} & {\small 0} &
{\small 1} & {\small 0}
\\
{\small 0} & {\small 1} & {\small 0} & {\small 1} & {\small 1} &
{\small 0} & {\small 0} & {\small 0} & {\small 1} & {\small 0} &
{\small 0} & {\small 1}
\\
{\small 1} & {\small 0} & {\small 2} & {\small 0} & {\small 0} &
{\small 1} & {\small 0} & {\small 1} & {\small 0} & {\small 1} &
{\small 1} & {\small 0}
\\
{\small 0} & {\small 1} & {\small 0} & {\small 0} & {\small 1} &
{\small 0} & {\small 0} & {\small 0} & {\small 1} & {\small 0} &
{\small 0} & {\small 0}
\\
{\small 0} & {\small 1} & {\small 0} & {\small 1} & {\small 1} &
{\small 0} & {\small 1} & {\small 0} & {\small 1} & {\small 0} &
{\small 0} & {\small 0}
\\
{\small 0} & {\small 0} & {\small 1} & {\small 0} & {\small 0} &
{\small 0} & {\small 0} & {\small 1} & {\small 0} & {\small 0} &
{\small 0} & {\small 0}
\\
{\small 0} & {\small -1} & {\small 0} & {\small -1} & {\small 0} &
{\small 0}
& {\small 0} & {\small 0} & {\small -1} & {\small 0} & {\small 0} & {\small 0%
} \\
{\small -1} & {\small 0} & {\small -2} & {\small 0} & {\small 0} &
{\small 0} & {\small 0} & {\small -1} & {\small 0} & {\small -1} &
{\small -1} &
{\small 0} \\
{\small 0} & {\small -2} & {\small 0} & {\small -1} & {\small -2} & {\small 0%
} & {\small -1} & {\small 0} & {\small -2} & {\small 0} & {\small 0}
&
{\small -1} \\
{\small -1} & {\small 0} & {\small -1} & {\small 0} & {\small 0} & {\small -1%
} & {\small 0} & {\small -1} & {\small 0} & {\small 0} & {\small -1}
&
{\small 0} \\
{\small 0} & {\small 0} & {\small -2} & {\small 0} & {\small 0} &
{\small -1} & {\small 0} & {\small -1} & {\small 0} & {\small -1} &
{\small -1} &
{\small 0} \\
{\small 0} & {\small 0} & {\small 0} & {\small -1} & {\small -1} &
{\small 0}
& {\small 0} & {\small 0} & {\small -1} & {\small 0} & {\small 0} & {\small 0%
}%
\end{array}%
\right)
\end{equation}%
giving the BPS spectrum%
\begin{equation}
\mathfrak{Q}_{9}^{E_{6}}:\left \{
\begin{tabular}{ll}
$b_{3}+c_{5},$ & $b_{2}+b_{4}+b_{5}+c_{3}+c_{6}\ $ \\
$b_{2}+b_{5}+c_{3},$ & $b_{1}+2b_{3}+b_{6}+c_{2}+c_{4}+c_{5}$ \\
$b_{3}+c_{2},$ & $b_{2}+b_{4}+b_{5}+c_{1}+c_{3}$%
\end{tabular}%
\right \}
\end{equation}

\item \emph{the matrix }$\mathcal{M}_{10}$

This matrix is given by%
\begin{equation}
\mathcal{M}_{10}=\left(
\begin{array}{cccccccccccc}
{\small 0} & {\small 0} & {\small -1} & {\small 0} & {\small 0} &
{\small 0}
& {\small 0} & {\small 0} & {\small 0} & {\small 0} & {\small -1} & {\small 0%
} \\
{\small 0} & {\small -1} & {\small 0} & {\small -1} & {\small -1} & {\small 0%
} & {\small 0} & {\small 0} & {\small -1} & {\small 0} & {\small 0}
&
{\small -1} \\
{\small -1} & {\small 0} & {\small -2} & {\small 0} & {\small 0} & {\small -1%
} & {\small 0} & {\small -1} & {\small 0} & {\small -1} & {\small
-1} &
{\small 0} \\
{\small 0} & {\small -1} & {\small 0} & {\small 0} & {\small -1} &
{\small 0}
& {\small 0} & {\small 0} & {\small -1} & {\small 0} & {\small 0} & {\small 0%
} \\
{\small 0} & {\small -1} & {\small 0} & {\small -1} & {\small -1} & {\small 0%
} & {\small -1} & {\small 0} & {\small -1} & {\small 0} & {\small 0}
&
{\small 0} \\
{\small 0} & {\small 0} & {\small -1} & {\small 0} & {\small 0} &
{\small 0}
& {\small 0} & {\small -1} & {\small 0} & {\small 0} & {\small 0} & {\small 0%
} \\
{\small 0} & {\small 0} & {\small 0} & {\small 0} & {\small 1} &
{\small 0} & {\small 0} & {\small 0} & {\small 0} & {\small 0} &
{\small 0} & {\small 1}
\\
{\small 0} & {\small 0} & {\small 1} & {\small 0} & {\small 0} &
{\small 1} & {\small 0} & {\small 0} & {\small 0} & {\small 0} &
{\small 1} & {\small 0}
\\
{\small 0} & {\small 1} & {\small 0} & {\small 1} & {\small 1} &
{\small 0} & {\small 0} & {\small 0} & {\small 1} & {\small 0} &
{\small 0} & {\small 0}
\\
{\small 0} & {\small 0} & {\small 1} & {\small 0} & {\small 0} &
{\small 0} & {\small 0} & {\small 0} & {\small 0} & {\small 1} &
{\small 0} & {\small 0}
\\
{\small 1} & {\small 0} & {\small 1} & {\small 0} & {\small 0} &
{\small 0} & {\small 0} & {\small 1} & {\small 0} & {\small 0} &
{\small 0} & {\small 0}
\\
{\small 0} & {\small 1} & {\small 0} & {\small 0} & {\small 0} &
{\small 0}
& {\small 1} & {\small 0} & {\small 0} & {\small 0} & {\small 0} & {\small 0}%
\end{array}%
\right)
\end{equation}%
The 6 new BPS states are as follows%
\begin{equation}
\mathfrak{Q}_{10}^{E_{6}}:\left \{
\begin{tabular}{ll}
$b_{5}+c_{6},$ & $b_{3}+b_{6}+c_{5}\ $ \\
$b_{3}+c_{4},$ & $b_{2}+b_{4}+b_{5}+c_{3}$ \\
$b_{2}+c_{1},$ & $b_{1}+b_{3}+c_{2}$%
\end{tabular}%
\right \}
\end{equation}

\item \emph{the matrix }$\mathcal{M}_{11}$

We have
\begin{equation}
\mathcal{M}_{11}=\left(
\begin{array}{cccccccccccc}
{\small 0} & {\small 0} & {\small 0} & {\small 0} & {\small 0} &
{\small 1} & {\small 0} & {\small 0} & {\small 0} & {\small 0} &
{\small 0} & {\small 0}
\\
{\small 0} & {\small 0} & {\small 0} & {\small 0} & {\small 1} &
{\small 0} & {\small 0} & {\small 0} & {\small 0} & {\small 0} &
{\small 0} & {\small 0}
\\
{\small 0} & {\small 0} & {\small 1} & {\small 0} & {\small 0} &
{\small 0} & {\small 0} & {\small 0} & {\small 0} & {\small 0} &
{\small 0} & {\small 0}
\\
{\small 0} & {\small 0} & {\small 0} & {\small 1} & {\small 0} &
{\small 0} & {\small 0} & {\small 0} & {\small 0} & {\small 0} &
{\small 0} & {\small 0}
\\
{\small 0} & {\small 1} & {\small 0} & {\small 0} & {\small 0} &
{\small 0} & {\small 0} & {\small 0} & {\small 0} & {\small 0} &
{\small 0} & {\small 0}
\\
{\small 1} & {\small 0} & {\small 0} & {\small 0} & {\small 0} &
{\small 0} & {\small 0} & {\small 0} & {\small 0} & {\small 0} &
{\small 0} & {\small 0}
\\
{\small 0} & {\small 0} & {\small 0} & {\small 0} & {\small -1} &
{\small 0}
& {\small 0} & {\small 0} & {\small 0} & {\small 0} & {\small 0} & {\small -1%
} \\
{\small 0} & {\small 0} & {\small -1} & {\small 0} & {\small 0} &
{\small -1}
& {\small 0} & {\small 0} & {\small 0} & {\small 0} & {\small -1} & {\small 0%
} \\
{\small 0} & {\small -1} & {\small 0} & {\small -1} & {\small -1} & {\small 0%
} & {\small 0} & {\small 0} & {\small -1} & {\small 0} & {\small 0}
&
{\small 0} \\
{\small 0} & {\small 0} & {\small -1} & {\small 0} & {\small 0} &
{\small 0}
& {\small 0} & {\small 0} & {\small 0} & {\small -1} & {\small 0} & {\small 0%
} \\
{\small -1} & {\small 0} & {\small -1} & {\small 0} & {\small 0} &
{\small 0}
& {\small 0} & {\small -1} & {\small 0} & {\small 0} & {\small 0} & {\small 0%
} \\
{\small 0} & {\small -1} & {\small 0} & {\small 0} & {\small 0} &
{\small 0}
& {\small -1} & {\small 0} & {\small 0} & {\small 0} & {\small 0} & {\small 0%
}%
\end{array}%
\right)
\end{equation}%
giving%
\begin{equation}
\mathfrak{Q}_{11}^{E_{6}}:\left \{
\begin{tabular}{ll}
$b_{6}\ ,$ & $b_{4}\ $ \\
$b_{5}\ ,$ & $b_{2}\ $ \\
$b_{3}\ ,$ & $b_{1}\ $%
\end{tabular}%
\right \}
\end{equation}%
which are not new BPS states as these are precisely the elementary
BPS\ states that appear in the quiver $\mathfrak{Q}_{0}$ given by
fig \ref{E6}.

\item \emph{the matrix }$\mathcal{M}_{12}$

This matrix reads as follows:%
\begin{equation}
\mathcal{M}_{12}=\left(
\begin{array}{cccccccccccc}
{\small 0} & {\small 0} & {\small 0} & {\small 0} & {\small 0} &
{\small -1} & {\small 0} & {\small 0} & {\small 0} & {\small 0} &
{\small 0} & {\small 0}
\\
{\small 0} & {\small 0} & {\small 0} & {\small 0} & {\small -1} &
{\small 0} & {\small 0} & {\small 0} & {\small 0} & {\small 0} &
{\small 0} & {\small 0}
\\
{\small 0} & {\small 0} & {\small -1} & {\small 0} & {\small 0} &
{\small 0} & {\small 0} & {\small 0} & {\small 0} & {\small 0} &
{\small 0} & {\small 0}
\\
{\small 0} & {\small 0} & {\small 0} & {\small -1} & {\small 0} &
{\small 0} & {\small 0} & {\small 0} & {\small 0} & {\small 0} &
{\small 0} & {\small 0}
\\
{\small 0} & {\small -1} & {\small 0} & {\small 0} & {\small 0} &
{\small 0} & {\small 0} & {\small 0} & {\small 0} & {\small 0} &
{\small 0} & {\small 0}
\\
{\small -1} & {\small 0} & {\small 0} & {\small 0} & {\small 0} &
{\small 0} & {\small 0} & {\small 0} & {\small 0} & {\small 0} &
{\small 0} & {\small 0}
\\
{\small 0} & {\small 0} & {\small 0} & {\small 0} & {\small 0} &
{\small 0}
& {\small 0} & {\small 0} & {\small 0} & {\small 0} & {\small 0} & {\small -1%
} \\
{\small 0} & {\small 0} & {\small 0} & {\small 0} & {\small 0} &
{\small 0}
& {\small 0} & {\small 0} & {\small 0} & {\small 0} & {\small -1} & {\small 0%
} \\
{\small 0} & {\small 0} & {\small 0} & {\small 0} & {\small 0} &
{\small 0}
& {\small 0} & {\small 0} & {\small -1} & {\small 0} & {\small 0} & {\small 0%
} \\
{\small 0} & {\small 0} & {\small 0} & {\small 0} & {\small 0} &
{\small 0}
& {\small 0} & {\small 0} & {\small 0} & {\small -1} & {\small 0} & {\small 0%
} \\
{\small 0} & {\small 0} & {\small 0} & {\small 0} & {\small 0} &
{\small 0}
& {\small 0} & {\small -1} & {\small 0} & {\small 0} & {\small 0} & {\small 0%
} \\
{\small 0} & {\small 0} & {\small 0} & {\small 0} & {\small 0} &
{\small 0}
& {\small -1} & {\small 0} & {\small 0} & {\small 0} & {\small 0} & {\small 0%
}%
\end{array}%
\right)
\end{equation}%
Notice that the entries of this matrix are either zero or negative
definite integers.\ The corresponding quiver
$\mathfrak{Q}_{12}^{E_{6}}$ involves
only anti-BPS states; and so should be viewed as the CPT conjugate of $%
\mathfrak{Q}_{0}^{E_{6}}$. Notice also the natural identity%
\begin{equation*}
\left( \mathcal{M}_{12}\right) ^{2}=I_{id}
\end{equation*}
\end{itemize}

\section{Appendix IV: Coxeter groups and Coxeter graphs}

In this appendix, we collect some useful tools on the building of
the Coxeter groups and their basic properties. We also give the
corresponding graphs and relation with the Dynkin diagrams of simple
Lie algebras.

\subsection{Coxeter groups}

A way to introduce a generic Coxeter group is by considering the two
following things: (\textbf{1}) a system on $n$ involutions $\left \{
s_{1},...,s_{n}\right \} $ satisfying $s_{i}^{2}=1$\ $\forall i$;
which has to be thought of as the group generators of the group.
(\textbf{2}) a symmetric $n\times n$ matrix $M$, known as the
Coxeter matrix, with integer
entries $\left( m_{ij}\right) $ constrained as%
\begin{equation}
m_{ii}=1,\qquad m_{ij}\geq 2\text{ \ for \ }i\neq j.
\end{equation}%
The Coxeter group associated with the $s_{i}$'s and the matrix $M$\
is generally denoted as $W\left( M\right) $ and called the Coxeter
group of
type $M$. This set is defined as:%
\begin{equation}
W\left( M\right) =\left \langle \left.
\begin{array}{cc}
s_{i}^{2} & =1 \\
\left( s_{i}s_{j}\right) ^{m_{ij}} & =1%
\end{array}%
\right. ;\text{ \ }m_{ij}\geq 2\right \rangle
\end{equation}%
Clearly according to the values of $m_{ij}$, the Coxeter groups
$W\left( M\right) $ may be finite or infinite discrete \ groups.
\newline Coxeter groups have several remarkable properties; one of
them is that giving the link between $M=\left( m_{ij}\right) $ and
the usual Cartan
matrix $K_{ij}$ of simple Lie algebras namely%
\begin{equation}
K_{ij}=-2\cos \frac{\pi }{m_{ij}}  \label{KC}
\end{equation}%
Besides the crucial role that play in approaching Lie algebras; this
identification is also important for classifying the Coxeter groups
$W\left( M\right) $ by using the determinant of the Cartan matrix
$K_{ij}$. We have \textrm{\cite{M,MM}}:

\begin{itemize}
\item finite $W\left( M\right) $ if $\det K>0$,

\item affine $W\left( M\right) $ if $\det K=0$,

\item hyperbolic $W\left( M\right) $ for $\det K<0$.
\end{itemize}

\  \  \  \newline Notice also that Coxeter groups are deeply
connected with reflection groups; including the Weyl group of simple
Lie algebras, and which can be thought of as a linear representation
of $W\left( M\right) $. Their group generators
given by invertible $n\times n$ matrices; and so form a a subgroup of $%
GL\left( n,R\right) $. To illustrate these kinds of discrete groups,
we give below three examples respectively concerning: the cyclic
group of order 2, the dihedral groups $Dih_{2m}$ with $m>2$; and the
groups $A_{n}$ associated with the $SU\left( n+1\right) $ gauge
symmetries.

\emph{Example 1: } \emph{cyclic group of order 2}\newline This is
the simplest Coxeter group with one generator $s$ and a Coxeter
matrix $M=\left( 1\right) $. Substituting, we get:
\begin{equation}
W\left( M\right) =\left \langle s|\text{ }s^{2}=1\right \rangle
\end{equation}%
which is the cyclic group $\mathbb{Z}_{2}$ of order 2. The elements
of this group are $\pm I_{id}$.

\emph{Example 2: dihedral group} $Dih_{2m}$\newline In this case, we
have two group generators, which we denote as $s$ and $t$,
and a Coxeter matrix M, denoted as $I_{2}\left( m\right) $, reads as follows,%
\begin{equation}
I_{2}\left( m\right) =\left(
\begin{array}{cc}
1 & m \\
m & 1%
\end{array}%
\right) ,\qquad m\geq 2
\end{equation}%
This group is precisely the dihedral group $Dih_{2m}$ given by,%
\begin{equation}
Dih_{2m}=\left \langle \left.
\begin{array}{c}
\\
\\
\end{array}%
\right. s,t|\text{ }s^{2}=t^{2}=1;\text{ }\left( st\right) ^{m}=1,\text{ }%
m\geq 2\right \rangle
\end{equation}%
Notice that depending on the choice of the integer $m$, we have
different groups.\ For the particular case $m=2$, the condition
$\left( st\right)
^{m}=\left( st\right) ^{2}=1$ leads as well to%
\begin{equation}
stst=1
\end{equation}%
This condition combined with $s^{2}=t^{2}=1$ leads to%
\begin{equation}
st=s\left( stst\right) t=ts
\end{equation}%
showing that the group $Dih_{4}$ is abelian. \newline In the case
$m\geq 3$, the corresponding group is non abelian; this property can
be obtained by using the involutions, we have $\left( st\right)
^{2}\neq 1$ which is also equal to $\left( st\right) ^{-1}$. The
term $\left( st\right) ^{2}$ can put as the group commutator
$sts^{-1}t^{-1}$ showing that $st\neq ts$.

\emph{Example 3}: \emph{groups} $A_{n}$\newline
We illustrate the construction of these groups on the cases A$_{3}$ and A$%
_{4}$ respectively associated with the $SU\left( 4\right) $ and
SU$\left( 5\right) $ gauge symmetries. The same thing is valid for
the general case. In the case of A$_{3}$, we have three generators
s, t, u \ and a Coxeter matrix M as
\begin{equation}
M=\left(
\begin{array}{ccc}
1 & 3 & 2 \\
3 & 1 & 3 \\
2 & 3 & 1%
\end{array}%
\right)
\end{equation}%
from which we read the group structure
\begin{equation}
A_{3}=\left \langle s,t,u\text{ \ }|\text{ }\left \{
\begin{array}{ccccc}
s^{2}=1 & , & t^{2}=1 & , & u^{2}=1 \\
\left( st\right) ^{3}=1 & , & \left( su\right) ^{2}=1 & , & \left(
tu\right)
^{3}=1%
\end{array}%
\right. \right \rangle
\end{equation}%
In the case of $SU\left( 5\right) $, the Coxeter matrix is given by%
\begin{equation}
M=\left(
\begin{array}{cccc}
1 & 3 & 2 & 2 \\
3 & 1 & 3 & 2 \\
2 & 3 & 1 & 3 \\
2 & 2 & 3 & 1%
\end{array}%
\right)
\end{equation}

\subsection{Coxeter diagrams}

\begin{figure}[tbph]
\begin{center}
\hspace{0cm} \includegraphics[width=12cm]{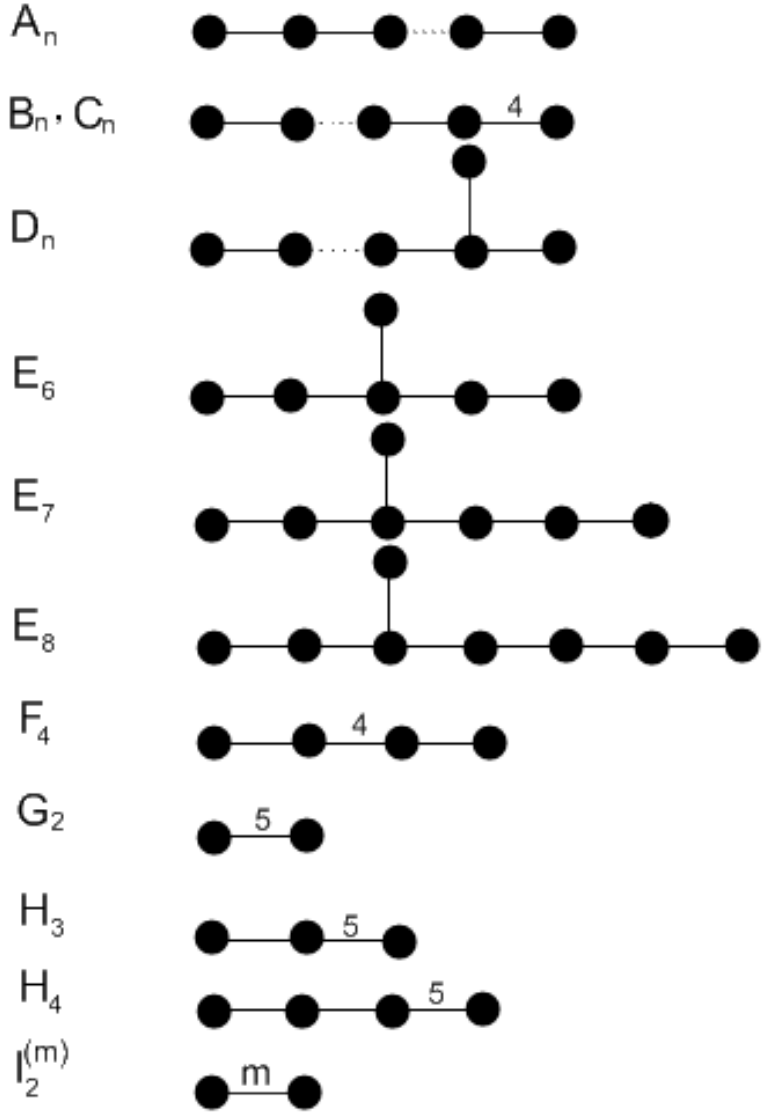}
\end{center}
\par
\vspace{-1 cm} \caption{Diagrams of irreducible finite Coxeter
systems; the diagrams $G_{2}$ and $I_{2}^{\left( 6\right) }$ are the
same. The diagrams $H_{2}$ and $B_{2}$ might be defined as
$I_{2}^{\left( 5\right) }$ and $I_{2}^{\left( 4\right) }$
respectively. The diagram $A_{2}$ coincides with $I_{2}^{\left(
3\right) }.$} \label{COX}
\end{figure}

To the symmetric $n\times n$ Coxeter matrix $M=\left( m_{ij}\right)
$, one associates a graph $\Gamma \left( M\right) $ having n
vertices in one to one with the generators of the group $W\left(
M\right) $. Two nodes $i$ and $j$ of the graph are joined by an edge
labeled $m_{ij}$ if $m_{ij}>2$. \newline For $m_{ij}=3$ the label 3
of the edge $\left \{ i,j\right \} $ is often omitted; and for
$m_{ij}=4$, one often draws a double bond instead of the label $4$
at the edge $\left \{ i,j\right \} $. \newline The data stored in
the Coxeter matrix $M$ can be reconstructed from the Coxeter
diagram, and so the Coxeter diagram and the Coxeter matrix can be
identified. \newline We end this appendix by the two following
things: First, the ADE Dynkin graphs of fig \ref{ADE} are particular
Coxeter diagrams; the relation between the two diagrams is given by
eq(\ref{KC}); see also fig \ref{COX}.
Second, the Coxeter number $h$ is the \emph{order} of a Coxeter
element of an irreducible Coxeter group. A Coxeter element of
$W\left( M\right) $ is
given by the product of all simple reflections. The number $h$ and its dual $%
\tilde{h}$ of finite dimensional Lie algebras are collected in the
following table
\begin{equation}
\begin{tabular}{l|l|l}
graphs & \  \ $\ h$ & $\  \  \tilde{h}$ \\ \hline
$A_{n}$ & $n$ & $n+1$ \\
$B_{n}$ & $2n$ & $2n-1$ \\
$C_{n}$ & $2n$ & $n+1$ \\
$D_{n}$ & $2n-2$ & $2n-2$ \\
$E_{6}$ & $12$ & $12$ \\
$E_{7}$ & $18$ & $18$ \\
$E_{8}$ & $30$ & $30$ \\
$F_{4}$ & $12$ & $9$ \\
$G_{2}$ & $6$ & $4$ \\ \hline
\end{tabular}%
\end{equation}

\end{document}